\documentclass[aps,prb,twocolumn,superscriptaddress]{revtex4-2}
\pdfoutput=1
\usepackage{diagbox}
\usepackage{tabularx}
\usepackage{booktabs}
\usepackage{makecell}
\usepackage{amsfonts}
\usepackage{amsmath}
\usepackage{graphicx}
\usepackage{dsfont}
\usepackage{diagbox}
\usepackage{array}
\usepackage{amssymb}
\usepackage{bbm}
\usepackage{float}
\usepackage{subfigure}
\usepackage{url}
\usepackage{hyperref}
\usepackage{array}
\usepackage{multirow}
\usepackage{bm}
\usepackage{booktabs}
\usepackage{bbm}
\usepackage{array}
\usepackage{color}
\usepackage[export]{adjustbox}
\usepackage{multirow}
\usepackage{makecell}
\usepackage{notes2bib}
\usepackage{float}
\usepackage{mathrsfs}
\usepackage{xcolor}
\usepackage{enumitem}
\newcommand{\PreserveBackslash}[1]{\let\temp=\\#1\let\\=\temp}
\newcolumntype{C}[1]{>{\PreserveBackslash\centering}p{#1}}
\newcolumntype{R}[1]{>{\PreserveBackslash\raggedleft}p{#1}}
\newcolumntype{L}[1]{>{\PreserveBackslash\raggedright}p{#1}}
\usepackage{color}

\begin{document}

\title{Complete characterization of non-Abelian topological phase transitions and detection of anyon splitting with projected entangled pair states}

\begin{abstract}
It is well-known that many topological phase transitions of intrinsic Abelian topological phases are accompanied by condensation and confinement of anyons. However, for non-Abelian topological phases, more intricate phenomena can occur at their phase transitions, because the multiple degenerate degrees of freedom of a non-Abelian anyon can change in different ways after phase transitions. In this paper, we study these new phenomena, including partial condensation, partial deconfinement and especially anyon splitting (a non-Abelian anyon splits into different kinds of the new anyon species) using projected entangled pair states (PEPS). First, we show that anyon splitting can be observed from the topologically degenerate ground states. Next, we construct a set of PEPS describing all possible degrees of freedom of the same non-Abelian anyon. From the overlaps of this set of PEPS of a given non-Abelian anyon with the ground state, we can extract the information of partial condensation. Then, we construct a central object, a matrix defined by the norms and overlaps among the PEPS in that set. The information of partial deconfinement can be extracted from this matrix. In particular, we use it to construct an order parameter which can directly detect anyon splitting.  We demonstrate the power of our approach by applying it to a range of non-Abelian topological phase transitions: From $D(S_3)$ quantum double to toric code,  from $D(S_3)$ quantum double to $D(Z_3)$ quantum double, from Rep$(S_3)$ string-net to toric code, and finally from double Ising string-net to toric code. 
\end{abstract}

\author{Wen-Tao Xu}
\affiliation{University of Vienna, Faculty of Physics, Boltzmanngasse 5, 1090 Wien, Austria}
\author{Jose Garre-Rubio}
\affiliation{University of Vienna, Faculty of Mathematics, Oskar-Morgenstern-Platz 1, 1090 Wien, Austria}
\author{Norbert Schuch}
\affiliation{University of Vienna, Faculty of Physics, Boltzmanngasse 5, 1090 Wien, Austria}
\affiliation{University of Vienna, Faculty of Mathematics, Oskar-Morgenstern-Platz 1, 1090 Wien, Austria}

\maketitle
\section{Introduction}
Since the discovery of the fractional quantum Hall effect\cite{FQHE,Laughlin_1983}, topological phases of matter have become an essential concept in modern condensed matter physics. A remarkable feature of topological phases is that they can support exotic excitations with fractional statistics: the so-called anyons. Importantly, non-Abelian anyons carry internal degrees of freedom (DOFs), which are essential to perform fault-tolerant quantum computation\cite{kitaev_toric_code_2003}. Recently, topological phases have been experimentally implemented on quantum processors\cite{TC_science_2021} and quantum simulators\cite{TC_science_2_2021}. Theoretically, topological phases can be realized by exactly solvable lattice models, such as the Kitaev's quantum double models\cite{kitaev_toric_code_2003} and the Levin-Wen string-net models\cite{Levin-Wen-string-net-2005}.

A powerful tool to study topologically ordered phases are projected entangled pair states (PEPS). On the one hand, PEPS can not only represent the ground states of the quantum double models and the string-net models exactly\cite{GuTensorNetwork_2009,buerschaper_explicit_2009}, but also characterize their topological order at the entanglement level of PEPS\cite{peps_degeneracy_2010,MPO_algebra_2017}. On the other hand, PEPS can be used to study topological phase transitions by either deforming the PEPS representing the ground states of these exactly solvable models\cite{Condensation_driven_2017,GaugingTNS_2015,haegeman_2015_shadows,xu_zhang_2018,zhu_gapless_2019,Xu_Zhang_Zhang_2020,zhang_xu_Wang_Zhang_2020,Xu_Schuch_2021}, or serving as an ansatz for numerical variational optimizations\cite{Bridging_PRL_2017,schotte_tensor-network_2019,Iqbal_2021_PRX}.

Through a topological phase transition, anyons can disappear from the excitation spectrum in two ways: First, anyons can condense into ground states, and second, anyons become confined into pairs. Due to the lack of local order parameters to characterize topological phases, it is challenging to study topological phase transitions. Among the various approaches to study topological phase transitions, the PEPS approach stands out as a very powerful one, because PEPS are not only capable of describing anyon condensation, confinement and identification through order parameters on the entanglement level of PEPS, but can also be used to study the nature of a topological phase transition via critical exponents extracted from these order parameters\cite{AnyonsCondensation_Z4_2017,AnyonsCondensation_Z4_2018}. Recently, some progress has been made to generalize these order parameters to non-Abelian topological phase transitions of string-net models\cite{Xu_Schuch_2021}. For non-Abelian theories, the ways in which the behavior of anyons changes after topological phase transitions are more intricate. First, a non-Abelian anyon can simultaneously condense into the vacuum and become identified with other deconfined anyons, which we dub partial condensation. Second, a non-Abelian anyon of the initial topological phase can turn into another anyon of the new topological phase with smaller quantum dimension, often into an Abelian anyon, through a mechanism called partial deconfinement. Third, a non-Abelian anyon might split into two or more distinct anyons in the new topological phase; such a phenomenon is dubbed anyon splitting. Unlike the condensation and deconfinement order parameters for Abelian anyons, there is no obvious way to construct order parameters directly detecting anyon splitting.

In this paper, we show that anyon splitting can be observed from the transfer operator spectrum of a topologically degenerate ground state. To define the non-Abelian anyonic order parameters, we construct a complete set of PEPS carrying all possible DOFs of a given non-Abelian anyon. This set of PEPS allows us to generalize the condensate and deconfinement order parameters for Abelian anyons, which are numbers, to two matrices $N$ and $M$ for non-Abelian anyons, which are the key objects in our analysis. The entries of $N$ are overlaps between the ground state and the aforementioned set of PEPS, while the entries of $M$ are norms or overlaps among PEPS in that set. From $N$ and $M$, we can extract the information of partial condensation and partial deconfinement respectively. We further demonstrate that if the anyon does not split, the matrix $M$ has a tensor product structure.  By taking the structure of $M$ into consideration, we then construct order parameters which allow to directly probe anyon splitting.

We apply the generalized condensation and deconfinement order parameters as well as the splitting order parameters to various topological phase transitions, including the phase transition between the $D(S_3)$ quantum double and the toric code, the phase transition between the Rep$(S_3)$ string-net and the toric code, the phase transition between the $D(S_3)$ quantum double and the $D(Z_3)$ quantum double, and finally the phase transition between the double Ising (DIsing) string-net and the toric code. We also propose a three-parameter phase diagram for the deformed $D(S_3)$ quantum double PEPS and a two-parameter phase diagram for the deformed DIsing string-net PEPS. In addition, we also show that the PEPS describing these topological phase transitions can be exactly  mapped to many well-known statistical mechanics models, including the Ising model, the Potts models, and the Ashkin-Teller model.

This paper is organized as follows. In Sec.~\ref{PEPS}, we introduce the PEPS description of topological states and anyon excitations, focusing the construction of non-Abelian anyons in PEPS. In Sec.~\ref{order_para}, we generalize condensate and deconfinement order parameters to the non-Abelian setting, and propose new order parameters for anyon splitting. In Sec.~\ref{Examples}, we apply our approach to the $D(S_3)$ quantum double model and the Rep$(S_3)$ string-net model. In Sec.~\ref{Example_2}, we apply our approach to the DIsing string-net model. Conclusions and discussions and are presented in Sec.~\ref{conclusion}. We provide technical details and proofs in Appendices.

\section{PEPS characterization of topological phases}\label{PEPS}

\subsection{PEPS and matrix product operators}

\begin{figure}
  \centering
  \includegraphics[width=8.5cm]{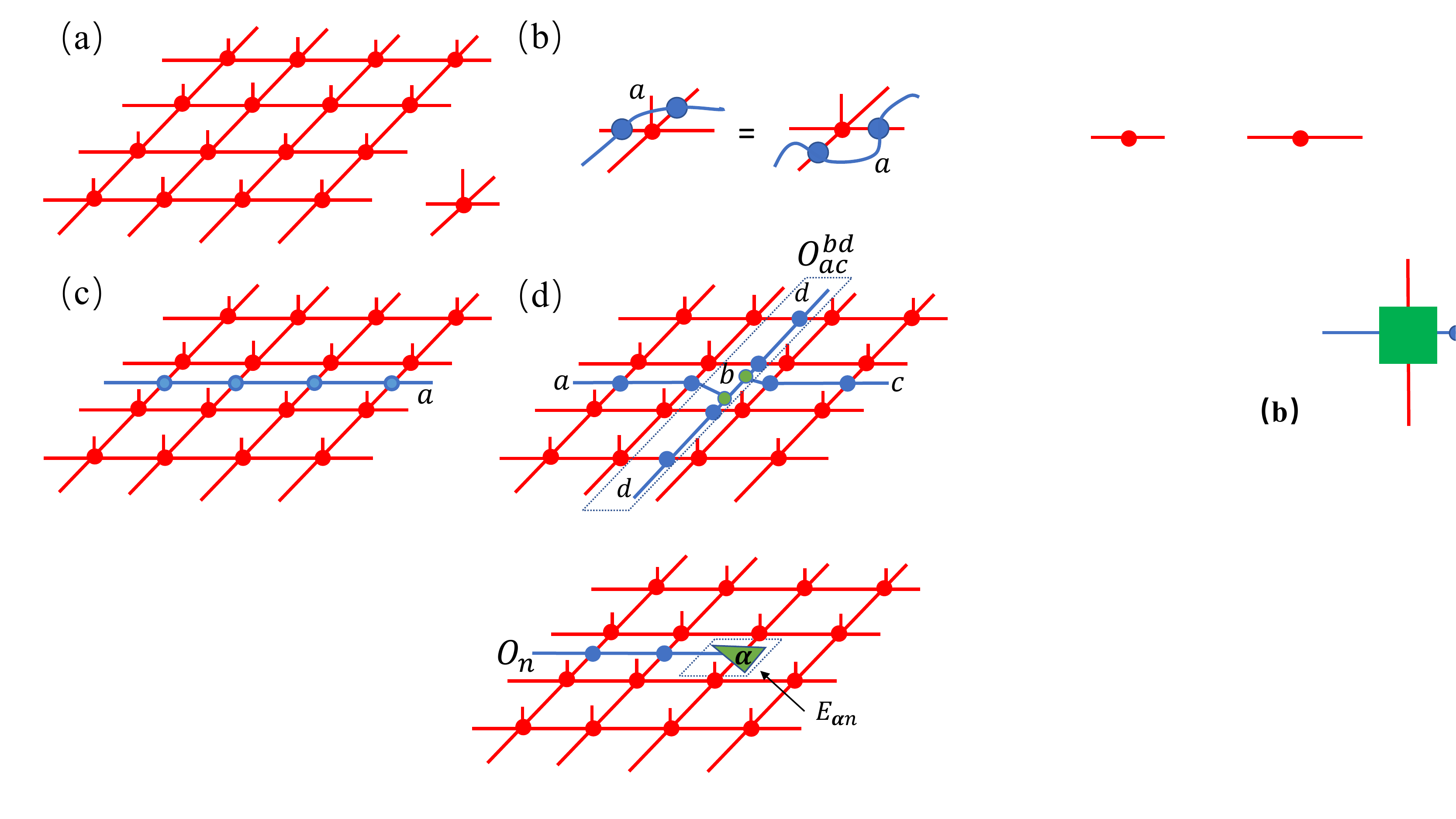}
  \caption{ (a) Graphical representation of a PEPS on a square lattice, local tensor depicted on the bottom-right. (b) The local tensor of a PEPS describing a topologically ordered state satisfies the pulling through condition, where the blue lines denotes the MPO $O_a$. (c) Inserting an MPO $O_a$ at the virtual level of the PEPS results in a new ground state. (d) General form of a ground state on the cylinder for topologically ordered PEPS with MPOs in the two directions cross forming the enlarged MPO  $O_{bd}^{ac}$, the green dots are tensors at the crossing.}\label{definition_of_MPOs}
\end{figure}

Projected entangled pair states (PEPS) are states defined by local tensors placed on the vertices of a two-dimensional lattice. An example of such a PEPS is illustrated in Fig.~\ref{definition_of_MPOs} (a) for the square lattice. The PEPS is given by the contraction of the indices corresponding to the edges of the lattice, the so-called virtual indices, of every local tensor. Throughout the paper we will use the standard graphical representation of tensors, where tensors are depicted as shapes and their indices are represented as legs. For the square lattice, the local tensor is a sphere with five legs: one leg corresponding to a physical index and the other four legs for the virtual indices.

Despite of its simplicity, the framework of PEPS is able to capture nontrivial phases of matter. In particular, the local structure characterizing PEPS with \emph{nonchiral} topological order has been developed \cite{peps_degeneracy_2010,buerschaper_twisted,MPO_algebra_2017,sahinoglu:mpo-injectivity}.  It turns out that topological order in PEPS is characterized by symmetries acting purely on the virtual level. Concretely, the input data of topological order are a fusion category $\mathcal{C}$, with simple objects $a, b, c, ....$ satisfying $a\otimes b = \bigoplus_c N_{ab}^c c$ where $N_{ab}^c$ is the fusion multiplicity. The virtual symmetry is realized by matrix product operators (MPOs) $\{O_a|a\in \mathcal{C}\}$, which form a representation of the fusion category $\mathcal{C}$: $O_a O_b = \sum_c N_{ab}^c O_c$. The MPO symmetry can be exactly derived from the PEPS with an analytic form, or extracted numerically from generic variational PEPS\cite{Anna_Abelian,Anna_2020}. The local tensors of the PEPS satisfy the so-called pulling through condition when acting with the MPOs on the virtual level, see Fig.~\ref{definition_of_MPOs} (b). In this case the PEPS with these symmetries are called MPO-injective PEPS \cite{sahinoglu:mpo-injectivity}.

The pulling through condition implies that any of these MPOs inserted at the virtual level of the PEPS can be moved freely, so that local operators, such as the local Hamiltonian terms, cannot detect them. Therefore, 
the PEPS with nontrivial MPOs at the virtual level naturally represent the different ground states of the Hamiltonian hosting a topological phase associated with $\mathcal{C}$,  see Fig.~\ref{definition_of_MPOs} (c). Fig.~\ref{definition_of_MPOs} (d) shows the most general ground state that can be constructed by inserting the MPOs along the two directions. In the crossing point of the MPOs, two new tensor different from the previous ones are placed. We denote the set of these enlarged MPOs, including the two new tensors plus the MPOs in one direction,  as $\left\{O_{a c}^{bd}\right\}$, which form a $C^*$-algebra\cite{Lan_Wen_2014,MPO_algebra_2017}.
That is, these MPOs form a closed algebra under multiplication and Hermitian conjugation. Moreover, $O_{a c}^{b d}$ reduces to the original MPO $O_d$ when $a=c=1$ since this implies that $b=d$, so that $O_d=O_{11}^{dd}$.

\subsection{Ground states and excited states in terms of PEPS}\label{SEC2B}

In the following, we will review the PEPS construction of the ground space and the anyonic excitations of MPO-injective PEPS\cite{MPO_algebra_2017,MPO_algebra_2017,sahinoglu:mpo-injectivity}. The construction will be given in terms of the idempotents of the algebra formed by the enlarged MPOs. We will also explicitly show how  the internal states of non-Abelian anyons are realized in the PEPS framework.

It is known that any $C^*$-algebra can be decomposed into its central idempotents, which we denote by ${\pmb{\alpha}}, {\pmb{\gamma}},\dots$; they are of the form
\begin{equation}\label{central_idempotent}
P^{\pmb{\alpha}}=\sum_{abd}C^{bd}_{aa}(\pmb{\alpha})O_{aa}^{bd},
\end{equation}
where the coefficients $C^{bd}_{aa}(\pmb{\alpha})$ are determined by the algebra, as shown in Appendix \ref{definition}. The central idempotents are Hermitian and orthogonal to each other:
\begin{equation}\label{property_of_CI}
(P^{\pmb{\alpha}})^\dagger=P^{\pmb{\alpha}},\quad P^{\pmb{\alpha}}P^{\pmb{\gamma}}=\delta_{\pmb{\alpha\gamma}}P^{\pmb{\alpha}}.
\end{equation}
Importantly, these central idempotents are in correspondence with the anyons of the model. Concretely,  $P^{\pmb{\alpha}}$ defines the subspace at the virtual level that corresponds to the topological sector in which the anyon $\pmb{\alpha}$ is supported.

There are algebras of enlarged MPOs (related to non-Abelian topological order) where the central idempotents defined in
Eq.~\eqref{central_idempotent} can be decomposed further into its simple components:
\begin{equation}\label{simpleidemdec}
P^{\pmb{\alpha}}=\sum_{a}P^{\pmb{\alpha}}_{aa},
\quad P^{\pmb{\alpha}}_{aa}=\sum_{bd}C_{aa}^{bd}(\pmb{\alpha})O_{aa}^{bd},
\end{equation}
where $P^{\pmb{\alpha}}_{aa}$ is a simple idempotent satisfying $(P^{\pmb{\alpha}}_{aa})^2=P^{\pmb{\alpha}}_{aa}$. The nilpotents of the algebra can also be defined in these terms\cite{MPO_algebra_2017}:
\begin{equation}
P^{\pmb{\alpha}}_{ac}=\sum_{bd}C_{ac}^{bd}(\pmb{\alpha})O_{ac}^{bd}, \quad a\neq c
\end{equation}
and they satisfy $(P^{\pmb{\alpha}}_{ac})^2 =0$.
The general relations between simple idempotents and the nilpotents are
\begin{equation}\label{property_of_simple_idemponent}
P^{\pmb{\alpha}}_{pq}P^{\pmb{\gamma}}_{mn}=\delta _{\pmb{\alpha \gamma}}\delta_{qm}P^{\pmb{\alpha}}_{pn},\quad (P^{\pmb{\alpha}}_{pq})^{\dagger
}=P^{\pmb{\alpha}}_{qp}.
\end{equation}

Fig.~\ref{Idempotent_excitation} (a) shows a PEPS on a cylinder with the simple idempotent or nilpotent $P^{\pmb{\alpha}}_{ac}$ inserted at the virtual level. On the torus, we must have $a=c$, so we obtain the ground state $|\Psi^{\pmb{\alpha}}_{aa}\rangle$. Using Eq.~\eqref{property_of_simple_idemponent}, it can be proven that $|\Psi^{\pmb{\alpha}}_{aa}\rangle$ with different $a$ labels are equal; we thus drop the label $a$ and denote $|\Psi^{\pmb{\alpha}}\rangle\equiv|\Psi^{\pmb{\alpha}}_{aa}\rangle$ from now on. These states correspond to the so-called minimally entangled states (MES)\cite{MES}.

In what follows we show how the anyons are constructed using the formalism of PEPS. Given a non-Abelian anyon ${\pmb{\alpha}}$ of the model, the state $|\bar{\pmb{\alpha}}_{x},\pmb{\alpha}_{y}\rangle_s$ on a torus with the internal DOFs $x, y$ of a pair of anyons $(\pmb{\alpha}, \bar{\pmb{\alpha}})$ can be constructed by inserting at the virtual level of the PEPS the open MPO string $O_s$ and dressing the ends with the tensors $(E^{\pmb{\alpha}}_{sy})^{\dagger}$ and $E^{\pmb{\alpha}}_{sx}$, see  Fig. \ref{Idempotent_excitation} (c). In order to construct a well-defined $\pmb{\alpha}$ anyon, the end point tensor $E^{\pmb{\alpha}}_{sx}$ should satisfy the following property under the action of the simple idempotents and nilpotents of the algebra:
\begin{equation}\label{relation_between_idempotents_and_exitations}
P^{\pmb{\gamma}}_{s^\prime t} E^{\pmb{\alpha}}_{sx} = \delta_{\pmb{\alpha\gamma}}\delta_{st} E^{\pmb{\alpha}}_{s^\prime x}\ .
\end{equation}
Fig. \ref{Idempotent_excitation} (b) shows the case when $\pmb{\alpha}=\pmb{\gamma}$ and $s=t$: the string attached to the anyon $\pmb{\alpha}$ is changed from $s$ to $s^\prime$ by the nilpotent $P^{\pmb{\alpha}}_{s^\prime s}$ when $s^\prime\neq s$. Using this property, it can be derived that the string $s$ between a pair of anyons is arbitrary, that is, $|\bar{\pmb{\alpha}}_{x},\pmb{\alpha}_{y}\rangle_s$ and $|\bar{\pmb{\alpha}}_{x},\pmb{\alpha}_{y}\rangle_{s^\prime}$ are the same state, so we simply denote the state by $|\bar{\pmb{\alpha}}_{x},\pmb{\alpha}_{y}\rangle$.

Notice that an isolated anyon can be considered in an infinite large system (with open boundary conditions). Specifically, the state $|\pmb{\alpha}_{x}\rangle_s$ can be constructed by inserting into the infinite PEPS the tensor $E^{\pmb{\alpha}}_{sx}$ with a semi-infinite MPO string $O_{s}$ attached to it, as shown in Fig.~\ref{Idempotent_excitation}~(d). In this case, $|\pmb{\alpha}_{x}\rangle_{s}$ and $|\pmb{\alpha}_{x}\rangle_{t}$ with different strings could be different states.

{The set of tensors $E^{\pmb{\alpha}}_{sx}$ satisfying property \eqref{relation_between_idempotents_and_exitations} can be constructed using the coefficients $C_{sx}^{bd}(\pmb{\alpha})$, as shown in Appendix \ref{definition}. At the renormalization fixed points of topological phases, these PEPS are orthogonal to each other:
\begin{equation}\label{orthogonality}
  \langle\bar{\pmb{\alpha}}_{x},\pmb{\alpha}_{y}|\bar{\pmb{\alpha}}_{x'},\pmb{\alpha}_{y'}\rangle\propto\delta_{xx^\prime}\delta_{yy^\prime},\quad {}_s\langle\pmb{\alpha}_{x}|\pmb{\alpha}_{y}\rangle_s\propto\delta_{xy},
\end{equation}
and we further impose the normalization condition for open boundary conditions:
\begin{equation}\label{orthonormal}
  {}_s\langle\pmb{\alpha}_{x}|\pmb{\alpha}_{y}\rangle_s=\delta_{xy},
\end{equation}
by multiplying the end tensor $E^{\pmb{\alpha}}_{sx}$ with a constant $m^{\pmb{\alpha}}_x$.

\begin{figure}
  \centering
  \includegraphics[width=8.5cm]{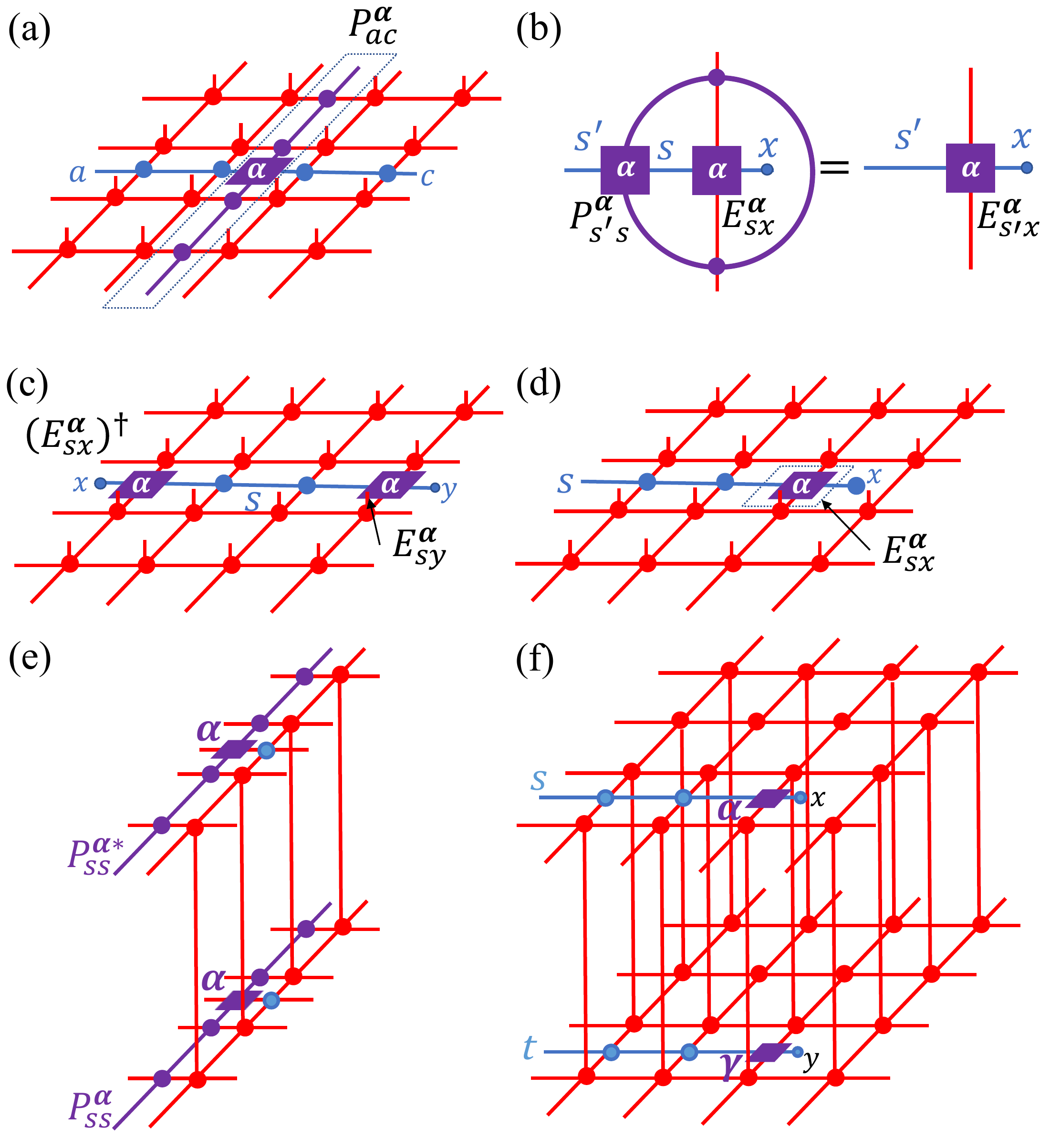}
  \caption{ (a) Inserting a simple idempotent $P^{\pmb{\alpha}}_{ac}$ and MPOs $O_a$ and $O_c$ into the PEPS on a cylinder, where the up and down sides connect. (b) A simple idempotent ($s^\prime=s$) or nilpotent $(s^\prime\neq s)$ acts on the tensor $E^{\pmb{\alpha}}_{sx}$ carrying an anyon $\pmb{\alpha}$, where $x$ labels the internal DOFs. (c) A PEPS $|\pmb{\alpha}_{x}^{\dagger},\pmb{\alpha}_{y}\rangle_s$ carrying two $\pmb{\alpha}$ excitations. (d) A PEPS  $|\pmb{\alpha}_{x}\rangle_s$ carrying an anyon excitation $\pmb{\alpha}$, with a semi-infinite MPO string $O_s$. (e) The transfer operator $\mathbb{T}^{\pmb{\alpha}}_{\pmb{\alpha}}$ of the MES norm $\langle\Psi^{\pmb{\alpha}}|\Psi^{\pmb{\alpha}}\rangle$; notice that we use the simple idempotents to define the transfer operator. (f)~The overlap $\langle\pmb{\alpha}_{x}|\pmb{\gamma}_{y}\rangle$ of the PEPS. }\label{Idempotent_excitation}
\end{figure}

\section{PEPS characterization of topological phase transitions}\label{order_para}

In this section, we describe the different ways in which anyons are affected by topological phase transitions, and show how they can be characterized through the transfer operator of the PEPS, as well as through anyonic order parameters. Importantly, we extend the framework developed in Ref.~\cite{AnyonsCondensation_Z4_2017,AnyonsCondensation_Z4_2018,Iqbal_2021_PRX} for Abelian theories to non-Abelian theories by generalizing the condensate and deconfinement order parameters and proposing a novel order parameter for anyon splitting.

\subsection{Anyons after topological phase transitions}\label{Sec_III_A}

After a topological phase transition, the anyons of the initial topological phase can suffer from condensation, identification, confinement and splitting. These changes in the nature of the anyons can be predicted by some general rules\cite{Bais_2002,Bais_Anyoncondensation_2009} using the properties of the anyons. Let us briefly review them here.

Only an anyon $\pmb{\gamma}$ with trivial self-statistics can condense.
When $\pmb{\gamma}$ is non-Abelian, its self-statistics depends on the fusion channels, and it can condense when its self-statistics is trivial in at least one fusion channel. Noncondensed anyons which differ between themselves just by a condensed anyon $\pmb{\gamma}$ become identical, i.e. they are identified as the same anyon. Any anyon $\pmb{\alpha}$ whose mutual statistics with a condensed anyon $\pmb{\gamma}$ is nontrivial becomes confined. The mutual statistics between non-Abelian anyons $\pmb{\alpha}$ and $\pmb{\gamma}$ depends on their fusion channels. It can thus happen that in some channels the non-Abelian anyon $\pmb{\alpha}$ has trivial statistics with the condensed anyon $\pmb{\gamma}$, while in other channels is has nontrivial statistics; then, in the former channels $\pmb{\alpha}$ is deconfined, while in the latter channels $\pmb{\alpha}$ is confined. We dub this scenario partial deconfinement. In such a situation, the non-Abelian anyon $\pmb{\alpha}$ becomes anther anyon with a smaller quantum dimension in the new topological phase.

Let us assume that after a phase transition the condensate is given by the set $S_{\text{cond}}=\{\pmb{1},\pmb{\gamma},\pmb{\delta},\cdots\}$, which consists of anyons of the initial topological order. If two identical non-Abelian anyons $\pmb{\alpha}$ can fuse into two or more different anyons in the condensate, i.e., $\pmb{\alpha}\otimes\pmb{\alpha}=\pmb{\gamma}\oplus\pmb{\delta}\oplus\cdots$ and $\pmb{\gamma},\pmb{\delta}\in S_{\text{cond}}$, then $\pmb{\alpha}$ splits into two or more distinguishable anyons $\pmb{\alpha}\rightarrow \pmb{x}\oplus \pmb{y}\oplus\cdots$, where $\pmb{x}, \pmb{y}$ etc. are the anyons of the new topological order. It could happen that a partially deconfined anyon $\pmb{\alpha}$ splits. In addition, when the non-Abelian anyon $\pmb{\alpha}$ splits into the trivial particle $\pmb{1}$, we say that the non-Abelian anyon $\pmb{\alpha}$ partially condenses.

\subsection{Transfer operator characterization of topological phase transitions}


The different changes of the nature of the anyons after a topological phase transition can be read off from the leading eigenvalue of the PEPS transfer operator\cite{TO_PEPS_2013}. We first review this characterization of condensation and confinement, and then demonstrate that also anyon splitting can be observed through the leading eigenvalue of the PEPS transfer operator.

We denote by $\mathbb{T}^{\pmb{\alpha}}_{\pmb{\gamma}}$ the transfer operator of the MES overlap  $\langle\Psi^{\pmb{\alpha}}|\Psi^{\pmb{\gamma}}\rangle$, as shown in Fig. \ref{Idempotent_excitation} (e). For Abelian theories, the MES can be obtained by using the central idempotents (they are also simple) defined in the previous section.
Let us denote the dominant eigenvalue of $\mathbb{T}^{\pmb{\alpha}}_{\pmb{\gamma}}$ as $t^{\pmb{\alpha}}_{\pmb{\gamma}}$, and define $\lambda^{\pmb{\alpha}}_{\pmb{\gamma}}=t^{\pmb{\alpha}}_{\pmb{\gamma}}/t^{\pmb{1}}_{\pmb{1}}$, where $\pmb{1}$ denotes to the trivial anyonic sector.
The following characterization for an Abelian anyon arises \cite{TO_PEPS_2013}: if $\pmb{\alpha}$ condenses, $\lambda_{\pmb{\alpha}}^{\pmb{1}}=1$; if $\pmb{\alpha}$ and $\pmb{\gamma}$ are identified, $\lambda_{\pmb{\alpha}}^{\pmb{\gamma}}=1$; if $\pmb{\alpha}$ is confined, $\lambda^{\pmb{\alpha}}_{\pmb{\alpha}}<1$.

This characterization can be generalized to non-Abelian theories. First of all, the definition of $\mathbb{T}^{\pmb{\alpha}}_{\pmb{\gamma}}$, and thus of $\lambda^{\pmb{\alpha}}_{\pmb{\alpha}}$, is more subtle in the non-Abelian case. This is because every central idempotent can be decomposed into its simple idempotents, see Eq.~\eqref{simpleidemdec}. It can be shown that the transfer operators defined by all different simple idempotents of $\pmb{\alpha}$ share the same spectrum, so that $\lambda^{\pmb{\alpha}}_{\pmb{\alpha}}$ from the central idempotent is naturally degenerate. However, restricting to one of the simple sectors lifts these irrelevant degeneracies. So we adopt the simple idempotents for non-Abelian anyons instead of central ones. Second, if $\pmb{\alpha}$ partially condenses and partially  identifies with $\pmb{\gamma}$, we can still observe that $\lambda_{\pmb{\alpha}}^{\pmb{1}}=1$ and $\lambda_{\pmb{\alpha}}^{\pmb{\gamma}}=1$. In summary:
\begin{itemize}[leftmargin=10pt]
\item[] (i) If $\pmb{\alpha}$ fully or partially condenses, $\lambda_{\pmb{\alpha}}^{\pmb{1}}=1$;
\item[] (ii)If $\pmb{\alpha}$ and $\pmb{\gamma}$ are fully or partially identified, $\lambda_{\pmb{\alpha}}^{\pmb{\gamma}}=1$;
\item[] (iii)If $\pmb{\alpha}$ is fully confined, $\lambda^{\pmb{\alpha}}_{\pmb{\alpha}}<1$.
\end{itemize}
We remark that MES change in the same way as the changes of anyons. For example, if anyons $\pmb{\alpha}$ and $\pmb{\gamma}$ are identified, $\lambda_{\pmb{\alpha}}^{\pmb{\gamma}}=1$ implies that MES $|\Psi^{\pmb{\alpha}}\rangle$ and $|\Psi^{\pmb{\gamma}}\rangle$ are the same state; if $\pmb{\alpha}$ is confined, $\lambda_{\pmb{\alpha}}^{\pmb{\alpha}}<1$ implies the MES $|\Psi^{\pmb{\alpha}}\rangle$ becomes a null state.  We also remark that if the non-Abelian anyon $\pmb{\alpha}$ is fully deconfined before a phase transition and partially deconfined after a phase transition, $\lambda_{\pmb{\alpha}}^{\pmb{\alpha}}$ is always one and its degeneracy does not change, so we cannot differentiate between partial and full deconfinement. The reason behind this deficiency is that there is only one MES for a given of non-Abelian anyon, which cannot capture the change of its internal property (quantum dimension). In the next subsection, we show that by considering the internal structure of a non-Abelian anyon excitation, we can differ full and partial deconfinement.

In the following Let us consider the characterization of anyon splitting. Assuming that the non-Abelian anyon $\pmb{\alpha}$ splits into various Abelian anyons $\pmb{x}$:  $\pmb{\alpha}\rightarrow\bigoplus_{\pmb{x}}c_{\pmb{a}\pmb{x}}\pmb{x}$, where $c_{\pmb{a}\pmb{x}}$ is the multiplicity, we claim that:
\begin{itemize}[leftmargin=10pt]
\item[] (iv) If $\pmb{\alpha}$  splits, $\lambda^{\pmb{\alpha}}_{\pmb{\alpha}}=1$ and deg$(\lambda^{\pmb{\alpha}}_{\pmb{\alpha}})\geqslant \sum_{\pmb{x}}c_{\pmb{\alpha x}}^2 $;
\end{itemize}
where deg$(\lambda^{\pmb{\alpha}}_{\pmb{\alpha}})$ stands for the degeneracy of $\lambda^{\pmb{\alpha}}_{\pmb{\alpha}}$. In Ref. \cite{D_4_PEPS} and all of the examples studied in Section \ref{Examples}, the degeneracies can be observed. Let us explain how anyon splitting contributes to this degeneracy. Because there is a one-to-one correspondence between the MES and the types of anyons, after splitting $|\Psi^{\pmb{\alpha}}\rangle$ becomes a superposition of the MES $|\Psi^{\pmb{x}}\rangle$ of the new anyons $\pmb{x}$:
\begin{equation}\label{GS_split}
  |\Psi^{\pmb{\alpha}}\rangle=\sum_{\pmb{x}}c_{\pmb{\alpha x}}|\Psi^{\pmb{x}}\rangle,
\end{equation}
 where $\langle\Psi^{\pmb{x}}|\Psi^{\pmb{x}}\rangle=\langle\Psi^{\pmb{y}}|\Psi^{\pmb{y}}\rangle, \forall \pmb{x}, \pmb{y}$ and  $\langle\Psi^{\pmb{x}}|\Psi^{\pmb{y}}\rangle=0,\forall\pmb{x}\neq\pmb{y}$.  We see that an MES splits in the same way as the splitting of corresponding anyon. Combine with the changes of the MES in the cases of identification and confinement, we can understand how ground space varies through phase transitions.

 From Eq.~\eqref{GS_split}, it can be found that the norm $\langle\Psi^{\pmb{\alpha}}|\Psi^{\pmb{\alpha}}\rangle$ equals to a sum of the norms of the $\sum_{\pmb{x}}c_{\pmb{\alpha x}}^2$ new MES, from which it can be derived that the degeneracy of $\lambda_{\pmb{\alpha}}^{\pmb{\alpha}}$ is at least $\sum_{\pmb{x}}c_{\pmb{\alpha x}}^2$.
 In Appendix \ref{GS_splitting}, we prove  Eq.~\eqref{GS_split} and derive this degeneracy for the quantum double models.
The degeneracy indicates that there should exist a correlation function with long range order associated with anyon splitting, and this long range order is just the order parameter of anyon splitting.

\subsection{Condensate, identification and deconfinement order parameters of non-Abelian anyons}

The changes in the topological nature of the anyons can also be detected through anyonic order parameters defined at the virtual level of PEPS \cite{AnyonsCondensation_Z4_2017,AnyonsCondensation_Z4_2018,Iqbal_2021_PRX}. Compared to the characterization through the transfer operator, anyonic order parameters are physically more tangible and easier to calculate. Importantly, we will see that the anyonic order parameters can detect partial confinement, which cannot be observed from the leading eigenvalue $\lambda_{\pmb{\alpha}}^{\pmb{\alpha}}$ of the transfer operator. Moreover, when the topological phase transitions are continuous, there are critical exponents $\beta$ associated with the anyonic order parameters $O\propto|k-k_c|^\beta$, where $k$ is the tuning parameter and $k_c$ is the critical point, which can be used to identify the universality class of the phase transition.

In essence, the anyonic order parameters detect the changes of the symmetry breaking patterns of the fixed point space of the transfer operator $\mathbb{T}$. For instance, suppose that the virtual symmetry of the PEPS is described by a group $G$ and the MPOs are $O_g,g\in G$, the transfer operator $\mathbb{T}$ has the symmetry $G\otimes G$. The fixed point space of $\mathbb{T}$ can be degenerate, implying the spontaneous symmetry breaking of $\mathbb{T}$. Let $H\subset G$ be a subgroup of $G$ and $Q\subset H$ be a normal subgroup of $H$. If a fixed point $\rho$ of $\mathbb{T}$ satisfies $O_h\rho O_h^\dagger=\rho, h\in H$ and $O_q \rho=\rho O^{\dagger}_q=\rho, q\in Q$, then  other fixed points of $\mathbb{T}$ have the symmetries isomorphic to that of $\rho$, i.e., the fixed point $\rho^\prime=\rho O_k,k\notin Q$ satisfies $O_h\rho^\prime O_{k^{-1}hk}^\dagger=\rho^\prime$ and $O_q \rho^\prime=\rho^\prime O^{\dagger}_{k^{-1}qk}=\rho^\prime$. So we denote the symmetry of the fixed point space of $\mathbbm{T}$ using the symbol $H\boxtimes Q$, which comes from Ref. \cite{AnyonsCondensation_Z4_2017,AnyonsCondensation_Z4_2018}, and $H$ is called diagonal symmetry while $Q$ is called off-diagonal symmetry. Particularly, the symmetry of the fixed point space of $\mathbb{T}$ in the original phase is always $G\boxtimes Z_1$. For the PEPS whose virtual MPO symmetry is described a unitary fusion category, the diagonal symmetry of the fixed point $\rho$ of $\mathbb{T}$ is $\{O_a|O_a\rho O_a^{\dagger}=\rho+\cdots\}$ and the off-diagonal symmetry is $\{O_b|O_b\rho=\rho O_b^{\dagger} =\rho+\cdots\}$, where "$\cdots$" denotes other possible fixed points of $\mathbb{T}$. In addition, we emphasize that for these non-Abelian PEPS, different transfer operator fixed points can have different symmetries which are also isomorphic to each other under the unitary fusion category.

In the following, we provide the definitions of the order parameters for condensation, identification and deconfinement in non-Abelian theories, which naturally extend the ones defined for Abelian anyons \cite{AnyonsCondensation_Z4_2018}. Let us denote the PEPS without any anyon as $|\pmb{1}\rangle$ on an infinite plane (open boundary conditions) and normalize it as $\langle\pmb{1}|\pmb{1}\rangle=1$. In this situation we can define a state $|\pmb{\alpha}_x\rangle_s$ with an isolated anyon, as shown in Fig. \ref{Idempotent_excitation} (d). The order parameters are based on the overlaps ${}_s\langle\pmb{\alpha}_{x} |\pmb{\gamma}_{y}\rangle_t$, shown in Fig.~\ref{Idempotent_excitation} (f).

To detect the condensation of a non-Abelian anyon $\pmb{\alpha}$, we can compute a matrix $N^{\pmb{1}}(\pmb{\alpha})$ with entries
\begin{equation}\label{N_1}
  N^{\pmb{1}}_{sx}(\pmb{\alpha})=\langle\pmb{1}|\pmb{\alpha}_x\rangle_s.
\end{equation}
If $\pmb{\alpha}$ does not condense, $N^{\pmb{1}}(\pmb{\alpha})=0$, and a nonzero $N(\pmb{\alpha})$ implies that $\pmb{\alpha}$ condenses.  We say that
\begin{itemize}[leftmargin=10pt]
  \item[] (i) $\pmb{\alpha}$ fully condenses if the rank of $N^{\pmb{1}}(\pmb{\alpha})=0$ is full;
  \item[] (ii) $\pmb{\alpha}$ partially condenses if $N^{\pmb{1}}(\pmb{\alpha})\neq 0$ and its rank is not full.
\end{itemize}
Since $N(\pmb{\alpha})$ is non-Hermitian in general, we use its singular values to characterize its rank. If we want to detect the identification of $\pmb{\alpha}$ with an Abelian anyon $\pmb{\gamma}$, we can proceed similarly and compute a matrix $N^{\pmb{\gamma}}(\pmb{\alpha})$ with entries
\begin{equation}\label{N_gamma}
  N^{\pmb{\gamma}}_{sx}(\pmb{\alpha})=\langle\pmb{\gamma}|\pmb{\alpha}_x\rangle_s
\end{equation}
and its singular values. In addition, from the nonzero singular values of $N^{\pmb{1}}(\pmb{\alpha})$ and $N^{\pmb{\gamma}}(\pmb{\alpha})$, the critical exponents associated with the partial condensation and identification can be defined; we will find later that they are universal for all models we study.

In order to detect the confinement of the non-Abelian anyon $\pmb{\alpha}$, we can consider the matrix $M(\pmb{\alpha})$ with entries
\begin{equation}\label{entries_of_M}
  M_{sx,ty}(\pmb{\alpha})={}_{s}\langle\pmb{\alpha}_{x}|\pmb{\alpha}_{y}\rangle_{t}.
\end{equation}
The matrix $M(\pmb{\alpha})$ is Hermitian and semi-positive definite. The space $\mathcal{V}^{\pmb{\alpha}}$ where $M(\pmb{\alpha})$ is supported is a tensor product of a space $\mathcal{V}_{\text{str}}^{\pmb{\alpha}}$ associated with the string labels and a space $\mathcal{V}_{\text{int}}^{\pmb{\alpha}}$ associated with the internal DOF labels: $\mathcal{V}_{\text{str}}^{\pmb{\alpha}}\otimes \mathcal{V}^{\pmb{\alpha}}_{\text{int}}$. If the anyon $\pmb{\alpha}$ is fully confined, the matrix $M(\pmb{\alpha})=0$. However, when the anyon $\pmb{\alpha}$ is fully or partially deconfined, $M(\pmb{\alpha})$ can be nonzero. To distinguish the two cases, we define the matrix
 $M_{\text{str}}(\pmb{\alpha})$ by tracing the subspace $\mathcal{V}_\text{int}$:
 \begin{equation}\label{M_str}
   M_{\text{str}}(\pmb{\alpha}):=\text{tr}_{\text{int}}[M(\pmb{\alpha})].
 \end{equation}
 Notice that for the string-net models, every label $x$ is associated with a quantum dimension $d_x$; when doing the partial trace we account for the quantum dimensions, i.e., $M_{\text{str}}(\pmb{\alpha})_{st}=\sum_{x}d_x M_{sx,tx}(\pmb{\alpha})$, and it is implicit in the partial trace. We say that
  \begin{itemize}[leftmargin=10pt]
    \item[] (iii) $\pmb{\alpha}$ is fully deconfined if the rank of $M_{\text{str}}(\pmb{\alpha})$ is full;
    \item[] (iv) $\pmb{\alpha}$ is partially deconfined if $M_{\text{str}}(\pmb{\alpha})\neq0$ and its rank is not full.
  \end{itemize}
Since $M_{\text{str}}(\pmb{\alpha})$ is Hermitian, we use its eigenvalues to characterize the rank.

Let us explain why our proposal works. Before phase transitions, there is a transfer operator fixed point $\rho$ with symmetry  $\mathcal{C}\boxtimes Z_1$, where $\mathcal{C}$ is the MPO symmetry of the PEPS ($Z_1$ is the trivial group with one element). Suppose that the MPO strings $\{O_s,O_t,\cdots\}=\mathcal{S}_{\pmb{\alpha}}\subset \mathcal{C}$ can be attached to $\pmb{\alpha}$, we must have $M_{sx,ty}(\pmb{\alpha})\propto \delta_{st}$, because the fixed point $\rho$ has no off-diagonal symmetry, see proof in Appendix.~\ref{prove_block_structure}. Therefore $M(\pmb{\alpha})$ is block diagonal in $\mathcal{V}_{\text{str}}^{\pmb{\alpha}}$, and for convenience, we define for  every diagonal block the matrix $B^{[s]}_\text{int}(\pmb{\alpha})$ supported in the subspace $\mathcal{V}^{\pmb{\alpha}}_\text{int}$ with entries $B^{[s]}_\text{int}(\pmb{\alpha})_{xy}=M_{sx,sy}(\pmb{\alpha})$ . Moreover, using Eq.~\eqref{relation_between_idempotents_and_exitations}, we prove that $\forall s, B^{[s]}_\text{int}(\pmb{\alpha})$ are equal in Appendix.~\ref{prove_block_structure}, so we can drop the superscript $s$. Then, $M(\pmb{\alpha})$ has the following tensor product structure:
\begin{equation}\label{form_M}
  M(\pmb{\alpha})=\mathbbm{1}_{\text{str}}\otimes B_{\text{int}}(\pmb{\alpha}),
\end{equation}
where $\mathbbm{1}_{\text{str}}$ is an identity matrix in the space $\mathcal{V}^{\pmb{\alpha}}_{\text{str}}$. So $M_{\text{str}}(\pmb{\alpha})$ defined in Eq.~\eqref{M_str} has $|\mathcal{S}_{\pmb{\alpha}}|$ (cardinal number of $\mathcal{S}_{\pmb{\alpha}}$) eigenvalues all equal to $\text{tr}[B_{\text{int}}(\pmb{\alpha})]$ before a phase transition.

In the following we consider what happens with $M(\pmb{\alpha})$ after topological phase transitions. For simplicity, we assume that the symmetry of the transfer operator fixed point $\rho$ breaks from $\mathcal{C}\boxtimes Z_1$ down to $\mathcal{D}\boxtimes Z_1$, where $\mathcal{D}\subset\mathcal{C}$. If $\tilde{\mathcal{S}}_{\pmb{\alpha}}\equiv\mathcal{S}_{\pmb{\alpha}}\cap\mathcal{D}=\varnothing$, where $\mathcal{S}_{\pmb{\alpha}}$ is the set of strings that can be attached to $\pmb{\alpha}$, then $M(\pmb{\alpha})=0$ and $\pmb{\alpha}$ is fully confined.

If $\tilde{\mathcal{S}}_{\pmb{\alpha}}=\mathcal{S}_{\pmb{\alpha}}$,  $M(\pmb{\alpha})$ has the form
\begin{equation}\label{full_deconfinement}
  M(\pmb{\alpha})=\bigoplus_{s\in \mathcal{S}_{\pmb{\alpha}}} B^{[s]}_\text{int}(\pmb{\alpha}).
\end{equation}
Notice that the blocks $B^{[s]}_\text{int}(\pmb{\alpha})$ are not necessarily equal. $M_{\text{str}}(\pmb{\alpha})$ has $|\mathcal{S}_{\pmb{\alpha}}|$ nonzero eigenvalues $\text{tr}[B^{[s]}_\text{int}(\pmb{\alpha})], \forall s\in \mathcal{S}_{\pmb{\alpha}}$, so $\pmb{\alpha}$ is fully deconfined.

If $\tilde{\mathcal{S}}_{\pmb{\alpha}}\neq \varnothing$ and $\tilde{\mathcal{S}}_{\pmb{\alpha}} \neq\mathcal{S}_{\pmb{\alpha}}$, we must have $B^{[s]}_\text{int}(\pmb{\alpha})=0, \forall s\in \tilde{\mathcal{S}}_{\pmb{\alpha}}^{\perp}\equiv(\mathcal{S}_{\pmb{\alpha}}-\tilde{\mathcal{S}}_{\pmb{\alpha}})$, and $M(\pmb{\alpha})$ has the form:
\begin{equation}\label{partial_deconfinement}
  M(\pmb{\alpha})=\bigoplus_{s\in \tilde{\mathcal{S}}_{\pmb{\alpha}}} B^{[s]}_\text{int}(\pmb{\alpha}).
\end{equation}
Then $M_{\text{str}}(\pmb{\alpha})$ has $|\tilde{\mathcal{S}}_{\pmb{\alpha}}|$ nonzero eigenvalues $\text{tr}[ B^{[s]}_\text{int}(\pmb{\alpha})], s\in \tilde{\mathcal{S}}_{\pmb{\alpha}}$, and $|\tilde{\mathcal{S}}_{\pmb{\alpha}}^{\perp}|$ zero eigenvalues. So $\pmb{\alpha}$ is partially deconfined and the quantum dimension of the anyon $\pmb{\alpha}$ reduces from $d_{\pmb{\alpha}}=\sum_{s\in \mathcal{S}_{\pmb{\alpha}}}d_s$ to $d_{\pmb{\alpha}}=\sum_{s\in \tilde{\mathcal{S}}_{\pmb{\alpha}}}d_s$.

There are some comments about partial deconfinement. First, when the transfer operator fixed points have an off-diagonal symmetry after phase transitions, $M(\pmb{\alpha})$ is not block diagonal. However we can always find an unitary transformation $U_{\text{str}}\otimes \mathbbm{1}_{\text{int}}$ to block diagonalize $M(\pmb{\alpha})$, as shown in Appendix~\ref{prove_block_structure}, so our approach can still distinguish partial and full deconfinement from the eigenvalues of $M_{\text{str}}(\pmb{\alpha})$. Second, one may wonder why we do not use the eigenvalues of $M(\pmb{\alpha})$ themselves to characterize the partial deconfinement of $\pmb{\alpha}$. The reason is that the eigenvalues of $M(\pmb{\alpha})$ can be zero even if $\pmb{\alpha}$ is fully deconfined, which could happen at the renormalization fixed point of the $\mathcal{D}\boxtimes Z_1$ phase. So the eigenvalues of $M_{\text{str}}(\pmb{\alpha})$ are better suited to characterize the partial deconfinement of $\pmb{\alpha}$.

\subsection{Order parameters of anyon splitting}\label{order_para_splitting}

After a phase transition, if the anyon $\pmb{\alpha}$ is fully or partially deconfined, it may become a non-simple anyon, i.e., it may split into various new anyons of the new topological order. We will discuss the construction of the anyon splitting order parameters in this subsection. First, Let us consider the case where the anyon $\pmb{\alpha}$ does not split. When $\pmb{\alpha}$ is fully deconfined, it can be seen that also after the phase transition $M(\pmb{\alpha})$ has the form shown in Eq.~\eqref{form_M}, see Appendix \ref{what_if}.
We also show that in Appendix \ref{what_if} when $\pmb{\alpha}$ is partially deconfined, $M(\pmb{\alpha})=P_{\text{str}}\otimes B_{\text{int}}(\pmb{\alpha})$, where $P_{\text{str}}$ is a projector. Therefore, if $\pmb{\alpha}$ does not split, $M(\pmb{\alpha})$ has a tensor product structure, which means the string DOFs and internal DOFs are not entangled. Equivalently, if $M(\pmb{\alpha})$ does not have such a tensor product structure, then $\pmb{\alpha}$ must split.

We now define the anyon splitting order parameters using the matrix $M(\pmb{\alpha})$. At the renormalization fixed point of the topological phase, because of the orthonormality shown in Eq.~\eqref{orthonormal}, $B_{\text{int}}=\mathbbm{1}$. However, if we perturb away from the renormalization fixed point, the orthonormality may be violated even before the phase transition. To define a well-behaved splitting order parameter, it is necessary to work in an orthonormal basis. Without loss of generality, we assume that $M(\pmb{\alpha})$ is already block diagonalized by $U_\text{str}\otimes \mathbbm{1}_{\text{int}}$. If there is partial deconfinement, then some diagonal blocks $B^{[s]}_{\text{int}}=0, \forall s\in \tilde{\mathcal{S}}_{\pmb{\alpha}}^{\perp}$ and we can always discard them.
There are two steps for constructing the splitting order parameter.

\emph{Step 1: finding an orthonormal basis.} Such a basis can be found by diagonalizing $M_{\text{int}}(\pmb{\alpha}):=\text{tr}_{\text{str}}M(\pmb{\alpha})$ using a unitary matrix $U_{\text{int}}$, $M_{\text{int}}(\pmb{\alpha})=U_{\text{int}}D U_{\text{int}}^{\dagger}$, where $D$ is a diagonal matrix. Notice that the quantum dimensions of the strings are implicit in the partial trace. Then, we define a new matrix
\begin{equation}
  M^{\prime}(\pmb{\alpha})=d_{\pmb{\alpha}}[\mathbbm{1}_{\text{str}}\otimes(D^{-\frac{1}{2}}U_{\text{int}}^{\dagger})] M(\pmb{\alpha}) [\mathbbm{1}_{\text{str}}\otimes(U_{\text{int}}D^{-\frac{1}{2}})],
\end{equation}
whose entries are $M^{\prime}_{sx,ty}={}_s\langle\pmb{\alpha}^{\prime}_{x}|\pmb{\alpha}^{\prime}_{y}\rangle_t$ and
\begin{equation}
  |\pmb{\alpha}^{\prime}_x\rangle_s=\sqrt{\frac{d_{\pmb{\alpha}}}{D_{xx}}}\sum_{y}(U_{\text{int}})_{yx}|\pmb{\alpha}_{y}\rangle_s.
\end{equation}
If $\pmb{\alpha}$ does not split, $M(\pmb{\alpha})$ has a tensor product structure and it can be diagonalized by $\mathbbm{1}_{\text{str}}\otimes U_{\text{int}}$, therefore the bases of $M^{\prime}(\pmb{\alpha})$ are orthonormal
\begin{equation}
  {}_{s}\langle\pmb{\alpha}^{\prime}_x|\pmb{\alpha}^{\prime}_y\rangle_{t}=\delta_{st}\delta_{xy}.
\end{equation}
However, if $\pmb{\alpha}$ splits, $M(\pmb{\alpha})$ does not have a tensor product structure and $\mathbbm{1}_{\text{str}}\otimes U_{\text{int}}$ can not diagonalize $M(\pmb{\alpha})$.
Therefore,  if $\pmb{\alpha}$ does not split, $M^{\prime}(\pmb{\alpha})$ is an identity matrix. Equivalently, if $M^\prime(\pmb{\alpha})$ is not an identity matrix, $\pmb{\alpha}$ splits.

\emph{Step 2: constructing splitting order parameters from the eigenvalues of $M^{\prime}(\pmb{\alpha})$.} If $\pmb{\alpha}$ splits, $M^{\prime}(\pmb{\alpha})$ can be diagonalized by another unitary matrix, and we denote the diagonalized matrix as $M^{\prime\prime}(\pmb{\alpha})$. Then we apply a discrete Fourier transformation $F_{\text{int}}$ to $M^{\prime\prime}(\pmb{\alpha})$ to measure the differences between the eigenvalues of $M^{\prime}(\pmb{\alpha})$:
\begin{equation}
  \tilde{M}(\pmb{\alpha})=(\mathbbm{1}_{\text{str}}\otimes F_{\text{int}})M^{\prime\prime}(\pmb{\alpha})(\mathbbm{1}_{\text{str}}\otimes F^\dagger_{\text{int}}).
\end{equation}
Then $\tilde{M}(\pmb{\alpha})$ is still an identity matrix if $\pmb{\alpha}$ does not split, and it is a nondiagonal matrix after splitting. Therefore the off-diagonal entries of $\tilde{M}(\pmb{\alpha})$ can serve as the splitting order parameters:
\begin{itemize}[leftmargin=10pt]
  \item[] (v) $\pmb{\alpha}$ splits if $\forall s, \exists x\neq y$ such that $\tilde{M}_{sx,sy}(\pmb{\alpha})\neq 0$.
\end{itemize}

There are some remarks.
First, along some fine-tuned paths of the phase transition, $M(\pmb{\alpha})=M^{\prime}(\pmb{\alpha})$ and they are diagonal.
Second, when the topological phase transition is continuous, the splitting order parameters exhibit a universal critical exponent. Next, as we mentioned before, the degenerate leading eigenvalue $\lambda_{\pmb{\alpha}}^{\pmb{\alpha}}$ of the transfer operator implies that there exists a correlation function with long range order, and it is just ${}_{s}\langle\bar{\tilde{\pmb{\alpha}}}_x,\tilde{\pmb{\alpha}}_x|\bar{\tilde{\pmb{\alpha}}}_y,\tilde{\pmb{\alpha}}_y\rangle_{s}$,
which reduces to the splitting order parameters $|\tilde{M}_{sx,sy}(\pmb{\alpha})|^2$ when the pair of anyons separate far apart.
In addition, for all anyon splitting examples considered in the next sections, $M(\pmb{\alpha})$ is $4\times 4$, and it can be checked that all nonzero off-diagonal entries of $\tilde{M}(\pmb{\alpha})$ satisfy $\tilde{M}_{11,12}(\pmb{\alpha})=\tilde{M}_{21,22}(\pmb{\alpha})=-\tilde{M}_{12,11}(\pmb{\alpha})=-\tilde{M}_{22,21}(\pmb{\alpha})$, so there is only one splitting order parameter and we denote it as $\tilde{M}_{\text{off}}(\pmb{\alpha})=|\tilde{M}_{11,12}(\pmb{\alpha})|$.

\section{Examples: $D(S_3)$ quantum double model and Rep$(S_3)$ string-net model}\label{Examples}

In this section we use the generalized condensate, deconfinement and anyon splitting order parameters to characterize the topological phase transitions of the $D(S_3)$ quantum double model\cite{kitaev_toric_code_2003} and the Rep$(S_3)$ string-net model\cite{Levin-Wen-string-net-2005}.
The input data of quantum double $D(G)$ models are categories called Vec$(G)$, in which the objects are group elements and the fusion is just the group multiplicity. The anyons in the $D(G)$ quantum double models are labeled by the conjugacy classes $K$ of $G$ and the irreducible representations (irreps) of the centralizers $C_k=\{g\in G| gk=kg\}$, where $k\in K$ is a representative of $K$ and $C_k \cong C_{k^\prime},\forall k,k^{\prime}\in K$.

Each $D(G)$ quantum double model has a Morita equivalent Rep$(G)$ string-net model; they share the same topological order\cite{Map_QD_to_string_net,Morita_mapping}. The input data of a Rep$(G)$ string-net model is a fusion category $\mathcal{C}=\text{Rep}(G)$, where the simple objects are irreps of $G$ and the fusion is given by the tensor product of irreps.

To obtain topological phase transitions, we act on the physical level of the PEPS the deformation operator $\bigotimes_{e}\exp(k O_e)$, where $k$ is a turning parameter and $O_e$ is a local Hermitian operator on the edge $e$. The topological order is stable for small deformation\cite{chen:topo-symmetry-conditions,Pollmann_Chen_2018}. However, when $k=\infty$, the deformation becomes a projector which filters out another simpler topological state, so a topological phase transition can be obtained by tuning $k$. The topological phase transition can also be obtained by adding a magnetic field term $-h\sum_{e} O_e$ to a fixed point Hamiltonian of a string-net model or quantum double model. For the sake of simplicity, we adopt the PEPS approach to obtain phase transitions in this paper.

\begin{figure}
  \centering
  \includegraphics[width=8cm]{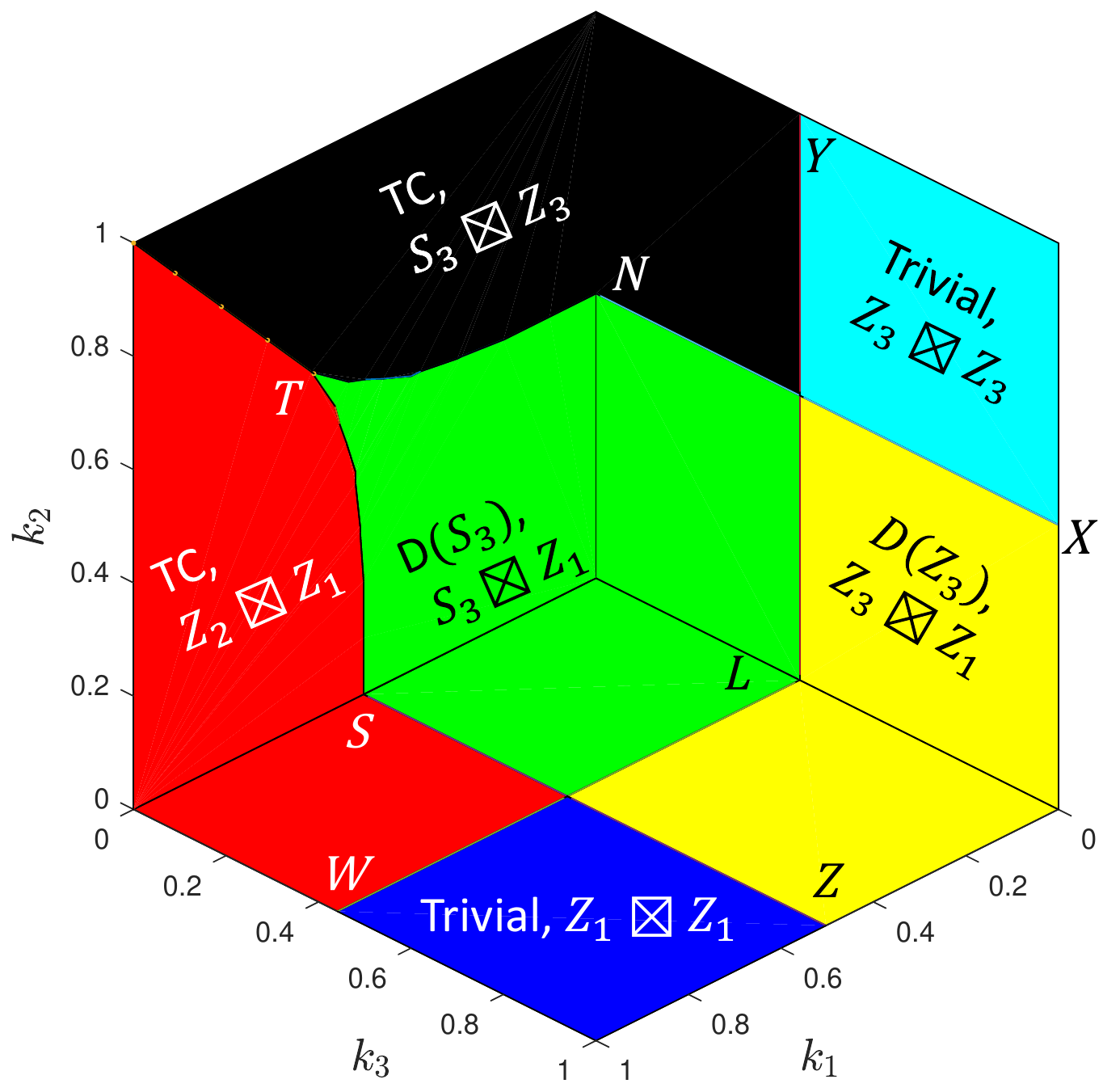}
  \caption{Phase diagram of the deformed $D(S_3)$ quantum double PEPS. Each phase is labeled by $H\boxtimes Q$, which is the symmetry of the transfer operator fixed points, and TC denotes toric code. The phase diagram is symmetric about the plane $k_1=k_2$. The phase transition points on three axes are $S=(\log(1+\sqrt{3})/2,0,0)$, $N=(0,\log(1+\sqrt{3})/2,0)$ and $L=(0,0,\log(1+\sqrt{2})/2)$. $SZ$, $WL$, $LY$ and $NX$ are straight lines parallel to the axes. }\label{S_3_phase_diagram}
\end{figure}

\begin{table}
\centering
\caption{Conjugacy classes $K_i$ of $S_3$, centralizers $C_k$ and irreps of the centralizers. We slightly abuse the notation and denote all trivial irreps as $1$. $\epsilon$ is the 1d nontrivial irrep of $S_3$, $\pi$ is the 2d irreps of $S_3$, $\omega$ and $\bar{\omega}$ are two nontrivial irreps of $Z_3$ and $-1$ is the nontrivial irrep of $Z_2$.}
\begin{tabularx}{\linewidth}{C{1.67cm}C{0.67cm}C{0.67cm}C{0.67cm}C{0.67cm}C{0.67cm}C{0.67cm}C{0.87cm}C{0.87cm}C{0.87cm}}
\toprule[1pt]
 $K_i$ & \multicolumn{4}{c}{$K_1=\{e\}$} & \multicolumn{3}{c}{$K_2=\{r,\bar{r}\}$} & \multicolumn{2}{c}{$K_3=\{s,sr,s\bar{r}\}$} \\
\midrule[1pt]
$C_k$ &\multicolumn{4}{c}{$S_3= C_e$} & \multicolumn{3}{c}{$Z_3\simeq C_r$} & \multicolumn{2}{c}{$Z_2\simeq C_s$} \\
irrep of ${C_k}$ & $1$ & $\epsilon$ & \multicolumn{2}{c}{$\pi$} & $1$ & $\omega$ & $\bar{\omega}$ & $1$ & $-1$ \\
\bottomrule[1pt]
\end{tabularx}
\label{irreps_of_S3}
\end{table}

\begin{table*}
\centering
\caption{Anyons of the $S_3$ topological order and condensation patterns through phase transitions}
\begin{tabular}{ccccccccccccc}
\toprule[1pt]
Anyon $\pmb{\alpha}$ & $\pmb{1}=\left(K_1,1\right)$ & $\pmb{A}=\left(K_1,\epsilon\right)$ & \multicolumn{2}{c}{$\pmb{C}=\left(K_1,\pi\right)$} & \multicolumn{2}{c}{$\pmb{B}=\left(K_2,1\right)$} & \multicolumn{2}{c}{$\pmb{D}=\left(K_2,\omega \right)$} &  \multicolumn{2}{c}{$\pmb{E}=\left(K_2,\omega ^*\right)$}
& $\pmb{F}=\left(K_3,1\right)$ &$\pmb{G}=\left(K_3,-1\right)$. \\
Type & Trivial& Chargon & \multicolumn{2}{c}{Chargon} & \multicolumn{2}{c}{Fluxon}& \multicolumn{2}{c}{Dyon} & \multicolumn{2}{c}{Dyon }& Fluxon & Dyon \\
\midrule[1pt]
Topo. spin & 1 & 1 & \multicolumn{2}{c}{1} & \multicolumn{2}{c}{1} & \multicolumn{2}{c}{$\exp \left(\frac{2\pi  i}{3}\right)$} & \multicolumn{2}{c}{$\exp \left(-\frac{2\pi  i}{3}\right)$} & 1 & -1 \\
Quantum dim. & 1 & 1 & \multicolumn{2}{c}{2} & \multicolumn{2}{c}{2} & \multicolumn{2}{c}{2} & \multicolumn{2}{c}{2} & 3 & 3 \\
$D(S_3)$-TC & $\pmb{1}$ & $\pmb{e}$ & $\pmb{1}$&$\pmb{e}$ & \multicolumn{2}{c}{conf.} & \multicolumn{2}{c}{conf.} & \multicolumn{2}{c}{conf.} & $\pmb{m}$ & $\pmb{f}$\\
Rep$(S_3)$-TC & $\pmb{1}$ &$\pmb{m}$ & \multicolumn{2}{c}{conf.}& $\pmb{1}$&$\pmb{m}$  & \multicolumn{2}{c}{conf.} & \multicolumn{2}{c}{conf.} & $\pmb{e}$ & $\pmb{f}$\\
$D(S_3)$-$D(Z_3)$ & $(e,1)$ &$(e,1)$ &$(e,\omega)$ &$(e,\bar{\omega})$& $(r,1)$&$(\bar{r},1)$  & $(r,\omega)$& $(\bar{r},\omega)$ & $(r,\bar{\omega})$& $(\bar{r},\bar{\omega})$& conf. & conf.\\
\bottomrule[1pt]
\end{tabular}
\label{Anyons_of_S3}
\end{table*}
\begin{figure*}
  \centering
  \includegraphics[width=18cm]{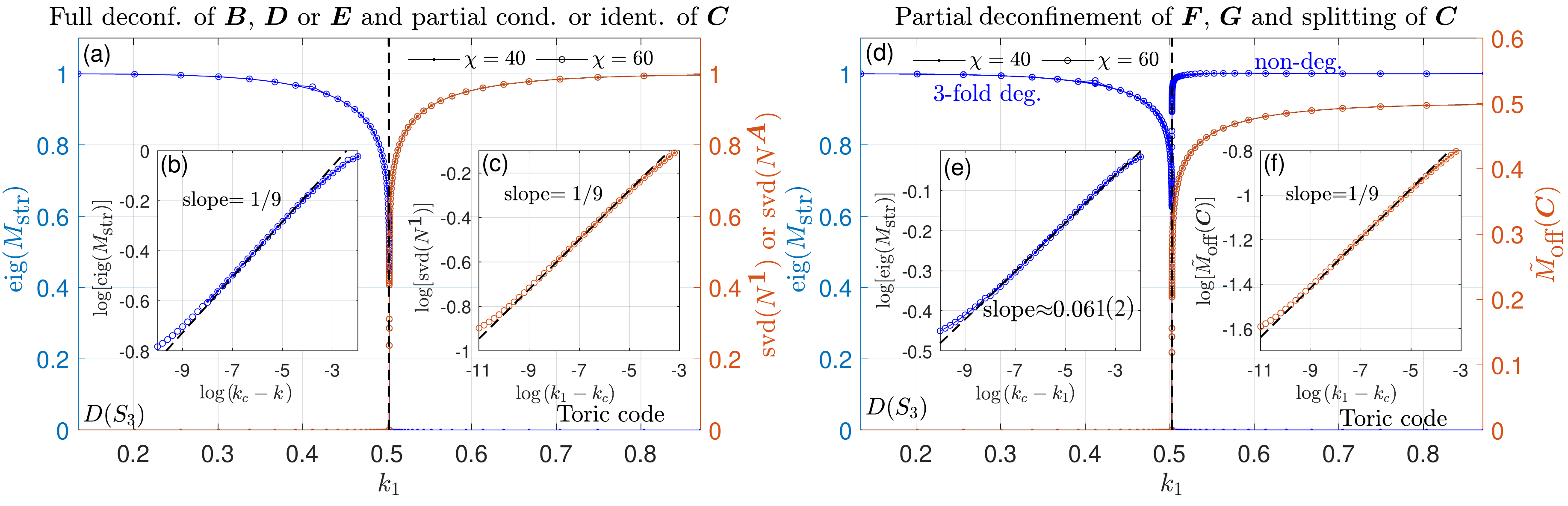}
  \caption{Anyonic order parameters for the phase transition between $D(S_3)$ phase and $Z_2\boxtimes Z_1$ toric code phase, and their critical exponents. The vertical dashed lines in (a) and (d) give the exact location of the critical point $k_c=\log(1+\sqrt{3})/2$. The dashed lines in the insets have the slope indicated in each inset.  $\chi$ is the bond dimension of the VUMPS. (a) Left: the deconfinement order parameters of $\pmb{B}$, $\pmb{D}$ or $\pmb{E}$, eig($M_{\text{str}}$) denotes the eigenvalues of $M_{\text{str}}(\pmb{B})$, $M_{\text{str}}(\pmb{D})$ or $M_{\text{str}}(\pmb{E})$. Right: the condensate (identification) order parameter, where svd$(N^{\pmb{1}})$ and svd$(N^{\pmb{A}})$ denote the singular values of $N^{\pmb{1}}(\pmb{C})$ and $N^{\pmb{A}}(\pmb{C})$. The insets (b) and (c) show the scaling of the corresponding anyonic order parameters. (d) Left: the deconfinement order parameters of $\pmb{F}$ or $\pmb{G}$, eig($M_{\text{str}}$) denotes the eigenvalues of $M_{\text{str}}(\pmb{F})$ or $M_{\text{str}}(\pmb{G})$. Right: the splitting order parameter of $\pmb{C}$. The insets (e) and (f) show the scaling of the corresponding anyonic order parameters.}\label{Fig_Vec_S3_Z2_all}
\end{figure*}
\subsection{Phase diagram of $D(S_3)$ quantum double model}
The simplest non-Abelian quantum double model is the $D(S_3)$ quantum double model, constructed using the simplest non-Abelian group $S_3=\{e,r,\bar{r},s,sr,s\bar{r}\}$, where the generators $s$ and $r$ satisfy $r^3=s^2=e, r^2=\bar{r},sr=\bar{r}s$. The physical DOFs are labeled by group elements of $S_3$. The conjugacy classes and irreps of the centralizers are shown in Table \ref{irreps_of_S3}, from which we can obtain the anyons in $D(S_3)$ quantum double, labeled by $\{\pmb{1},\pmb{A},\cdots,\pmb{G}\}$, see Table~\ref{Anyons_of_S3}.

All possible topological phase transitions from the $D(S_3)$ quantum double to Abelian topological phases can be obtained by deforming the ground state $|\Psi_0\rangle$ of the fixed point $D\left(S_3\right)$ quantum double model using a three-parameter deformation operator $Q(k_1,k_2,k_3)$:
\begin{equation}\label{deformed_PEPS_S3_Z2}
|\Psi(k_1,k_2,k_3)\rangle =Q^{\otimes N}(k_1,k_2,k_3)|\Psi_0\rangle,
\end{equation}
where the on-site deformation operator is a $6\times6$ matrix:
\begin{eqnarray}\label{deformation_Q}
  Q(k_1,k_2,k_3)&=&\left(
                     \begin{array}{cc}
                       q(k_1,k_2) & 0_{3\times3} \\
                       0_{3\times 3} & e^{-k_3}q(k_1,k_2) \\
                     \end{array}
                   \right),\notag\\
  q(k_1,k_2)&=&\left(
               \begin{array}{ccc}
                 1+2f(k_1) & f(k_2) & f(k_2) \\
                 f(k_2) & 1-f(k_1) & f(k_2) \\
                 f(k_2) & f(k_2) & 1-f(k_1) \\
               \end{array}
             \right), \notag \\
             f(k)&=&\frac{1-e^{-k}}{1+2 e^{-k}},\notag
\end{eqnarray}
and the ordering of the bases supporting $Q(k_1,k_2,k_3)$ is ${|e\rangle,|r\rangle,|\bar{r}\rangle,|s\rangle,|sr\rangle,|s\bar{r}\rangle}$. 
The phase diagram of the deformed PEPS is shown in Fig.~\ref{S_3_phase_diagram}. Every phase is labeled by the symmetry of the transfer operator fixed points. There are two different toric code phases, which are dual to each other. We focus on phase transitions along three axes. As shown in Appendix \ref{map_D_S3_to_stat_model_1}, along the $k_1$ and $k_2$ axes, the deformed PEPS can be mapped to the 3-state Potts model, from which we can infer that the critical points are at $k_c=\log(1+\sqrt{3})/2\approx0.5025$ and are described by the 3-state Potts universality class. In addition, for the deformation along the $k_2$ axis, the deformed PEPS can be mapped to the Rep$(S_3)$ string-net model deformed by a string tension; we discuss this case in subsection~\ref{example_rep_s3}. Finally,  along the $k_3$ axis, the deformed PEPS can be mapped to the Ising model, from which we infer that the critical point is at $k_c=\log(1+\sqrt{2})/2\approx 0.4407$ and belongs to the Ising universality class.

Next, we calculate the anyonic order parameters along the three axes. We compute every entry of $N^{\pmb{\gamma}}(\pmb{\alpha})$ and $M(\pmb{\alpha})$ defined in Eqs.~\eqref{N_1}, \eqref{N_gamma} and \eqref{entries_of_M}. These entries are contractions of norms or overlaps of PEPS, which can be performed by the variational uniform matrix product state (VUMPS) algorithm\cite{VUMPS_2018,VUMPS_2019}; the basic ideas are shown in Appendix.~\ref{prove_block_structure}.

\subsection{Phase transition from $D\left(S_3\right)$ quantum double to toric code}\label{example_1}

We consider the phase transition from the $D(S_3)$ quantum double to the $Z_2\boxtimes Z_1$ toric code along the $k_1$ axis.
The anyon condensation pattern through this phase transition has already been predicted\cite{Bais_Anyoncondensation_2009}, we summarize it in Table \ref{Anyons_of_S3}, where we label the trivial particle, electric charge, magnetic flux and fermion of toric code using the standard notions $\pmb{1}$, $\pmb{e}$, $\pmb{m}$ and $\pmb{f}=\pmb{e}\otimes\pmb{m}$, respectively. The Abelian chargon $\pmb{A}$ becomes $\pmb{e}$, the non-Abelian chargon $\pmb{C}$ splits to $\pmb{1}$ and $\pmb{e}$. The fluxon $\pmb{B}$ and the dyons $\pmb{D}$ and $\pmb{F}$ are fully confined. The fluxon $\pmb{F}$ and dyon $\pmb{G}$ are partially deconfined, and they become $\pmb{m}$ and $\pmb{f}$, respectively.

First we compute the condensate and identification order parameters of $\pmb{C}$, which comes from the singular values of the $2\times2$ matrices $N^{\pmb{1}}(\pmb{C})$ and $N^{\pmb{A}}(\pmb{C})$.  In the toric code phase, both $N^{\pmb{1}}(\pmb{C})$ and $N^{\pmb{A}}(\pmb{C})$ have one zero singular value and one nonzero singular value, separately. 
Therefore, $\pmb{C}$ partially condenses and partially identifies with $\pmb{A}$ in the toric code phase. 
The nonzero singular values of $N^{\pmb{1}}(\pmb{C})$ and $N^{\pmb{A}}(\pmb{C})$ are $1/9$, as shown in Fig. \ref{Fig_Vec_S3_Z2_all} (c); this is consistent with the 3-state Potts universality class expected from the duality mapping.

Next, we consider the deconfinement order parameters of the fluxon $\pmb{B}$ and the dyons $\pmb{D}$ and $\pmb{F}$. It can be found that the $4\times4$ matrices $M(\pmb{B})=M(\pmb{D})=M(\pmb{F})$.
Fig. \ref{Fig_Vec_S3_Z2_all} (a) shows the deconfinement order parameters of $\pmb{B}$, $\pmb{D}$ or $\pmb{E}$, which are the eigenvalues of the $2\times2$ matrices $M_{\text{str}}(\pmb{B})$, $M_{\text{str}}(\pmb{D})$ and $M_{\text{str}}(\pmb{F})$, and each of them are nonzero and two-fold degenerate. So, in the $D(S_3)$ phase, $\pmb{B}$, $\pmb{D}$ or $\pmb{E}$ are fully deconfined, and in the toric code phase $\pmb{B}$, $\pmb{D}$ or $\pmb{E}$ are fully confined. The critical exponent of these deconfinement order parameters is also $1/9$, as shown in Fig. \ref{Fig_Vec_S3_Z2_all} (b).

Then, we consider the splitting order parameter. The $2\times 2$ matrix $M_\text{str}(\pmb{C})$ always has two nonzero eigenvalues in both the $D(S_3)$ phase and the toric code phase, which indicates that $\pmb{C}$ is always deconfined. We compute the order parameters of splitting according to procedures shown in subsection \ref{order_para_splitting}. In Fig. \ref{Fig_Vec_S3_Z2_all} (d), we show the splitting order parameter $\tilde{M}_{\text{off}}(\pmb{C})$, it can be found that its critical exponent is also $1/9$, as displayed in Fig. \ref{Fig_Vec_S3_Z2_all} (f).

Finally we consider the fluxon $\pmb{F}$ and the dyon $\pmb{G}$, it can be found that the $9\times9$ matrices $M(\pmb{F})=M(\pmb{G})$.
Fig.~\ref{Fig_Vec_S3_Z2_all} (d) shows the eigenvalues of the $3\times 3$ matrices $M_{\text{str}}(\pmb{F})$ and $M_{\text{str}}(\pmb{G})$. It can also be found that $M_{\text{str}}(\pmb{F})$ ($M_{\text{str}}(\pmb{G})$) has one three-fold degenerate nonzero eigenvalue in the $D(S_3)$ phase and one nondegenerate nonzero eigenvalue in the toric code phase. The eigenvalues imply that the quantum dimensions of $\pmb{F}$ and $\pmb{G}$ reduce from 3 to 1. Moreover, there is a critical exponent $\beta\approx 0.061(2)$ associated with the partial deconfinement order parameters, as shown in Fig.~\ref{Fig_Vec_S3_Z2_all} (e); to the best of our knowledge, this critical exponent does not match any known critical exponents of the 3-state Potts model.

In order to test if the critical exponents of these order parameters are universal, we perturb the path of the phase transition along the $k_1$ axis by changing the deformation operator $Q(k_1,0,0)$ in Eq.~\eqref{deformed_PEPS_S3_Z2} to $Q(k_1,0,0)+A_{\text{rand}}$, where $A_{\text{rand}}$ is a small random matrix, such that we have another path of phase transition between the $D(S_3)$ quantum double and the $Z_2\boxtimes Z_1$ toric code. We observe numerically that on this new path of the phase transition, all critical exponents of the generalized condensate and deconfinement order parameters as well as the splitting order parameters do not change beyond the accuracy of the method, which suggests that these critical exponents could be universal.

\subsection{Phase transition from $\text{Rep}\left(S_3\right)$ string-net to toric code}\label{example_rep_s3}

Along the $k_2$ axis, there is a transition from the $D(S_3)$ phase to another $S_3\boxtimes Z_3$ toric code phase. Because the $k_2$ axis is dual to the $k_1$ axis, this phase transition is very similar to the previous one, the only difference is that in the $S_3\boxtimes Z_3$ toric code, $\pmb{C}$ is fully confined and $\pmb{B}$ splits into $\pmb{1}$ and $\pmb{e}$ of the toric code. The anyonic order parameters we proposed work perfectly for the phase transition along the $k_2$ axis. Instead of repeating the calculation of these order parameters for the $D(S_3)$ model, we will map the deformed $D(S_3)$ PEPS along the $k_2$ axis to the deformed Rep$(S_3)$ string-net PEPS, and thereby show that the anyonic order parameters also work for the string-net model. In particular, this will allow us to compare similarities and differences of the deformed Rep$(S_3)$ and $D(S_3)$ models. The condensation pattern of the phase transition between the Rep$(S_3)$ and the toric code phase is  summarized in Table \ref{Anyons_of_S3}.
Compared to the phase transition of the $D(S_3)$ model along the $k_2$ axis, the difference is that the Abelian chargon $\pmb{A}$ becomes $\pmb{m}$ of the toric code, and the non-Abelian fluxon $\pmb{B}$ splits into $\pmb{1}$ and $\pmb{m}$ of the toric code.

The physical DOFs of the Rep$(S_3)$ model are labeled by the irreps $\{1,\epsilon,\pi\}$ of $S_3$ . As shown in Appendix~\ref{Map_Rep(S3)_to stat_model}, the deformed $D(S_3)$ PEPS in Eq.~\eqref{deformed_PEPS_S3_Z2} along the $k_2$ axis is mapped to the following deformed Rep$(S_3)$ PEPS:
\begin{eqnarray}\label{deformed_PEPS_Rep_S3}
|\Psi(k_\pi)\rangle &=&Q_\text{rep}^{\otimes N}(k_{\pi})|\Psi_0\rangle,\notag\\
Q_\text{rep}(k_{\pi})&=&|1\rangle\langle1|+|\epsilon\rangle\langle\epsilon|+e^{-k_{\pi}}|\pi\rangle\langle\pi|,
\end{eqnarray}
where $|\Psi_0\rangle$ is the ground state of the fixed point Rep$(S_3)$ string-net model.
Here we only act the string tension operators $Q_\text{rep}(k_{\pi})$ on a subset of edges of honeycomb lattice, as shown in Appendix. \ref{Map_Rep(S3)_to stat_model}, such that the deformed Rep$(S_3)$ string-net model on a honeycomb lattice can be mapped to the 3-state Potts model on a square lattice, which are more convenient for studying the phase transitions\cite{Xu_Zhang_Zhang_2020,Xu_Schuch_2021}. Therefore, the deformed PEPS \eqref{deformed_PEPS_Rep_S3} will undergo a phase transition to the toric code phase at the critical point $k_c=\log(1+\sqrt{3})/2\approx 0.5025$, which is still described by the 3-state Potts universality class. Interestingly, there are two kinds of transfer operator fixed points in the toric code phase, the first kind has the symmetry $Z_2\boxtimes Z_1$, while the second kind has the symmetry $\mathcal{C}\boxtimes Z_2$, where $\mathcal{C}=\{O_1,O_{\epsilon},O_{\pi}\}$ and $Z_2=\{O_1,O_{\epsilon}\}$. Our approach works for both kinds of fixed points.
\begin{figure*}
  \centering
  \includegraphics[width=18cm]{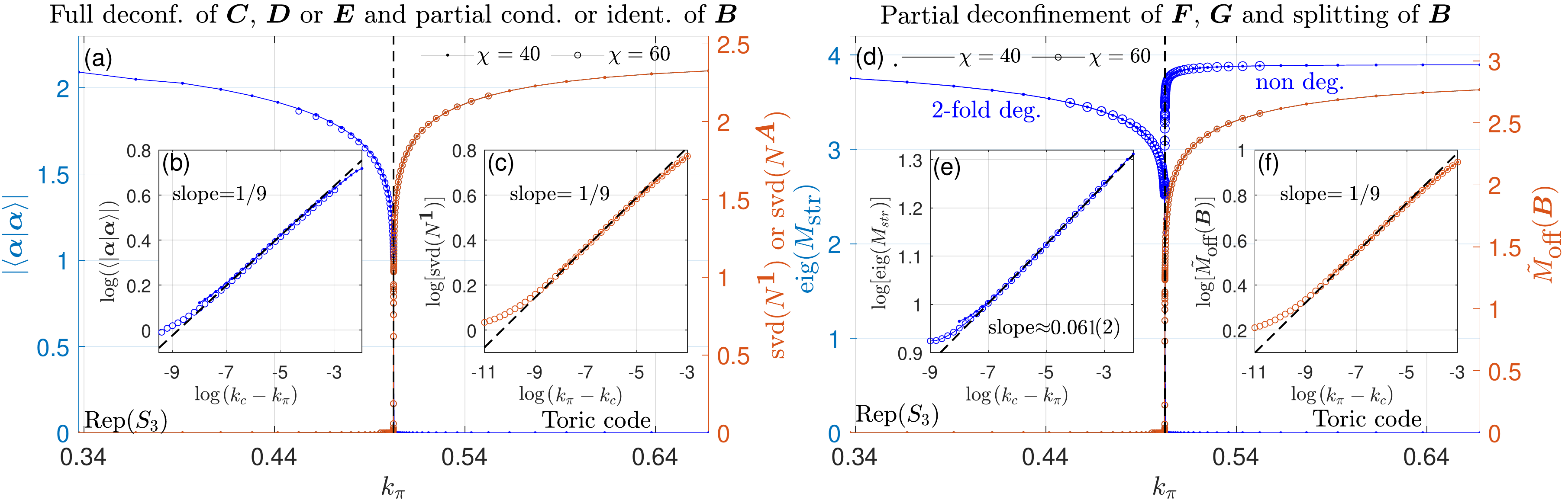}
  \caption{Anyonic order parameters of the Rep$(S_3)$ model and their critical exponents. The vertical dashed lines in (a) and (d) represent the exact location of the critical point $k_c=\log(1+\sqrt{3})/2$. The dashed lines in the insets have the slope given in each inset. $\chi$ is the bond dimension of the VUMPS. (a) Left: the deconfinement order parameters, where $\pmb{\alpha}$ can be $\pmb{C}$, $\pmb{D}$ or $\pmb{E}$. Right: the condensate (identification) order parameter, where svd$(N^{\pmb{1}})$ and svd$(\pmb{A})$ are the singular values of $N^{\pmb{1}}(\pmb{B})$ and $N^{\pmb{A}}(\pmb{B})$. The insets (b) and (c) show the critical scaling of the corresponding anyonic order parameters. (d) Left: the partial deconfinement order parameters of $\pmb{F}$ or $\pmb{G}$, eig($M_{\text{str}}$) denotes the eigenvalues of $M_{\text{str}}(\pmb{F})$ or $M_{\text{str}}(\pmb{G})$. Right: the splitting order parameter of $\pmb{B}$. The insets (e) and (f) show the critical scaling of the corresponding anyonic order parameters. }\label{Fig_All_conf_of_Rep_S3}
\end{figure*}

First we compute the condensate and identification order parameters, which come from the singular values of $2\times2$ matrices $N^{\pmb{1}}(\pmb{B})$ and $N^{\pmb{A}}(\pmb{B})$. In the toric code phase, $N^{\pmb{1}}(\pmb{B})$ and $N^{\pmb{A}}(\pmb{B})$ have one zero singular value and one nonzero singular value respectively, as shown in Fig.~\ref{Fig_All_conf_of_Rep_S3} (a).
Therefore, $\pmb{B}$ partially condenses and
partially identify with $\pmb{A}$ in the toric code phase.
We find that the critical exponent of these nonzero singular values is 1/9, as shown in Fig.~\ref{Fig_All_conf_of_Rep_S3} (c); this is consistent with the mapping to the 3-state Potts model.

Next, consider the deconfinement order parameters of the chargon $\pmb{C}$, dyons $\pmb{D}$ and $\pmb{E}$. In the Rep$(S_3)$ string-net model, the string and internal DOF labels are fixed to $\pi$ and the matrices $M(\pmb{C})$, $M(\pmb{D})$ and $M(\pmb{E})$ are $1\times 1$ and reduce to usual deconfinement order parameters, and they satisfy $M(\pmb{C})=M(\pmb{D})=M(\pmb{E})$.
Fig.~\ref{Fig_All_conf_of_Rep_S3} (a) shows these order parameters, and Fig.~\ref{Fig_All_conf_of_Rep_S3} (b) shows their scaling at criticality, which yet again yields a critical exponent of 1/9 for these these deconfinement order parameters.

Then, we consider the splitting order parameter. The
$2\times2$ matrix $M_{\text{str}}(\pmb{B})$ always has two nonzero eigenvalues in both the Rep$(S_3)$ phase and the toric code phase, which indicates that $\pmb{B}$ is always deconfined. In Fig. \ref{Fig_All_conf_of_Rep_S3} (d), we show the splitting order parameter $\tilde{M}_{\text{off}}(\pmb{B})$, and we observe that its critical exponent is also $1/9$, as shown in Fig. \ref{Fig_All_conf_of_Rep_S3} (f).

Finally, we consider the fluxons $\pmb{F}$ and dyon $\pmb{G}$. The matrices $M(\pmb{F})=M(\pmb{G})$ are $4\times 4$, the string and internal DOF labels of $\pmb{F}$ ($\pmb{G}$) can take $\{1,\pi\}$ ($\{\epsilon,\pi\}$). When calculating the partial trace, we should notice that the quantum dimension $d_{\pi}=2$.
Fig.~\ref{Fig_All_conf_of_Rep_S3} (d) shows the eigenvalues of the $2\times 2$ matrices $M_{\text{str}}(\pmb{F})$ and $M_{\text{str}}(\pmb{G})$. In the Rep($S_3)$ phase, there is a two-fold degenerate nonzero eigenvalue. In the toric code phase, $M_{\text{str}}(\pmb{F})$ ($M_{\text{str}}(\pmb{G})$) has one nonzero eigenvalue labeled by the string $1(\epsilon)$, and one zero eigenvalue labeled by $\pi$, so the quantum dimensions of $\pmb{F}$ and $\pmb{G}$ reduce from $3$ to $1$.} The critical exponent of the deconfinement order parameters is $\beta\approx 0.061(2)$, see Fig.~\ref{Fig_All_conf_of_Rep_S3} (e); this critical exponent is very close to the one found in the preceding subsection~\ref{example_1}.

\subsection{Phase transition from $D\left(S_3\right)$ quantum double to $D(Z_3)$ quantum double}
 In this subsection, we study the phase transition from the $D(S_3)$ quantum double to the $D(Z_3)$ quantum double along the $k_3$ axis; this phase transition has been investigated in Ref.~\cite{williamson_2017_SET}. We summarize the condensation pattern in Table \ref{Anyons_of_S3}. The Abelian charge $\pmb{A}$ condenses, the non-Abelian charge $\pmb{C}$ splits into two $Z_3$ chargons. The fluxon $\pmb{B}$ splits into two $Z_3$ fluxons. The two dyons $\pmb{D}$ and $\pmb{F}$ split into two $Z_3$ dyons, separately. The fluxon $\pmb{F}$ and the dyon $\pmb{G}$ are fully confined.
 Because $Z_3$ is a normal subgroup of $S_3$, there are some essential differences compared to the previous examples. First, there is no partial condensation (identification) and deconfinement. Second, unlike previous examples, the splitting of anyons can not be inferred from the partial condensation and identification, so the splitting order parameters are indispensable in this case.

\begin{figure}
  \centering
  \includegraphics[width=8.5cm]{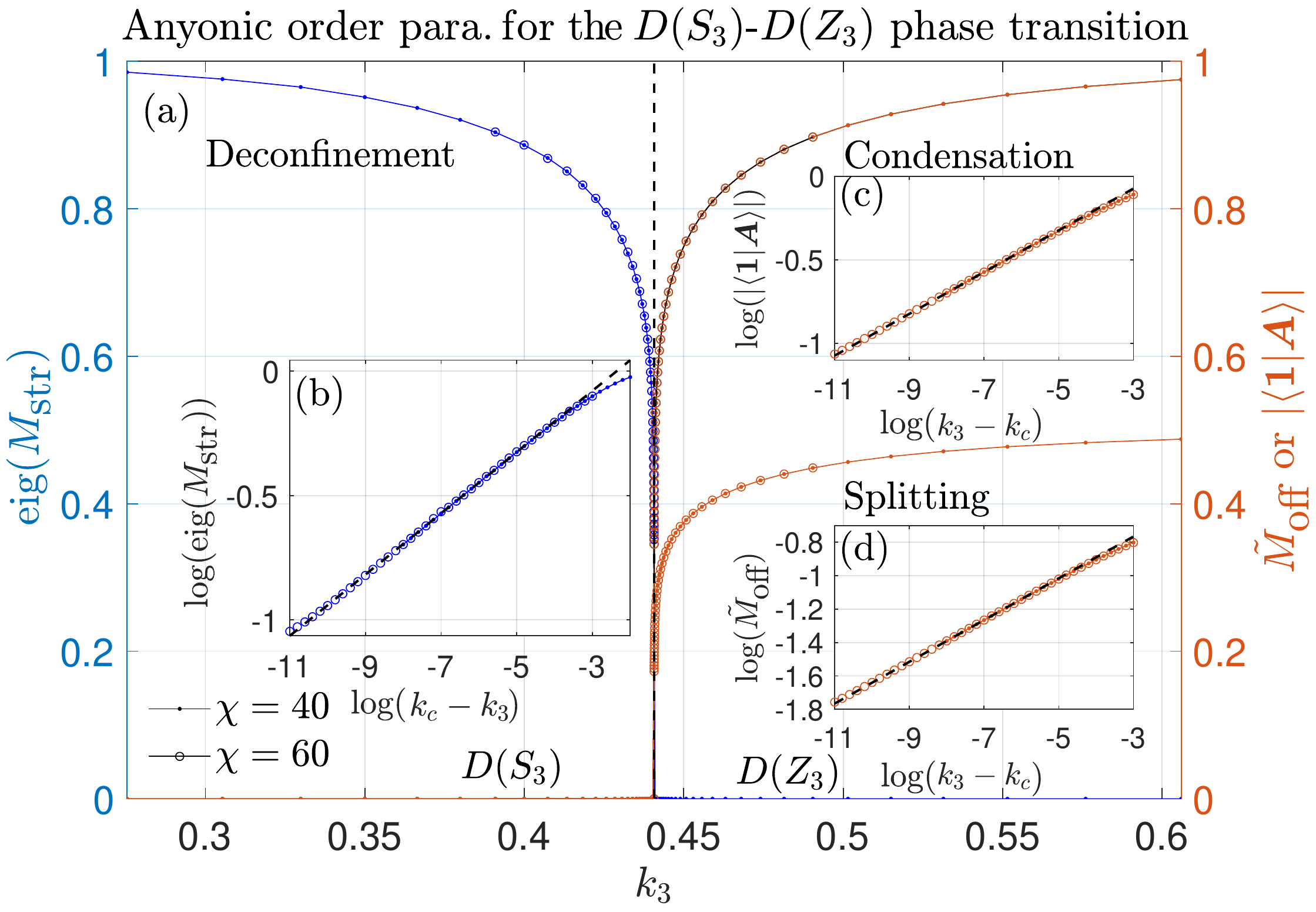}
  \caption{Anyonic order parameters for the phase transition between the $D(S_3)$ phase and the $D(Z_3)$ phase, and their critical exponents. The vertical dashed line in (a) indicates the location of the critical point $k_c=\log(1+\sqrt{2})/2$. The slope of the dashed lines in the insets is $1/8$. $\chi$ is the bond dimension of the VUMPS. (a) Left: the deconfinement order parameters of $\pmb{F}$ and $\pmb{G}$, and eig$(M_{\text{str}})$ denotes the eigenvalues of $M_{\text{str}}(\pmb{F})$ or $M_{\text{str}}(\pmb{G})$. Top right: the condensate order parameter of $\pmb{A}$. Bottom right: the splitting order parameters for $\pmb{C}$, $\pmb{B}$ $\pmb{D}$ or $\pmb{E}$. The insets (b), (c) and (d) show the scaling of the corresponding anyonic order parameters.}\label{Fig_All_conf_of_S3_Z3}
\end{figure}

  First, because $\pmb{A}$ is an Abelian anyon, $N^{\pmb{1}}(\pmb{A})=\langle\pmb{1}|\pmb{A}\rangle$ reduces to the usual condensate order parameter, which is shown in Fig. \ref{Fig_All_conf_of_S3_Z3} (a).
   The condensate order parameter can be mapped to the local order parameter of the
Ising model, so its critical exponent is  $1/8$, as displayed in Fig. \ref{Fig_All_conf_of_S3_Z3} (c).
Next we consider the splitting of $\pmb{C}$, $\pmb{B}$, $\pmb{D}$ and $\pmb{E}$, it can be found that the $4\times4$ matrices $M(\pmb{C})=M(\pmb{B})=M(\pmb{D})=M(\pmb{E})$. Fig. \ref{Fig_All_conf_of_S3_Z3} (a) also shows the splitting order parameters of $\pmb{C}$, $\pmb{B}$, $\pmb{D}$ and $\pmb{E}$, which are $\tilde{M}_{\text{off}}(\pmb{C})=\tilde{M}_{\text{off}}(\pmb{B})=\tilde{M}_{\text{off}}(\pmb{D})=\tilde{M}_{\text{off}}(\pmb{E})$.
Their critical exponents are also $1/8$, as shown in Fig.~\ref{Fig_All_conf_of_S3_Z3} (d).
Finally we consider the deconfinement order parameters of the fluxon $\pmb{F}$ and the dyon $\pmb{G}$. It can be found that the $9\times9$ matrices $M(\pmb{F})=M(\pmb{G})$.
Fig. \ref{Fig_All_conf_of_S3_Z3} (a) shows the deconfinement order parameters of $\pmb{F}$ and $\pmb{G}$, which are the eigenvalues of the $3\times3$ matrices $M_{\text{str}}(\pmb{F})$ and $M_{\text{str}}(\pmb{G})$, and each of them are three-fold degenerate. In $D(S_3)$ phase, because all eigenvalues are zero, $\pmb{F}$ and $\pmb{G}$ are fully confined. The critical exponent of these eigenvalues is also $1/8$, as displayed in Fig. \ref{Fig_All_conf_of_S3_Z3} (b). Again, these critical exponents are what we expect from the mapping to the 2D Ising model.

\section{Example: phase transition from DIsing string-net to toric code}\label{Example_2}
 In this section, we study anyon splitting through the phase transition from the DIsing string-net model to the toric code model. The input data of the DIsing string-net model are the Ising category with three objects $\{1,\sigma,\psi\}$, and the local physical DOFs of the DIsing string-net model are labeled by these objects. There are nine anyons in the DIsing topological order, given by the pairs in $\{\pmb{1},\pmb{\sigma},\pmb{\psi}\}\otimes\{\pmb{1},\bar{\pmb{\sigma}},\bar{\pmb{\psi}}\}$, as displayed in Table \ref{DIsing_anyon}. The anyon condensation pattern through this phase transition has been proposed in Ref.~\cite{Bais_Anyoncondensation_2009}.
 By condensing $\pmb{\psi}\bar{\pmb{\psi}}$, it can be found that $\pmb{\sigma}$, $\bar{\pmb{\sigma}}$, $\pmb{\sigma}\bar{\pmb\psi}$ and $\pmb{\psi}\bar{\pmb\sigma}$ are confined, $\pmb{\sigma}\bar{\pmb{\sigma}}$ splits into $\pmb{e}$ and $\pmb{m}$ of the toric code, $\pmb{\psi}$ and $\pmb{\bar{\psi}}$ become $\pmb{f}$ of the toric code. We summarize this condensation pattern in Table \ref{DIsing_anyon}.

 The phase transition from the DIsing phase to the toric code phase can be achieved by deforming the ground state $|\Psi_0\rangle$ of the fixed point DIsing string-net model using a two-parameter string tension operator $Q(k_{\sigma},k_{\psi})$:
\begin{eqnarray}\label{deformed_DIsing_PEPS}
|\Psi(k_{\sigma},k_{\psi})\rangle &=&Q^{\otimes N}(k_{\sigma},k_{\psi})|\Psi_0\rangle,\notag\\
Q(k_{\sigma},k_{\psi})&=&|1\rangle\langle1|+e^{-k_{\sigma}}|\sigma\rangle\langle\sigma|+e^{-k_{\psi}}|\psi\rangle\langle\psi|.
\end{eqnarray}
Similar to the example of the Rep$(S_3)$ string-net, we only act the deformation operator $Q(k_{\sigma},k_\psi)$ on a subset of edges of a honeycomb lattice, as shown in Appendix~\ref{Map_DIsing_to stat_model}. 
\begin{table*}
\centering
\caption{Anyons of the DIsing topological order and the condensation pattern}
\begin{tabularx}{\linewidth}{C{3cm}C{1.4cm}C{1.4cm}C{1.4cm}C{1.4cm}C{1cm}C{1cm}C{1.4cm}C{1.4cm}C{1.4cm}C{1.4cm}}
\toprule[1pt]
Anyon $\pmb{\alpha}$& $\pmb{1}$ & $\bar{\pmb{\sigma}}$ & $\bar{\pmb{\psi}}$ & $\pmb{\sigma}$ & \multicolumn{2}{c}{$\pmb{\sigma}\bar{\pmb{\sigma}}$}& $\pmb{\sigma}\bar{\pmb{\psi}}$ & $\pmb{\psi} $ & $\pmb{\psi} \bar{\pmb{\sigma}}$ & $\pmb{\psi}\bar{\pmb{\psi}}$ \\
\midrule[1pt]
Quantum dim. & 1 & $\sqrt{2}$ & 1 & $\sqrt{2}$ & \multicolumn{2}{c}{2} & $\sqrt{2}$ & 1 & $\sqrt{2}$ & 1 \\
Topo. spin & 1 & $\exp \left(-\frac{\pi i }{8}\right)$ & -1 & $\exp \left(\frac{\pi i}{8}\right)$ & \multicolumn{2}{c}{1} & $\exp \left(-\frac{7\pi  i}{8}\right)$
& -1 & $\exp \left(\frac{7\pi  i}{8}\right)$ & 1 \\
Cond. pattern & $\pmb{1}$ & conf. & $\pmb{f}$ & conf.  & $\pmb{e}$&$\pmb{m}$ & conf. & $\pmb{f}$ & conf. & $\pmb{1}$ \\
\bottomrule[1pt]
\end{tabularx}\label{DIsing_anyon}
\end{table*}

\begin{figure}
  \centering
  \includegraphics[width=7.5cm]{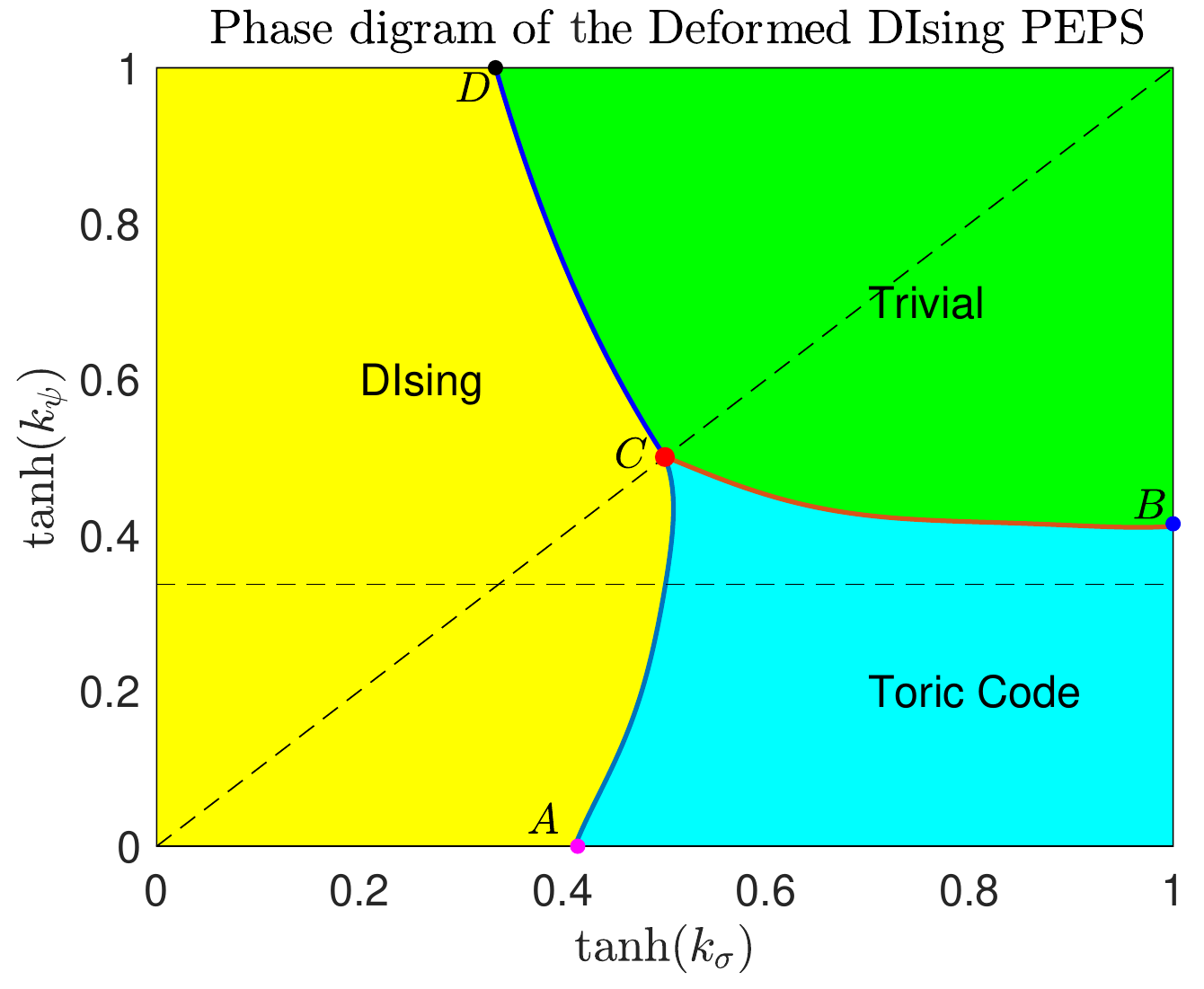}
  \caption{Phase diagram of the deformed DIsing PEPS. The two dashed lines are $k_\sigma=k_\psi$ and $k_\psi=0.35$. If we express the coordinates using $(k_\sigma,k_\psi)$, the special points are  $A=(\log(1+\sqrt{2})/2,0)$, $B=(+\infty,\log(1+\sqrt{2})/4)$, $C=(\log\sqrt{3},\log\sqrt{3})$ and $D=(\log\sqrt{2},+\infty)$.}\label{DIsing_phase_diagram}
\end{figure}

Inspired by Ref.~\cite{Gils_2009_Ashkin_Teller_DIsing}, we find that the deformed PEPS can be mapped to the 2-dimensional classical Ashkin-Teller model on a square lattice, as shown in Appendix \ref{Map_DIsing_to stat_model}, from which the phase diagram of the deformed DIsing PEPS can be obtained, as displayed in Fig.~\ref{DIsing_phase_diagram}. The phase diagram is similar to that obtained by deforming the Hamiltonian in Ref.~\cite{Ising_Hamitonian_phase_diagram}. Here we restrict the range of parameter to $k_\sigma\geqslant 0$ and $k_\psi\geqslant 0$ for simplicity. There are a DIsing phase, a toric code phase and a trivial phase. The phase boundaries $AC$ and $BC$ belong to the 2d Ising universality class. The phase boundary $DC$ belongs to the Ashkin-Teller universality class with continuously varying critical exponents. The tricritical point $C$ is described by the 4-state Potts universality class.

\begin{figure}
  \centering
  \includegraphics[width=8.8cm]{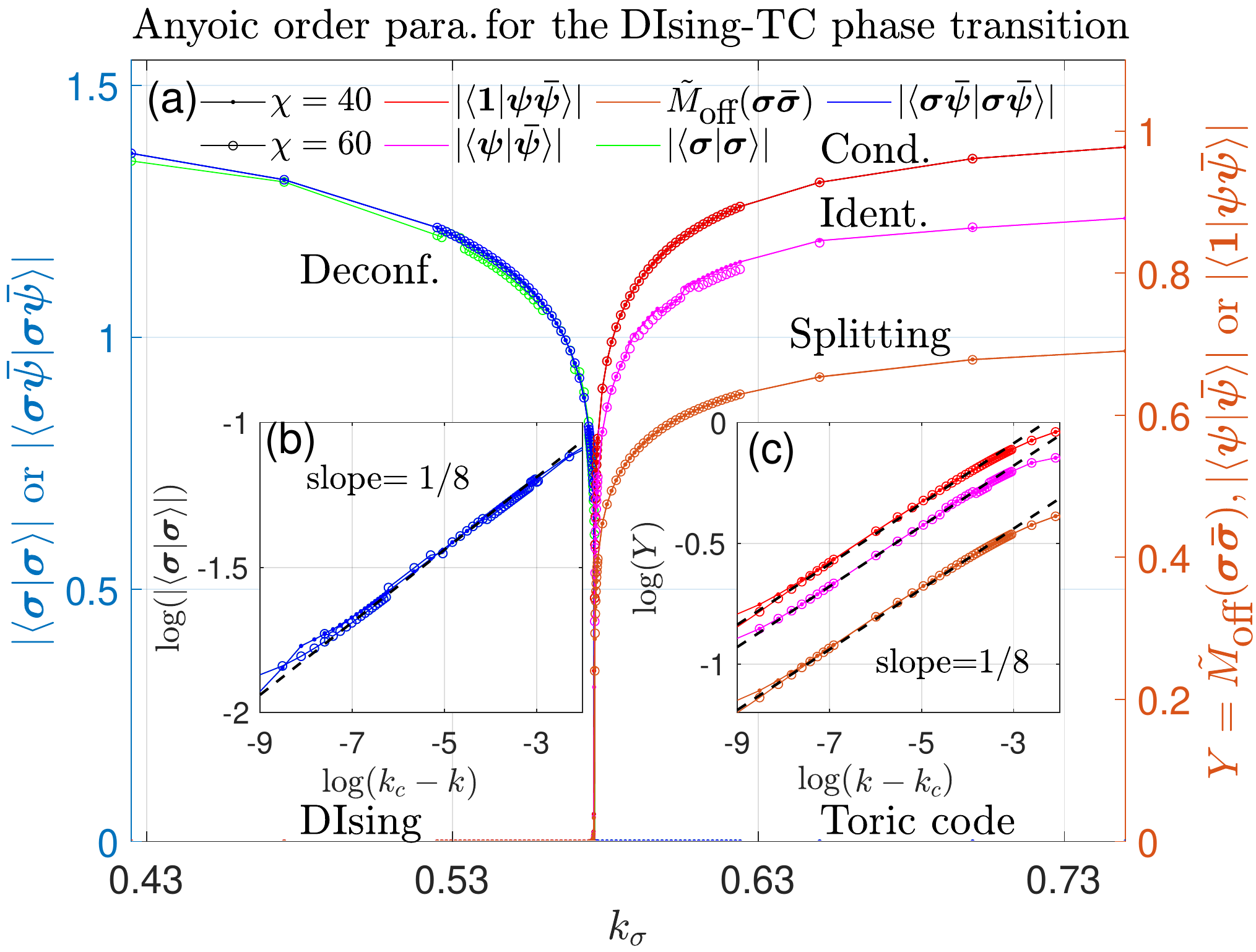}
  \caption{Anyonic order parameters for the phase transition between the DIsing phase and the toric code phase, and their critical exponents. The slope of dashed lines in the insets is $1/8$. $\chi$ is the bond dimension of VUMPS. (a) The condensate order parameter $|\langle\pmb{1}|\pmb{\psi}\bar{\pmb{\psi}}\rangle|$, identification order parameter $|\langle\pmb{\psi}|\bar{\pmb{\psi}}\rangle|$,  deconfinement order parameters $|\langle\pmb{\sigma}|\pmb{\sigma}\rangle|$ and $|\langle\pmb{\sigma}\bar{\pmb{\psi}}|\pmb{\sigma}\bar{\pmb{\psi}}\rangle|$, and splitting order parameter $\tilde{M}_{\text{off}}(\pmb{\sigma}\bar{\pmb{\sigma}})$ . The insets (b) and (c) shows their critical scaling.}\label{DIsing_TC_order_para}
\end{figure}

We compute the anyonic order parameters along the line $k_\psi=0.35$, which is a path of the phase transition between the DIsing phase and the toric code phase, the critical point $(k_\psi,k_\sigma)\approx(0.35, 0.5765)$ is described by the Ising university class. First we consider the condensate order parameter of $\pmb{\psi}\bar{\pmb{\psi}}$ and the identification order parameter of $\pmb{\psi}$ and $\bar{\pmb{\psi}}$, these anyons are Abelian and they have usual anyonic order parameters. Fig.~\ref{DIsing_TC_order_para} (a)
shows the condensate order parameter $\langle\pmb{1}|\pmb{\psi}\bar{\pmb{\psi}}\rangle$ and the identification order parameter $\langle\pmb{\psi}|\bar{\pmb{\psi}}\rangle$. The critical scaling shown in Fig.~\ref{DIsing_TC_order_para} (c) yields that their
critical exponents are $1/8$, consistent with the expected Ising universality class. Then we consider the the deconfinement order parameters of $\pmb{\sigma},\bar{\pmb{\sigma}},\pmb{\sigma}\bar{\pmb{\psi}}$ and $\pmb{\psi}\bar{\pmb{\sigma}}$. Since the string and internal DOF labels are fixed to $\sigma$ for these anyons, their $M$ matrices are $1\times 1$. In addition, it can be found that $M(\pmb{\sigma})=M(\pmb{\bar{\sigma}})=M(\pmb{\sigma}\bar{\pmb{\psi}})=M(\pmb{\psi}\bar{\pmb{\sigma}})$.
Fig. \ref{DIsing_TC_order_para} (a) also displays the deconfinement order parameters of $\pmb{\sigma}$ and $\pmb{\sigma}\bar{\pmb{\psi}}$, and their critical exponents are $1/8$, see Fig. \ref{DIsing_TC_order_para} (d).

Finally we consider the splitting order parameter of $\pmb{\sigma}\bar{\pmb{\sigma}}$. $M(\pmb{\sigma}\bar{\pmb{\sigma}})$ is $4\times4$, and the $2\times 2$ matrix $M_{\text{str}}(\pmb{\sigma}\bar{\pmb{\sigma}})$ always has two nonzero eigenvalues, implying that $\pmb{\sigma}\bar{\pmb{\sigma}}$ is always deconfined. In Fig.~\ref{DIsing_TC_order_para} (a), we show the computed splitting order parameters $\tilde{M}_{\text{off}}(\pmb{\sigma}\bar{\pmb{\sigma}})$; we find that
its critical exponent is also $1/8$, see Fig.~\ref{DIsing_TC_order_para} (c).

\section{Conclusions and discussions}\label{conclusion}

In this paper, we have extended the PEPS characterization of condensation and deconfinement of Abelian anyons to include partial condensation, partial deconfinement, and splitting of non-Abelian anyons in the PEPS characterization. We showed that anyon splitting can be observed from the topologically degenerate ground states in terms of MES. We constructed a complete set of PEPS $|\pmb{\alpha}_x\rangle_s$ carrying a non-Abelian anyon $\pmb{\alpha}$ with different strings $s$ and internal DOFs $x$. We generalized the usual condensate order parameter to a matrix $N^{\pmb{1}}(\pmb{\alpha})$ given by the overlaps between the set of PEPS $|\pmb{\alpha}_x\rangle_s$ and the ground state $|\pmb{1}\rangle$. We also generalized the usual deconfinement order parameter to a matrix $M(\pmb{\alpha})$ defined by the norms and overlaps among the PEPS  $|\pmb{\alpha}_x\rangle_s$. From the singular values of $N^{\pmb{1}}(\pmb{\alpha})$, we can distinguish full and partial condensation. By partially tracing $M(\pmb{\alpha})$,   $M_{\text{str}}(\pmb{\alpha})=\text{tr}_{\text{int}}[M(\pmb{\alpha})]$, we can identify partial and full deconfinement from the eigenvalues of $M_{\text{str}}(\pmb{\alpha})$. Moreover, we show that if the anyon $\pmb{\alpha}$ does not split, $M(\pmb{\alpha})$ has a tensor product structure. By expressing $M(\pmb{\alpha})$ in a suitable orthonormal basis, we obtain a matrix $M^{\prime}(\pmb{\alpha})$ which allows to construct the order parameters $\tilde{M}_{\text{off}}(\pmb{\alpha})$ detecting the splitting of the non-Abelian anyon $\pmb{\alpha}$ from its eigenvalues.

  We also propose a three-parameter phase diagram of the deformed $D(S_3)$ quantum double PEPS, where there are a $Z_2\boxtimes Z_1$ toric code phase and a dual $S_3\boxtimes Z_3$ toric code phase, as well as a $D(Z_3)$ quantum double phase. The deformed $D(S_3)$ PEPS along the $k_2$ axis is also dual to the string tension deformed Rep$(S_3)$ string-net PEPS. 
  Through the phase transitions from the $D(S_3)$ or the Rep$(S_3)$ phase to the toric code phase, partial condensation and partial deconfinement occurs, which we observe through the singular values of $N^{\pmb{1}}(\pmb{\alpha})$ and eigenvalues of $M_{\text{str}}(\pmb{\alpha})$. We demonstrate that through all phase transitions, the splitting order parameters $\tilde{M}_{\text{off}}(\pmb{\alpha})$ are fully capable of detecting anyon splitting. In addition, we also propose a two-parameter phase diagram of the deformed DIsing PEPS, and we detect the splitting of the anyon $\pmb{\sigma}\bar{\pmb{\sigma}}$ using the splitting order parameter $\tilde{M}_{\text{off}}(\pmb{\sigma}\bar{\pmb{\sigma}})$ through the phase transition to the toric code phase.

There are also some remaining open questions. First, we prove our statements about anyon splitting using the language of group theory; it could be better to rigorously describe anyon splitting using the language of category theory\cite{kong_anyon_cond_category} or of Hopf algebras~\cite{molnar_hopf}, such that all nonchiral topological states are included. Second, when partial deconfinement happens, we cannot observe it from the leading eigenvalue of the transfer operator, and the critical exponent of the deconfinement order parameter cannot be derived from the universality class of the critical point (similar to the critical exponent for deconfinement found in Ref.~\cite{Iqbal_2021_PRX}), so a better understanding of partial deconfinement is necessary. Third, one can easily design Hamiltonians with tuning parameters realizing all transitions in our examples, and variantionally optimizing PEPS to approach the ground states of the Hamiltonians \cite{Iqbal_2021_PRX}. However, it is very challenging to construct the generalized condensate and deconfinement order parameters as well as the splitting order parameters for the variationally optimized PEPS. These problems deserve further explorations.

\begin{acknowledgments}
W.-T. Xu thanks Mingru Yang for helpful discussion and Qi Zhang for providing the VUMPS codes. This work has been supported by the European Research Council (ERC) under the
European Union's Horizon 2020 research and innovation programme through the
ERC-CoG SEQUAM (Grant Agreement No.~863476). The computational results
presented have been achieved using the Vienna Scientific Cluster (VSC).
\end{acknowledgments}

\bibliography{refs}
\newpage
\appendix
\section{Definition of PEPS tensors, idempotents, nilpotents and endpoint tensors}\label{definition}
\subsection{Quantum double model}

The local tensor generating a ground state PEPS of the quantum double of $D(G)$ is\cite{peps_degeneracy_2010}
\begin{equation}\label{tensor_of_S3}
   \includegraphics[width=2.5cm,valign=c]{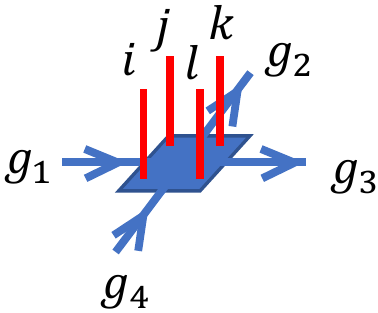}=\delta_{j,g_1g_2^{-1}}\delta_{k,g_2g_3^{-1}}\delta_{l,g_3g_4^{-1}}\delta_{i,g_4g_1^{-1}},
\end{equation}
where both the physical and virtual indices are group elements of $G$. It can be checked that the pulling through condition of the tensor is
\begin{equation}
   \includegraphics[width=5cm,valign=c]{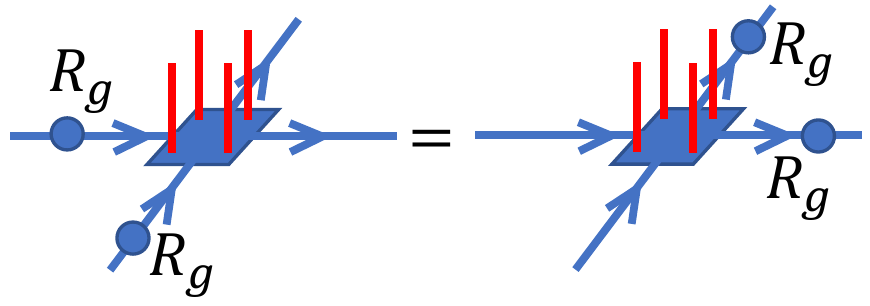},
\end{equation}
where $R_g$ is the right regular representation, i.e., $R_g|h)=|hg^{-1})$. And the round brackets denote the vector space of virtual level. Applying the gauge transformation $\sum_{g}|g^{-1})( g|$ to the PEPS, the virtual symmetry operator can be changed to the left regular representation, which matches with the convention in other papers, i.e., Ref.~\cite{D_4_PEPS}. The enlarged MPO is\cite{TNS_RMP_2021}
\begin{equation}
  O_{g,k}=R_g^{\otimes N}\otimes | gkg^{-1})(k|.
\end{equation}
It corresponds to the MPO $O_{ac}^{bd}$ in Fig. \ref{definition_of_MPOs} (d) with $a=gkg^{-1}$, $c=k$, $d=g$ and there is no index $b$. They satisfy
\begin{eqnarray}
  O_{g,k}O_{g^\prime,k^{\prime}}&=&\delta_{k,g^\prime k^\prime g^{\prime -1}}O_{gg^{\prime},k^\prime},\nonumber\\
  O^{\dagger}_{g,k}&=&O_{g^{-1},gkg^{-1}}.
\end{eqnarray}
We define the following notations: conjugacy classes of $G$ as $K$, centralizers of $k$ as $C_k$ and matrices of irreps $\gamma$ of $C_k$ as $I^{C_k}_{\gamma}$. Then, the central idempotents can be expressed as
\begin{equation}
  P^{(K,\gamma)}=\sum_{k\in K}\frac{\dim \gamma}{|C_k|}\sum_{g\in C_k} \text{tr}[I_{\gamma}^{C_k}(g)] O_{g,k}.
\end{equation}
They satisfy properties (\ref{property_of_CI}):
\begin{equation}
  P^{(K,\gamma)}P^{(K^\prime,\delta)}=\delta_{K,K^\prime}\delta_{\gamma,\delta}P^{(K,\gamma)}.
\end{equation}
The simple idempotents $(p=q$ and $h\in C_k)$ and nilpotents $(p\neq q$ or $h\notin C_k)$ are\cite{williamson_2017_SET}:
\begin{equation}\label{idem_nil_of_QD}
  P^{(K,\gamma)}_{(p,hkh^{-1}),(q,k)}=\frac{\dim \gamma}{|C_k|}\sum_{g\in hC_k} [I^{C_k}_{\gamma}(g)_{pq}] O_{g,k},
\end{equation}
where $h$ is a representative of the coset: $G/C_k=\{C_k,\cdots,hC_k,\cdots\}$. So the dimension of the central idempotent $P^{(K,\gamma)}$ is $d_{(K,\gamma)}=\dim(I^{C_k}_{\gamma})|K|$, and $\sum_{K,\gamma}d_{K,\gamma}^2=|G|^2$ is the dimension of the quantum double algebra $D(G)$.

The excited state $|(K,\gamma)_{(q,hkh^{-1})}\rangle_{(p,k)}$ in Fig.~\ref{Idempotent_excitation} (d) can be created by inserting at the virtual level of the PEPS the operator\cite{Pepe_2017}
\begin{eqnarray}\label{anyon_creation_operator}
\left(\bigotimes_{\text{str}} R_k\right)\otimes E^{(K,\gamma)}_{(p,k),(q,hkh^{-1})},
\end{eqnarray}
where the tensor product is over a path of a semi-infinity long string, and the end point tensor is
\begin{equation}\label{end_tensor_QD}
 E^{(K,\gamma)}_{(p,k),(q,hkh^{-1})}=\sum_{g\in C_k}I^{C_k}_{\gamma}(g^{-1})_{pq} |hg)(hg|.
\end{equation}

\subsection{String-net model}\label{app_string_net}
The PEPS tensors for the string-net models can be defined using the data $\{d_i,N_{ij}^k, F^{ijk}_{tsu}\}$ from the input fusion category, which are the quantum dimensions, the fusion multiplicities and the $F$-symbols, respectively\cite{GuTensorNetwork_2009,buerschaper_explicit_2009}. The nonzero entries of the $F$ tensor are determined by $N_{ij}^k$,
i.e., $F^{ijk}_{tsu}$ is nonzero if $N_{ij}^kN_{ts}^kN_{is}^uN_{jt}^u\neq0$. For the DIsing string-net, $(d_1,d_\sigma,d_\psi)=(1,\sqrt{2},1)$. The nonzero entries of the fusion multiplicity tensor $N$ are
\begin{equation}
  N_{11}^{1}=N_{\sigma 1}^{\sigma}=N_{\psi 1}^{\psi}=N_{\sigma \sigma}^{\psi}=1,
\end{equation}
 up to the permutation of the indices. The nontrivial entries of $F$ are
\begin{eqnarray}
  F_{\sigma\sigma1}^{\sigma\sigma1}&=&F_{\sigma\sigma1}^{\sigma\sigma\psi}=F_{\sigma\sigma\psi}^{\sigma\sigma1}
  =F_{\sigma\sigma\psi}^{\sigma\sigma\psi}=\frac{1}{\sqrt{2}},\nonumber\\
  F_{\sigma\psi\sigma}^{\sigma\psi\sigma}&=&F_{\psi\sigma\sigma}^{\psi\sigma\sigma}=-1,
\end{eqnarray}
and other nonzeros entries are $1$.
For the Rep$(S_3)$ string-net, $(d_1,d_\epsilon,d_\pi)=(1,1,2)$. The nonzero entries of the fusion multiplicity tensor $N$ are
\begin{equation}
  N_{11}^{1}=N_{\epsilon 1}^{\epsilon}=N_{\pi 1}^{\pi}=N_{\pi \epsilon}^{\pi}=N_{\pi \pi}^{\pi}=1,
\end{equation}
 up to the permutation of the indices. The nontrivial entries of $F$ allowed are
\begin{eqnarray}
  F_{\pi\pi\pi}^{\epsilon\pi\pi}&=&F_{\pi\pi\pi}^{\pi\epsilon\pi}=F^{\pi\pi\pi}_{\epsilon\pi\pi}
  =F^{\pi\pi\pi}_{\pi\epsilon\pi}=-1,\nonumber\\
   F_{\pi\pi1}^{\pi\pi1}&=&F_{\pi\pi1}^{\pi\pi\epsilon}=F_{\pi\pi\epsilon}^{\pi\pi1}
  =F_{\pi\pi\epsilon}^{\pi\pi\epsilon}=\frac{1}{2},\nonumber\\
  F_{\pi\pi1}^{\pi\pi\pi}&=&F_{\pi\pi\pi}^{\pi\pi1}=\frac{1}{\sqrt{2}},\nonumber\\
  F_{\pi\pi\epsilon}^{\pi\pi\pi}&=&F_{\pi\pi\pi}^{\pi\pi\epsilon}=-\frac{1}{\sqrt{2}},  F_{\pi\pi\pi}^{\pi\pi\pi}=0,
\end{eqnarray}
and other allowed entries are $1$.

 From the $F$ tensor, it is convenient to define the $G$ tensor:
\begin{equation}
  G^{ijk}_{\alpha\beta\gamma}=F^{ijk}_{\alpha\beta\gamma}/\sqrt{d_kd_\gamma}.
\end{equation}
A triple-line local tensor generating the PEPS of a string-net ground state can be expressed as\cite{GuTensorNetwork_2009,buerschaper_explicit_2009}:
\begin{equation}\label{string-net_tensor}
 \includegraphics[width=2cm,valign=c]{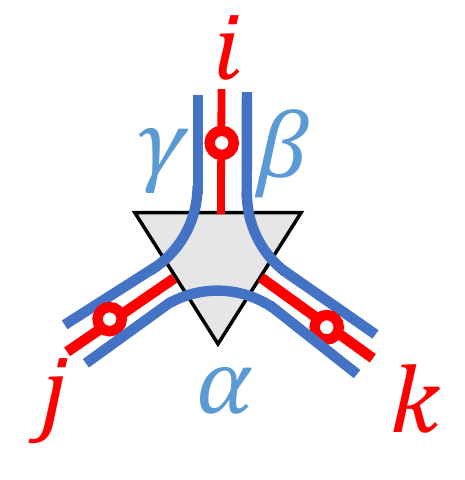}=(d_{i}d_{j}d_{k})^\frac{1}{4}G^{ijk}_{\alpha\beta\gamma},
\end{equation}
where the open circles represent the physical DOFs and the lines are the virtual DOFs.
 In the PEPS representation of the string-net wavefunctions,
 there is a convention that when contracting virtual DOFs, we should at first assign the quantum dimensions to the DOFs that will be contracted. The local tensor on a square lattice is obtained
  by contracting two of the previous tensors:
\begin{equation}\label{double_tensor_string_net}
 \includegraphics[width=3cm,valign=c]{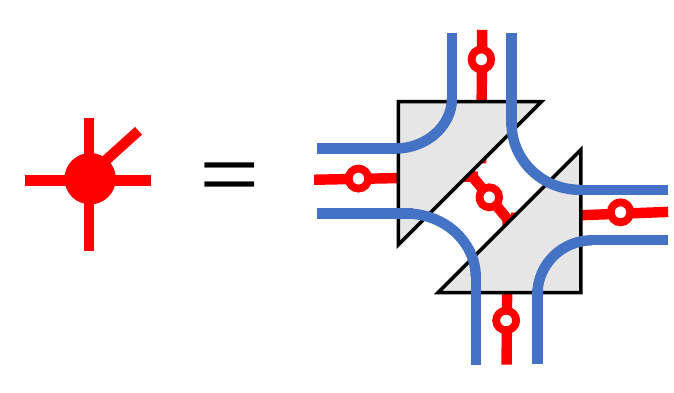}.
\end{equation}
Besides, the local tensor has five physical indices denoted by the red circles, the square root of deformation operators $\sqrt{Q}$ denoted by the red squares act on four of five physical indices:
\begin{equation}
 \includegraphics[width=4cm,valign=c]{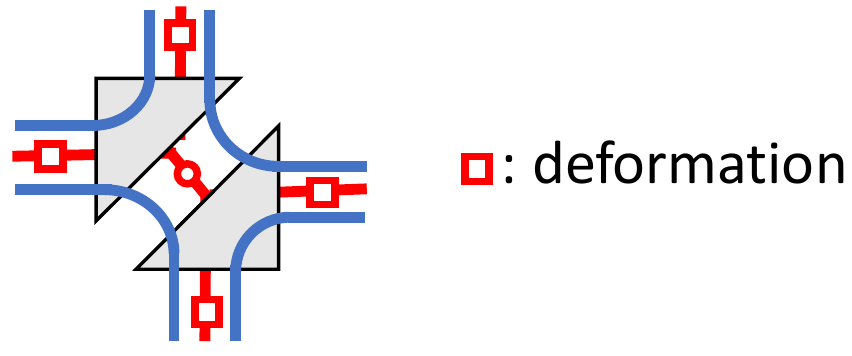}.\notag
\end{equation}
In the main text, we have deformed both Rep$(S_3)$ and DIsing string-net on a honeycomb lattice in this way such that they can be mapped to the statistical mechanics models on a square lattice, as shown in Appendices \ref{Map_Rep(S3)_to stat_model}, \ref{Map_DIsing_to stat_model} and Fig.~\ref{QD_honey} (b).

The local tensor of the MPO $O_n$ is\cite{MPO_algebra_2017}:
\begin{equation}\label{MPO_tensor}
 \includegraphics[width=3cm,valign=c]{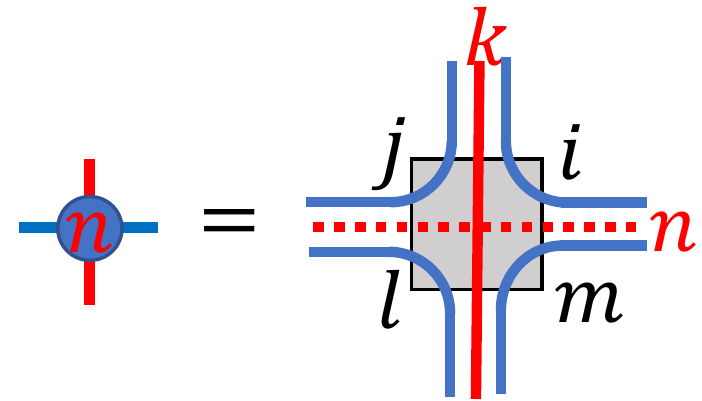}=G_{lmn}^{ijk},
\end{equation}
where $n$ is a fixed index of the input category. In the abbreviated graphs, the red (blue) lines represent
the triple-line (double-line) in the original graphs.
Furthermore, by defining the tensor with the fixed indices $i,j,k$ and $l$:
\begin{equation}
 \includegraphics[width=3.5cm,valign=c]{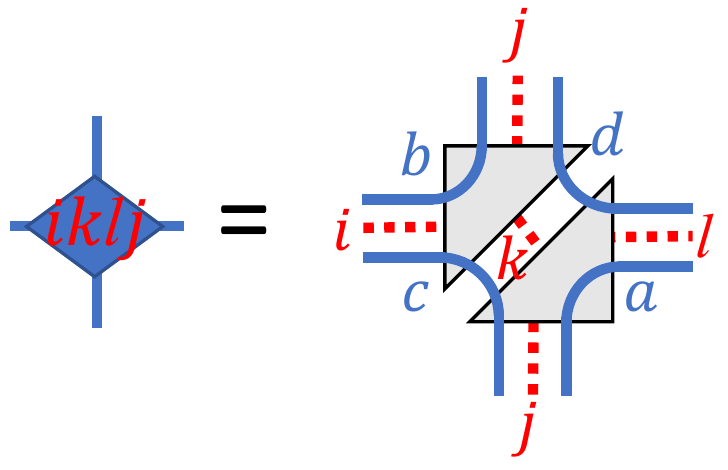}=G^{ijk}_{dcb}G^{jkl}_{dac},
\end{equation}
the enlarged MPO $O_{il}^{kj}$ in Fig.~\ref{definition_of_MPOs} (d) can be generated together with the tensor (it should be rotated by $\pi/2$) in Eq. \eqref{MPO_tensor}:
\begin{equation}
 O_{il}^{kj}=\includegraphics[width=1cm,valign=c]{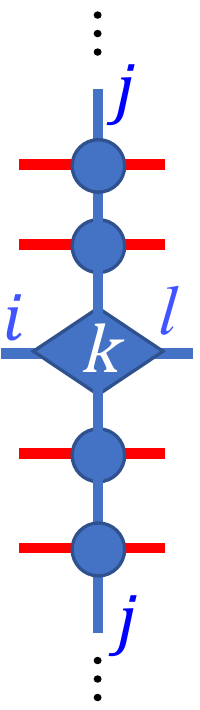},
\end{equation}
where the up and down legs are connected.

For the string-net models whose input categories are modular, the anyons can be labeled as $\pmb{a}\bar{\pmb{b}}$, where $\bar{\pmb{b}}$ is the time reversal counterpart of $\pmb{b}$. The simple idempotents and nilpotents are
\begin{eqnarray}\label{idempotents}
P^{\bm{a}\pmb{\bar{b}}}_{il}&=&\frac{d_a d_b }{\sum_{n}d_n^2}\sum_{kj}d_j d_kC_{il}^{kj}(\pmb{a}\bar{\pmb{b}})O_{il}^{kj},\nonumber\\
  C_{il}^{kj}(\pmb{a}\pmb{\bar{b}})&=&\sum_{\gamma\delta} d_\gamma d_\delta R^{aj}_{\gamma}  R^{j b}_{\delta}G^{i\delta\gamma}_{j a b}G^{k a \delta}_{b j l}G^{i k j}_{a \gamma \delta},
\end{eqnarray}
where indices $(a,b)$ are determined by $\pmb{a}\bar{\pmb{b}}$, and the tensor $R^{ij}_k$ characterizes the braiding of the strings $i$ and $j$ in the fusion channel $k$. The nonzero entries of the $R$ tensor for the DIsing string-net are
\begin{eqnarray}
R^{11}_{1}&=&R^{1\sigma}_{\sigma}=R^{\sigma1}_{\sigma}=R^{\psi1}_{\psi}=R^{1\psi}_{\psi}=1,\nonumber\\
R^{\sigma\psi}_{\sigma}&=&R^{\psi\sigma}_{\sigma}=-i, \quad R^{\psi\psi}_{1}=-1,\nonumber\\
R^{\sigma\sigma}_{1}&=&\exp(-\frac{\pi i}{8}),\quad R^{\sigma\sigma}_{\psi}=\exp(\frac{3\pi i}{8}).
\end{eqnarray}
The end tensor carrying anyonic excitations with a string $k$ and an internal DOF $l$ is defined as\cite{Turaev-Viro_2022}
\begin{eqnarray}\label{end_tensor}
&& E^{\bm{\alpha}}_{kl}=\includegraphics[width=4cm,valign=c]{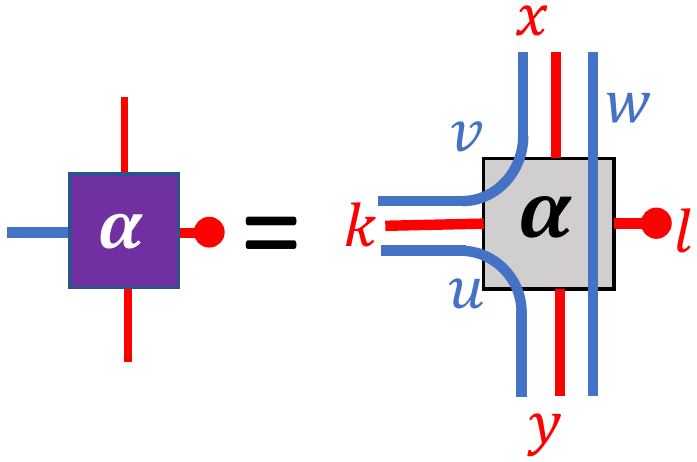}\notag\\
 &&=d_a^{\frac{1}{4}}d_b^{\frac{1}{4}}d_x^{\frac{1}{4}}d_y^{\frac{1}{4}}d_k\sum_{\beta}d_\beta C_{k l}^{\beta w}(\pmb{\alpha})G^{u\beta x}_{w vk}G_{x\beta l}^{wyu}.
\end{eqnarray}

However, because the Rep$(G)$ is not a modular category, we do not have the $R$ tensor hence we can not use Eqs. \eqref{idempotents} and \eqref{end_tensor} to obtain the idempotents, nilpotents and end tensors. According to Refs. \cite{MPO_algebra_2017,extended_Levin_wen_2018}, the idempotents and nilpotents of the Rep$(S_3)$ model are given by:
\begin{equation}
P^{\pmb{\alpha}}_{il}=\sum_{kj}d_k^{3/2}d_jC_{il}^{kj}(\pmb{\alpha})O_{il}^{kj},
\end{equation}
where the nonzero entries of $C^{kj}_{il}(\pmb{\alpha})$ are
\begin{widetext}
\begin{eqnarray}
  C^{11}_{11}(\pmb{1})&=&C^{\epsilon\epsilon}_{11}(\pmb{1})=C^{\pi\pi}_{11}(\pmb{1})=\frac{1}{6};\quad
 C^{\epsilon 1}_{\epsilon\epsilon}(\pmb{A})=C^{1\epsilon}_{\epsilon\epsilon}(\pmb{A})=\frac{1}{6}, \quad C_{\epsilon\epsilon}^{\pi\pi}(\pmb{A})=-\frac{1}{6};\nonumber\\
  C^{\pi1}_{\pi\pi}(\pmb{C})&=&C^{1\pi}_{\pi\pi}(\pmb{C})=C^{\pi\pi}_{\pi\pi}(\pmb{C})=\frac{1}{6},\quad C^{\pi\epsilon}_{\pi\pi}(\pmb{C})=C^{\epsilon\pi}_{\pi\pi}(\pmb{C})=-\frac{1}{6};\nonumber\\
   C^{11}_{11}(\pmb{B})&=&C^{\epsilon1}_{\epsilon\epsilon}(\pmb{B})=C^{\epsilon\epsilon}_{11}(\pmb{B})=C^{1\epsilon}_{\epsilon\epsilon}(\pmb{B})=\frac{1}{3},\quad C^{\pi\pi}_{11}(\pmb{B})=-\frac{1}{6},\quad  C^{\pi\pi}_{\epsilon\epsilon}(\pmb{B})=\frac{1}{6},\quad C^{\pi\pi}_{\epsilon1}(\pmb{B})=-\frac{i}{2\sqrt{3}},\,  C^{\pi\pi}_{1\epsilon}(\pmb{B})=\frac{i}{2\sqrt{3}};\nonumber\\
   C^{\pi1}_{\pi\pi}(\pmb{D})&=&\frac{1}{6}, \,C^{\pi\epsilon}_{\pi\pi}(\pmb{D})=-\frac{1}{6},\quad  C^{1\pi}_{\pi\pi}(\pmb{D})=\frac{e^{\frac{2\pi i}{3}}}{6},\quad \quad  C^{\epsilon\pi}_{\pi\pi}(\pmb{D})=\frac{e^{-\frac{\pi i}{3}}}{6},\quad C^{\pi\pi}_{\pi\pi}(\pmb{D})=\frac{e^{-\frac{2\pi i}{3}}}{6};\nonumber\\
   C^{\pi1}_{\pi\pi}(\pmb{E})&=&\frac{1}{6}, \quad C^{\pi\epsilon}_{\pi\pi}(\pmb{E})=-\frac{1}{6}, \quad C^{1\pi}_{\pi\pi}(\pmb{E})=\frac{e^{-\frac{2\pi i}{3}}}{6},\quad  C^{\epsilon\pi}_{\pi\pi}(\pmb{E})=\frac{e^{\frac{\pi i}{3}}}{6},\quad C^{\pi\pi}_{\pi\pi}(\pmb{E})=\frac{e^{\frac{2\pi i}{3}}}{6};\nonumber\\
   C^{\pi1}_{\pi\pi}(\pmb{F})&=&C^{\pi\epsilon}_{\pi\pi}(\pmb{F})=C^{1\pi}_{\pi\pi}(\pmb{F})=C^{\epsilon\pi}_{\pi\pi}(\pmb{F})=\frac{1}{4},\quad
   C^{11}_{11}(\pmb{F})=\frac{1}{2}, \quad C^{\epsilon\epsilon}_{11}(\pmb{F})=-\frac{1}{2},\quad C^{\pi\pi}_{\pi 1}(\pmb{F})=\frac{1}{ 2^{5/4}}, \quad C^{\pi\pi}_{1\pi}(\pmb{F})=\frac{1}{2^{7/4}};\nonumber\\
   C^{\pi1}_{\pi\pi}(\pmb{G})&=&C^{\pi\epsilon}_{\pi\pi}(\pmb{G})=\frac{1}{4},\quad C^{1\pi}_{\pi\pi}(\pmb{G})=C^{\epsilon\pi}_{\pi\pi}(\pmb{G})=-\frac{1}{4},\quad
   C^{\epsilon1}_{\epsilon\epsilon}(\pmb{G})=\frac{1}{2}, \quad C^{1\epsilon}_{\epsilon\epsilon}(\pmb{G})=-\frac{1}{2},\, C^{\pi\pi}_{\pi\epsilon}(\pmb{G})=\frac{i}{ 2^{5/4}}, \quad C^{\pi\pi}_{\epsilon\pi}(\pmb{G})=\frac{i}{2^{7/4}}.\nonumber
\end{eqnarray}
\end{widetext}
 The end tensors of the Rep$(S_3)$ model can also be constructed by simply replacing the coefficients in Eq. \eqref{end_tensor} with the above ones.

\section{Anyon splitting in the ground state}\label{GS_splitting}

In this Appendix, using the quantum double models we prove Eq.~\eqref{GS_split}, and we show it again here:
\begin{equation}
  |\Psi^{\pmb{\alpha}}\rangle=\sum_{\pmb{x}}c_{\pmb{\alpha x}}|\Psi^{\pmb{x}}\rangle.
\end{equation}
We consider the phase transition from the $D(G)$ quantum double to the $D(H)$ quantum double, where the virtual symmetry of the PEPS breaks from $G$ down to $H\subset G$, and we assume that $H$ is an Abelian group for simplicity. $G$ can be divided into the cosets $Hq$, where $q$ is a representative of the coset $Hq$. On the virtual level of PEPS, because symmetry $G$ breaks down to $H$, the PEPS becomes a superposition:
\begin{equation}\label{GS_SB}
  |\Psi\rangle=\sum_{q}|\Psi\rangle_{q}.
\end{equation}
Notice that $|\Psi\rangle_{q}, \forall q$ are the same wavefunction, but they are PEPS with different virtual symmetries, $|\Psi\rangle_{q}$ has the virtual symmetry $H^\prime={q^{-1}}Hq$.

Then we consider the splitting of a chargon: $\pmb{\alpha}\rightarrow \bigoplus_{\pmb{x}}c_{\pmb{\alpha}\pmb{x}}\pmb{x}$, where $\pmb{\alpha}=(K_1,\alpha)$ and $K_1=\{e\}$. In other words:
\begin{equation}\label{subduced_rep}
  I^{G}_{\alpha}(h)=\bigoplus_{x} c_{\pmb{\alpha x}} I^{H}_{x}(h),\quad h\in H.
\end{equation}
 Notice that here we choose a special irrep $\alpha$ of $G$ according to $H$. Since we assume that $H$ is Abelian, its irreps are one dimensional and $I^{G}_{\alpha}(h)$ is diagonal with nonzero entries $I^{G}_{\alpha}(h)_{kk}=I^{H}_{x}(h)$. A map $f$ from an index $k$ of $I^{G}_{\alpha}(h)$ to an irrep $x$ of $H$ can be defined: $x=f(k)$.

Consider a simple idempotent of the chargon $\pmb{\alpha}$:
\begin{equation}
  P^{\pmb{\alpha}}_{kk}=\frac{d_{\alpha}}{|G|}\sum_{g\in G}I^{G}_{\alpha}(g)_{kk} R^{\otimes N}(g).
\end{equation}
In the PEPS $|\Psi\rangle_{q}$, the strings $g\notin q^{-1}Hq$ are confined, therefore inserting $P^{\pmb{\alpha}}_{kk}$ into $|\Psi\rangle_{q}$ is equivalent to inserting
\begin{eqnarray}
  P^{\pmb{\alpha},q}_{kk}&=&\frac{d_{\alpha}}{|G|}\sum_{g\in q^{-1}Hq}I^{G}_{\alpha}(g)_{kk} R^{\otimes N}(g)\notag\\
 &=&\frac{d_{\alpha}}{|G|}\sum_{h\in H}\sum_{l}|I^{G}_{\alpha}(q)_{lk}|^2I^{H}_{f(l)}(h) R^{\otimes N}(q^{-1}hq),\notag\\
  &=&\frac{d_{\alpha}}{|G|}\sum_{h\in q^{-1}Hq}\sum_{l}|I^{G}_{\alpha}(q)_{lk}|^2I^{q^{-1}Hq}_{f(l)}(h) R^{\otimes N}(h),\notag
\end{eqnarray}
which gives rise to
\begin{equation}
  \frac{d_{\alpha}|H|}{|G|}\sum_{l}|I^{G}_{\alpha}(q)_{lk}|^2|\Psi^{f(l)}\rangle.
\end{equation}
Considering Eq.~\eqref{GS_SB}, we sum over $q$ and obtain
\begin{equation}
  |\Psi^{\pmb{\alpha}}\rangle=\sum_{l}|\Psi^{f(l)}\rangle=\sum_{\pmb{x}}c_{\pmb{\alpha x}}|\Psi^{\pmb{x}}\rangle,
\end{equation}
where we use the relation $\sum_{q}|I^{G}_{\alpha}(q)_{lk}|^2=|G|/(|H|d_{\alpha})$. So we have proven Eq.~\eqref{GS_split} for chargons.

Next let us consider fluxons. To split the fluxon $\pmb{\alpha}=(K,1)$, we require that $K\subset q^{-1}Hq, \forall q$ and $H$ is Abelian. Then each $k\in K$ becomes a conjugacy class when the symmetry breaks from $G$ down to $q^{-1}Hq$, hence the fluxon $\pmb{\alpha}$  splits. In addition, since $q^{-1}Hq$ is an Abelian group, the centralizer $C_k^{q^{-1}Hq}=\{h\in q^{-1}Hq|hk=kh\}=q^{-1}Hq$. Moreover, because $C_k^{q^{-1}Hq}\subset C_k^{G}$, we have $q^{-1}Hq\subset C_k^{G}$.

After the phase transition, we can insert a global horizontal string $k$ into both sides of Eq.~\eqref{GS_SB}, which gives rise to
\begin{equation}\label{B1}
  |\Psi_k\rangle=\sum_{q}|\Psi_k\rangle_{q},
\end{equation}
where $|\Psi_k\rangle_{q}$ still has the virtual symmetry $q^{-1}Hq$.
 Consider a simple idempotent of the fluxon $\pmb{\alpha}$:
 \begin{equation}
  P^{\pmb{\alpha}}_{kk}=\frac{1}{|C^{G}_k|}\sum_{g\in C^{G}_k} R^{\otimes N}_{g}\otimes |k)( k|.
\end{equation}
Inserting $P^{\pmb{\alpha}}_{kk}$ into the PEPS $|\Psi_k\rangle_p$ is equivalent to inserting
 \begin{equation}
  P^{\pmb{\alpha}}_{kk}=\frac{1}{|C^{G}_k|}\sum_{g\in qHq^{-1}} R^{\otimes N}_{g}\otimes |k)( k|,
\end{equation}
which results in
 \begin{equation}
  \frac{|H|}{|C_k^G|}|\Psi^{(q^{-1}kq,1)}\rangle.
\end{equation}
Since $H\subset C_k^G\subset G$, $C_k^G$ can be divided as the cosets $Hp$ and $G$ can be divided as the cosets $C_k^Gr$, and we can write $q=pr$ where $[p,k]=0$.
According to Eq.~\eqref{B1}, we sum over $q$ and obtain
\begin{eqnarray}
  |\Psi^{\pmb{\alpha}}\rangle&=&\sum_{q}\frac{|H|}{|C_k^G|}|\Psi^{(q^{-1}kq,1)}\notag\rangle\\
  &=&\sum_{r}|\Psi^{(r^{-1}kr,1)}\rangle=\sum_{\pmb{x}}c_{\pmb{\alpha x}}|\Psi^{\pmb{x}}\rangle,
\end{eqnarray}
where we define the mapping $\pmb{x}=f(r)$ via $\pmb{x}=(r^{-1}kr,1)$ and $c_{\pmb{\alpha x}}=\sum_r\delta_{\pmb{x},f(r)}$. So we have proven Eq.~\eqref{GS_split} for fluxons. Combining the proofs for chargons and fluxons, Eq.~\eqref{GS_split} can be proven for dyons straightforwardly.

Next, we show that if $\pmb{\alpha}$ splits, the dominant eigenvalues $\lambda_{\pmb{\alpha}}^{\pmb{\alpha}}$ are degenerate.
The MES norm can be expressed in terms of the MES transfer operators $\mathbb{T}^{\pmb{\gamma}}_{\pmb{\gamma}}$:
\begin{equation}
  \langle\Psi^{\pmb{\gamma}}|\Psi^{\pmb{\gamma}}\rangle=\text{tr}[\lim_{N\rightarrow \infty}(\mathbb{T}^{\pmb{\gamma}}_{\pmb{\gamma}})^N].
\end{equation}
The transfer operator of the left-hand side of Eq.~\eqref{GS_split} is $\mathbb{T}_{\pmb{\alpha}}^{\pmb{\alpha}}$ and that of the right-hand side is $\bigoplus_{\pmb{x}}c^2_{\pmb{\alpha x}}\mathbb{T}_{\pmb{x}}^{\pmb{x}}$. $\mathbb{T}_{\pmb{\alpha}}^{\pmb{\alpha}}$ and  $\bigoplus_{\pmb{x}}c^2_{\pmb{\alpha x}}\mathbb{T}_{\pmb{x}}^{\pmb{x}}$ differ by a similar transformation and they share the spectrum. If each $\mathbb{T}_{\pmb{x}}^{\pmb{x}}$ has a nondegenerate dominant eigenvalue, then, the degeneracy of $\lambda^{\pmb{\alpha}}_{\pmb{\alpha}}$ is $\sum_{\pmb{x}}c^2_{\pmb{\alpha x}}$.

\section{The block diagonal structure of $M(\pmb{\alpha})$}\label{prove_block_structure}
In this subsection, we show the structure of $M(\pmb{\alpha})$ both before and after the phase transitions using the quantum double models. We at first prove that before the phase transition, $M(\pmb{\alpha})$ has the tensor product structure:
\begin{equation}\label{form_of_M_matrix}
  M(\pmb{\alpha})=\mathbbm{1}_{\text{str}}\otimes B_{\text{int}}(\pmb{\alpha}).
\end{equation}
We start from the double tensor
\begin{equation}
\includegraphics[width=5cm,valign=c]{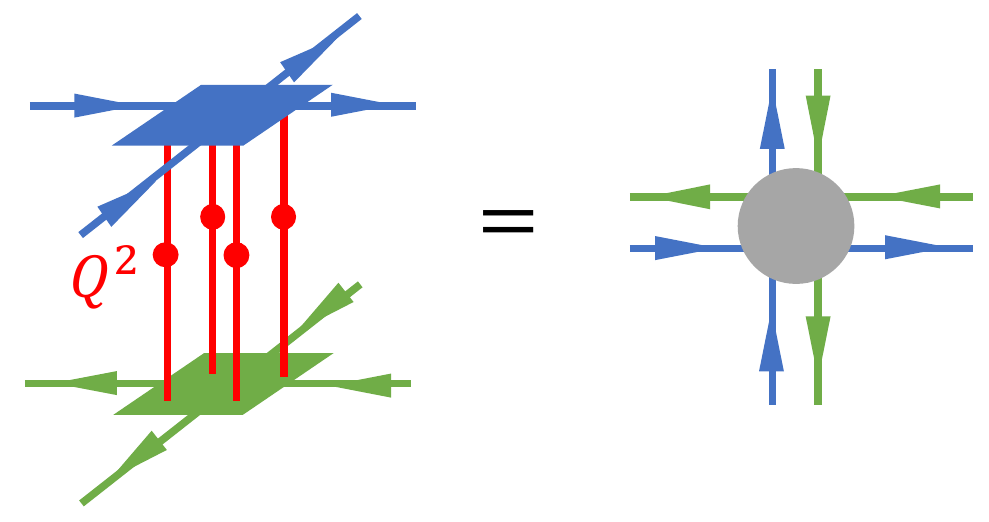},
\end{equation}
where $Q$ is the deformation operator in Eq.~\eqref{deformation_Q}. Let us consider a general transfer operator $\mathbbm{T}_k^{k^\prime}$ generated by the double tensor and the symmetry operators:
\begin{equation}\label{different_strings}
\mathbb{T}_k^{k^\prime}=\includegraphics[width=7.5cm,valign=c]{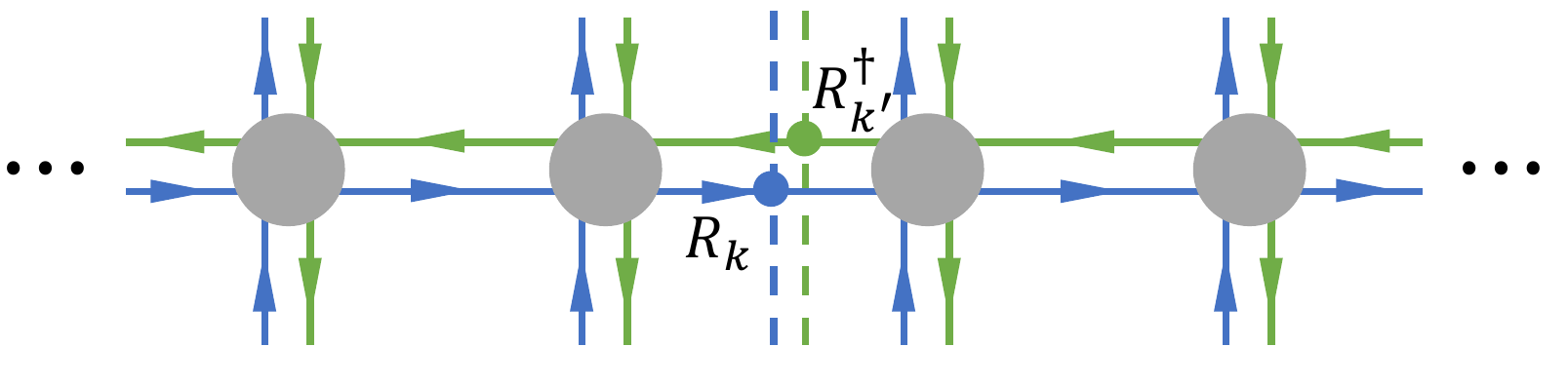}.
\end{equation}
The bond dimension of the dash lines is $1$. When $k=k^{\prime}=e$, the identity element of group $G$, we denote $\mathbb{T}_e^{e}=\mathbb{T}$. Its fixed points $\rho$ can be approximated by an matrix product state (MPS):
 \begin{equation}\label{fixed_point_rho}
\includegraphics[width=7cm,valign=c]{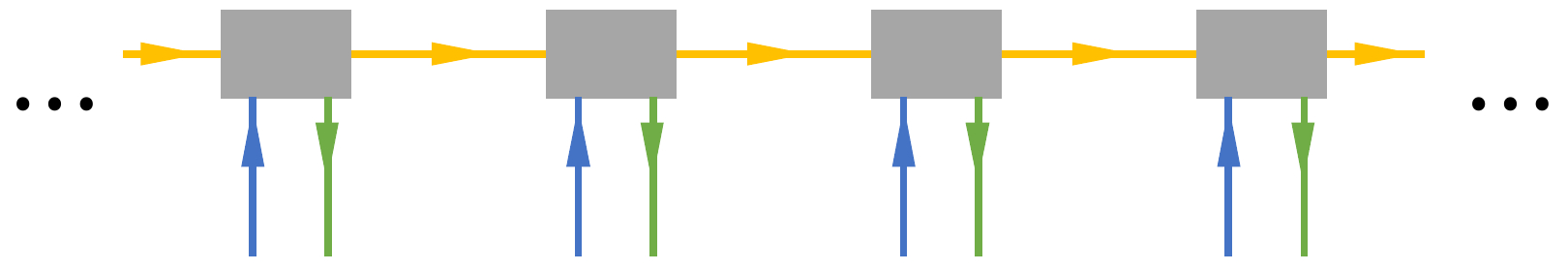},
\end{equation}
where the orange lines represent the virtual DOFs of the MPS. We can calculate this fixed point MPS using the VUMPS algorithm. It is known that in the original topological phase, the fixed point $\rho$ has the symmetry $G\boxtimes Z_1$\cite{AnyonsCondensation_Z4_2017,AnyonsCondensation_Z4_2018}, namely,
\begin{equation}\label{symmetry of_fixed point}
  R_g^{\otimes N} \rho=\rho R_g^{\otimes N} , g\in G.
\end{equation}
Notice that for non-Abelian groups, there is a group automorphism: $g\rightarrow hgh^{-1}, h\in G$, therefore $\rho$ may have the symmetry $R_g^{\otimes N} \rho=\rho R_{hgh^{-1}}^{\otimes N}$. But we can redefine $\rho\leftarrow\rho R_h^{\otimes N}$ such that we change the symmetry of $\rho$ to Eq.~\eqref{symmetry of_fixed point}. Following Ref.~\cite{AnyonsCondensation_Z4_2017}, it is reasonable to assume that the MPS $\rho$ is injective. Because different representations of an injective MPS are related by a gauge transformation\cite{MPDO_2017}, the MPS tensor satisfies:
\begin{equation}\label{symmetry_of_local_tensor}
\includegraphics[width=5cm,valign=c]{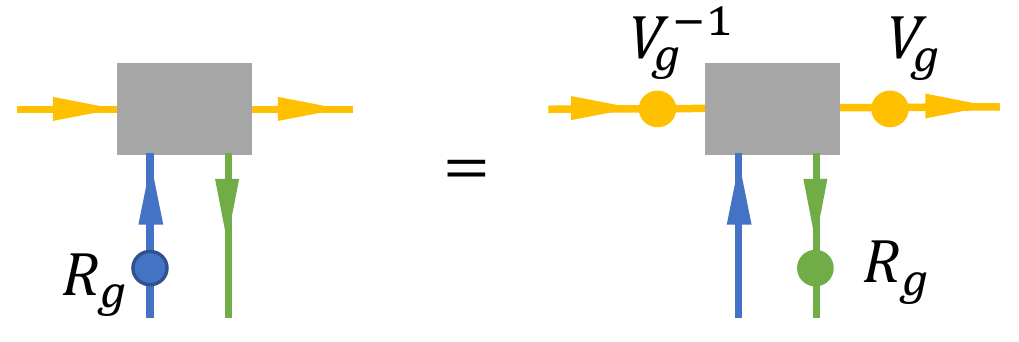}.
\end{equation}
In general, $V_g$ can be a projective representation of $g$, here we assume that it is a linear representation of $g$ for simplicity. Moreover, we can define the channel operator $T$ and its fixed point $F$:
\begin{equation}\label{channel_fixed_point}
\includegraphics[width=6cm,valign=c]{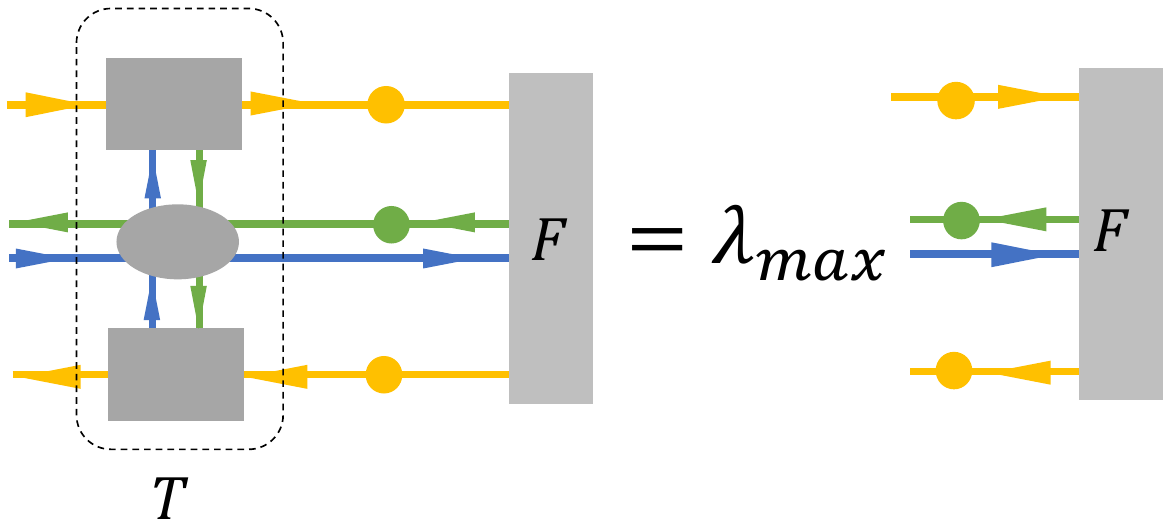},
\end{equation}
where $\lambda_{max}$ is the largest eigenvalue of the channel operator. The channel fixed point $F$ has the symmetry
\begin{equation}\label{symmetry_of_fixed_points}
\includegraphics[width=4.5cm,valign=c]{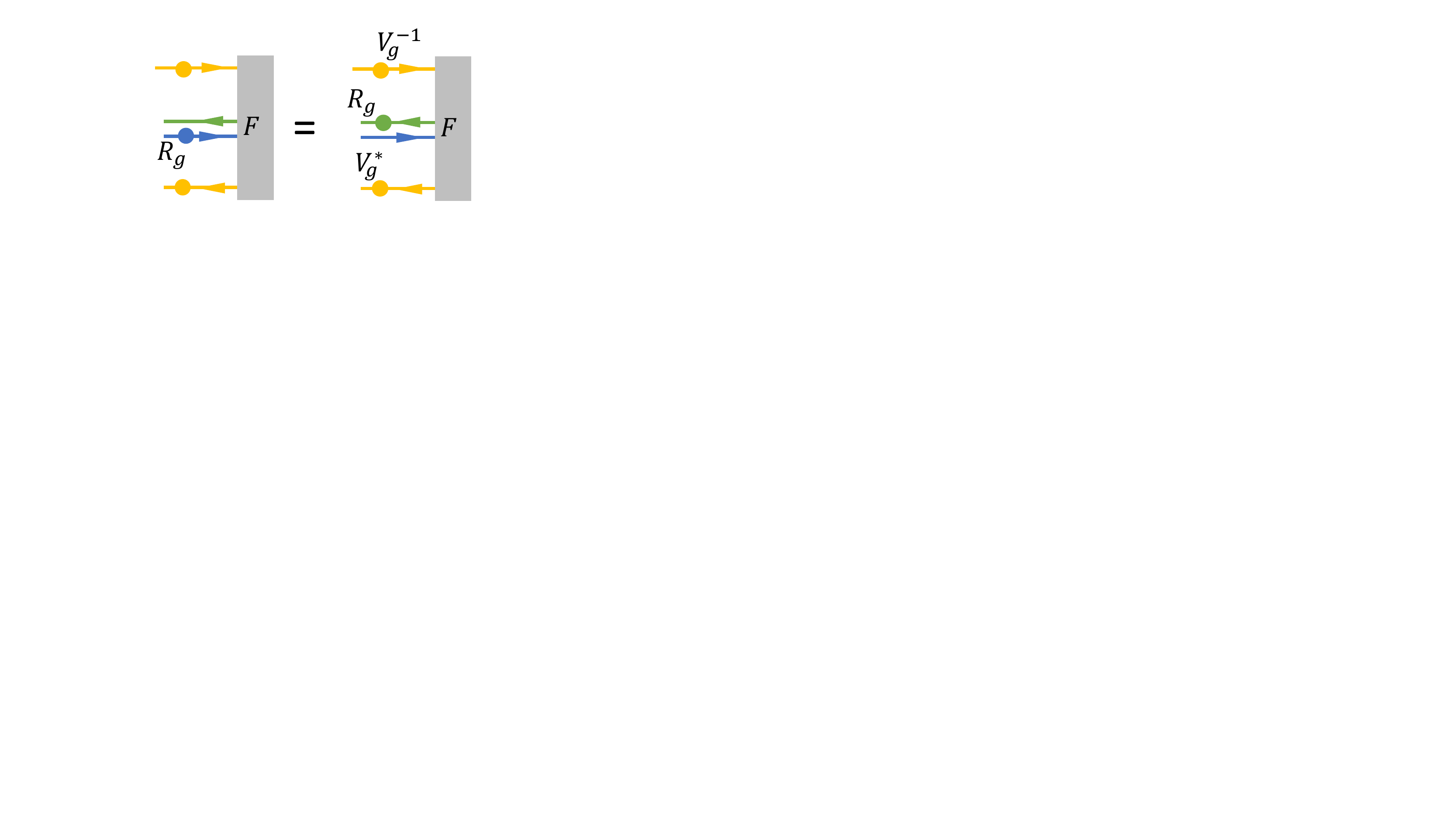}.
\end{equation}

Then we consider the general transfer operator $\mathbb{T}_{k}^{k^\prime}$. Because $\rho$ has the symmetry $G\boxtimes Z_1$, it can be found that
\begin{equation}\label{string_orthogonal}
 \lim_{N\rightarrow +\infty}\frac{\text{tr}[(\mathbb{T}_k^{k^\prime})^N]}{\text{tr}[(\mathbb{T})^N]}=\lim_{N\rightarrow +\infty}\frac{\text{tr}(R^{\otimes N}_k\rho R_{k^\prime}^{\otimes N\dagger}\rho^\dagger)}{\text{tr}(\rho \rho^{\dagger})}=\delta_{k,k^\prime},
\end{equation}
because $R^{\otimes N}_k\rho R_{k^\prime}^{\otimes N\dagger}$ is another fixed point of $\mathbb{T}$ orthogonal to $\rho$ if $k\neq k^{\prime}$. So we only consider the transfer operator $\mathbb{T}_k^k$.
It can be proven that its fixed point is
\begin{equation}\label{fixed_point_rho_k}
\rho_k=\includegraphics[width=7.5cm,valign=c]{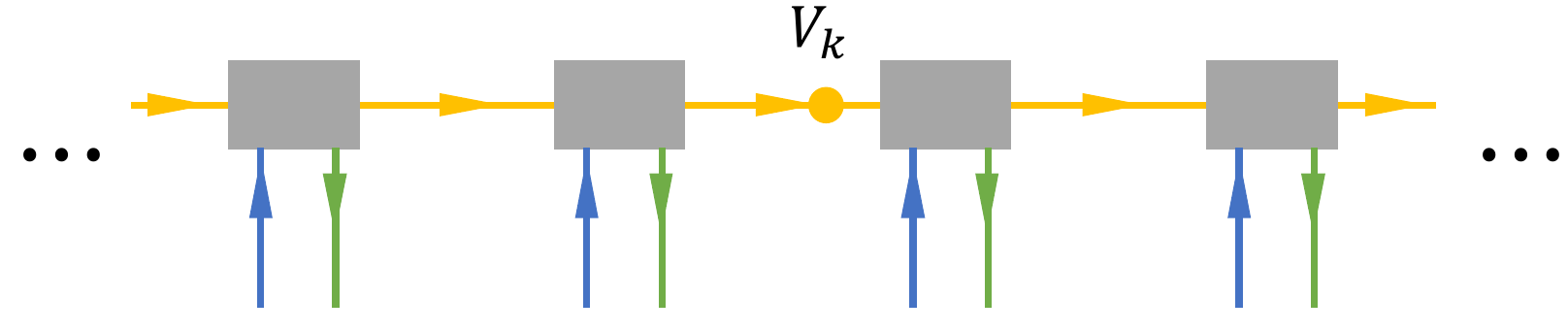}.
\end{equation}
Using Eq.~\eqref{symmetry_of_local_tensor}, it can be derived that
\begin{equation}\label{sym_of_rho_k}
  R_g^{\otimes N} \rho_k=\rho_{gkg^{-1}}R_g^{\otimes N} .
\end{equation}
When $g\in C_k$, $\rho_k$ is invariant under the action of $R_g^{\otimes N}$. In other words, $\rho_k$ has the $C_k\boxtimes Z_1$ symmetry. %

Next, we contract the tensor network of ${}_s\langle\pmb{\alpha}_z|\pmb{\alpha}_y\rangle_t$. For the quantum double models, recall that $\pmb{\alpha}=(K,\gamma)$, $t=(p,k)$, $y=(q,x)$, $s=(\tilde{p},k)$ and $z=(\tilde{q},\tilde{x})$ according to Eq.~\eqref{end_tensor_QD}, and we denote $x=hkh^{-1}$ and $\tilde{x}=\tilde{h}k\tilde{h}^{-1}$ for simplicity. With the transfer operator fixed points \eqref{fixed_point_rho} and \eqref{fixed_point_rho_k} and the channel fixed point \eqref{channel_fixed_point}, the tensor network can be contracted:
\begin{eqnarray}
&&\includegraphics[width=6cm,valign=c]{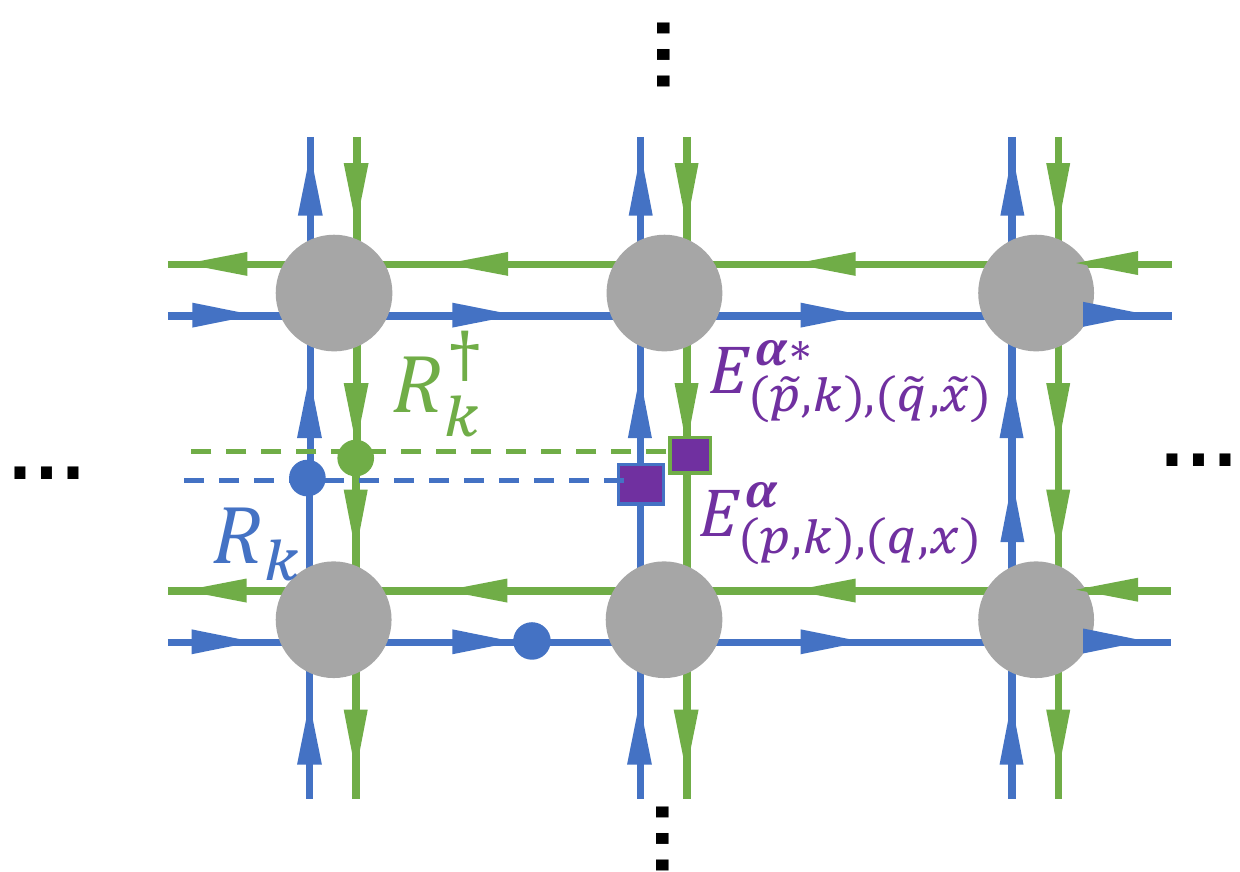}\\\notag
&=&\includegraphics[width=7cm,valign=c]{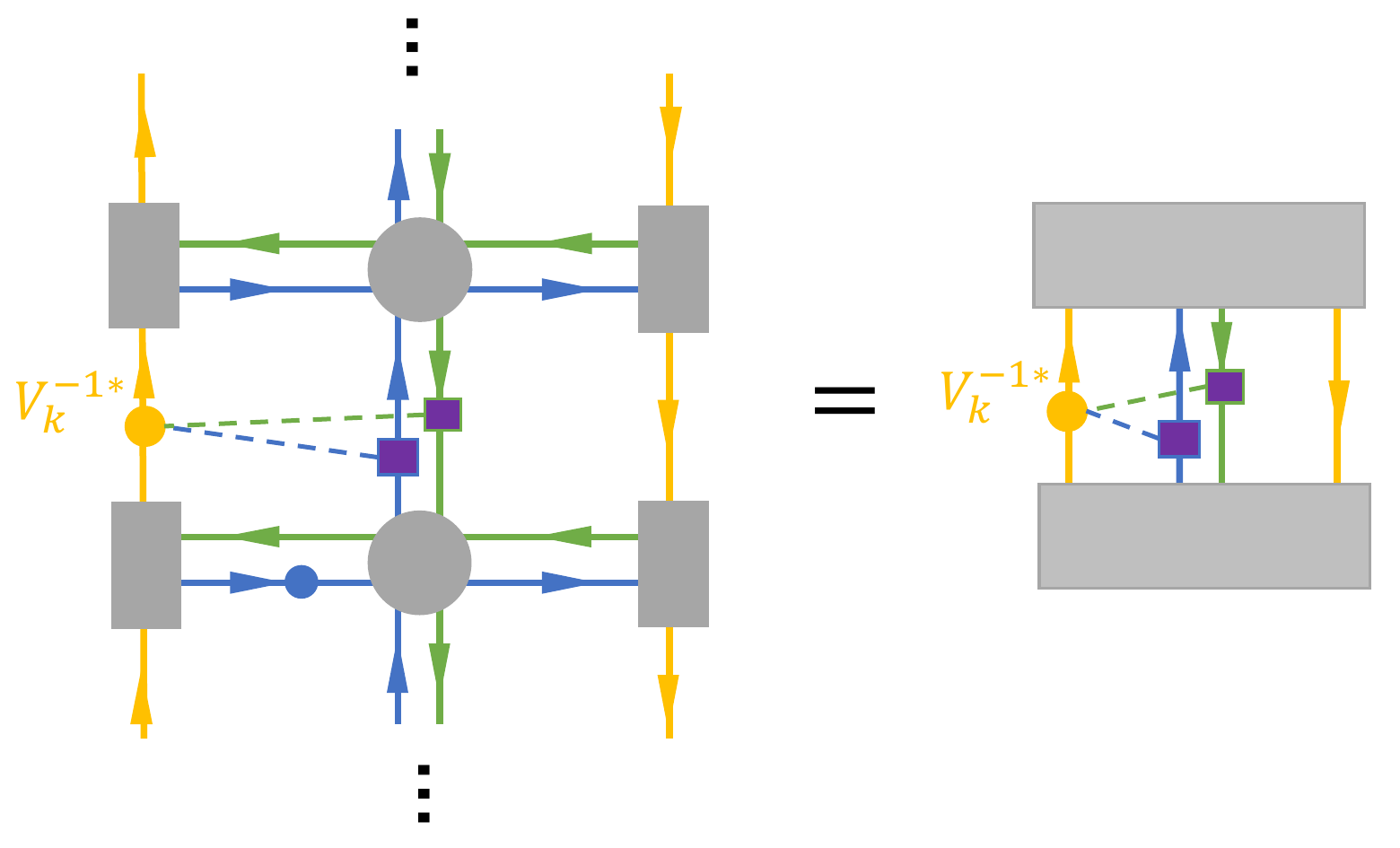}.\\\notag
\end{eqnarray}
The last expression in the above equation can be viewed as the contraction of two end tensors in bra and ket layers and their environment, defined by two channel fixed points and $V_k^{-1*}$. Using the symmetry of the channel fixed points shown in Eq.~\eqref{symmetry_of_fixed_points} and the simple idempotents and nilpotents shown in Eq.~\eqref{idem_nil_of_QD}, it can be proved that the environment of the end tensors satisfies:
\begin{equation}\label{pull_through_env}
\includegraphics[width=8cm,valign=c]{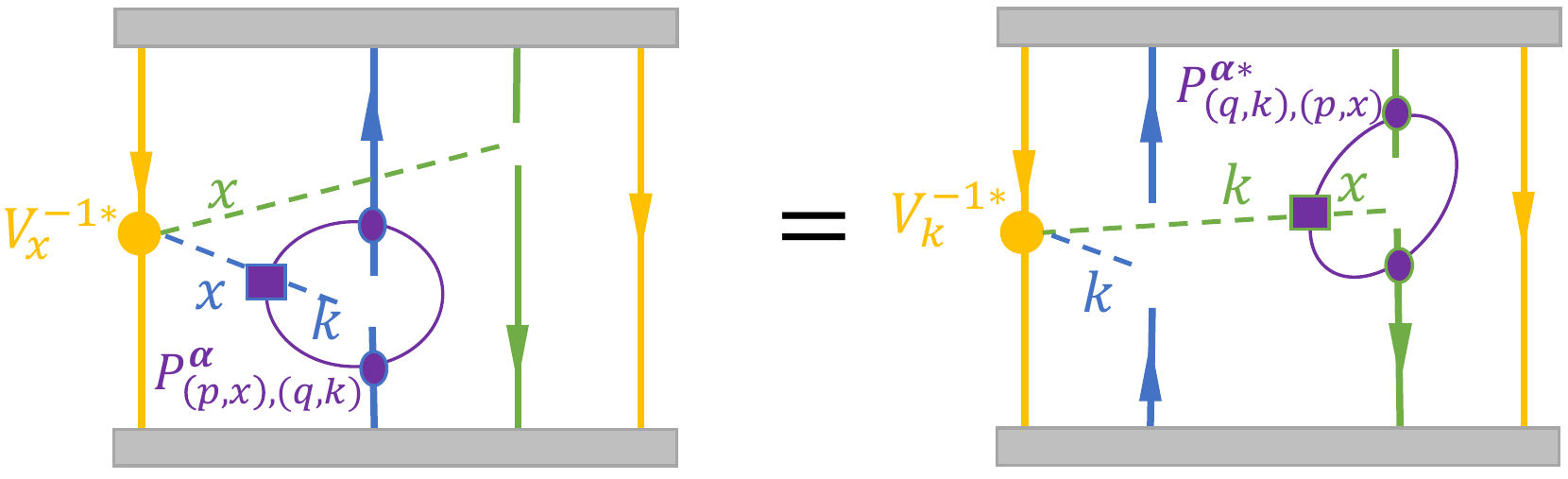},
\end{equation}
which means that we can pull through a simple idempotent or a nilpotent between the bra and ket layers of the environment. With this property of the environment, we can finally prove the structure of the $M$ matrix. Consider the contraction of ${}_s\langle\pmb{\alpha}_z|\pmb{\alpha}_y\rangle_t$:
\begin{equation}
\includegraphics[width=8cm,valign=c,page=1]{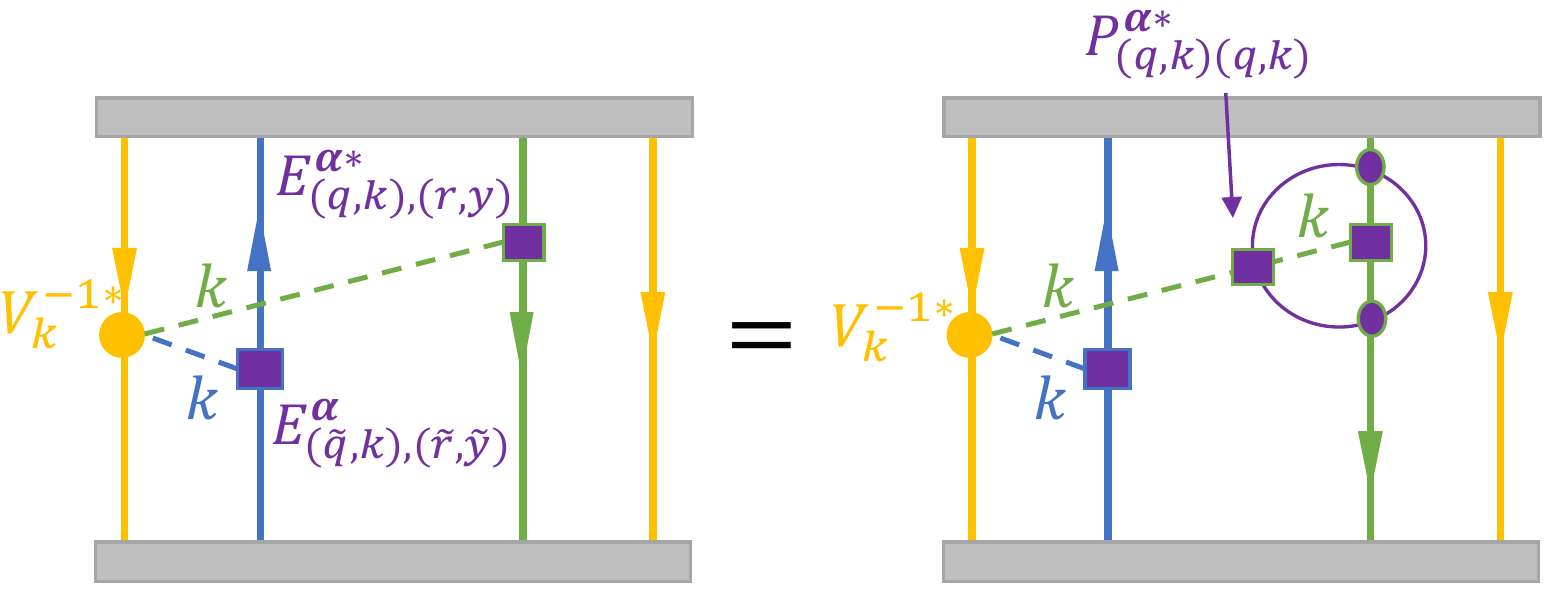},
\end{equation}
where we use the property of an end tensor shown in Eq.~\eqref{relation_between_idempotents_and_exitations} and Fig.~\ref{Idempotent_excitation} (b). Next we can split the idempotent into two nilpotents according to Eq.~\eqref{property_of_simple_idemponent}, and pull through one nilpotent from the bra layer to the ket layer using Eq.~\eqref{pull_through_env}:
\begin{equation}
\includegraphics[width=8.5cm,valign=c,page=2]{Derive_block_structure.pdf}.
\end{equation}
In the last, by absorbing the nilpotents into the end tensors using Eq.~\eqref{relation_between_idempotents_and_exitations} again, we obtain
\begin{equation}\label{Derive_block_structure_3}
\includegraphics[width=3.7cm,valign=c,page=3]{Derive_block_structure.pdf}=\delta_{q,\tilde{q}}
\includegraphics[width=3.7cm,valign=c,page=4]{Derive_block_structure.pdf}.
\end{equation}
If $q=\tilde{q}$, the above equation implies that
\begin{equation}
  {}_s\langle\pmb{\alpha}_z|\pmb{\alpha}_y\rangle_s= {}_t\langle\pmb{\alpha}_z|\pmb{\alpha}_y\rangle_t,\forall s,t.
\end{equation}
If $q\neq\tilde{q}$, taking Eq.~\eqref{string_orthogonal} into consideration, we have
\begin{equation}
  {}_s\langle\pmb{\alpha}_z|\pmb{\alpha}_y\rangle_t\propto \delta_{st}.
\end{equation}
Therefore $M(\pmb{\alpha})$ has the tensor product structure: $M(\pmb{\alpha})=\mathbbm{1}_{\text{str}}\otimes B_{\text{int}}(\pmb{\alpha})$.

In the following, Let us also consider the structure of $M(\pmb{\alpha})$ after a phase transition, assuming that the symmetry of the fixed points $\rho$ breaks down to $H\boxtimes Z_1$, $H\subset G$. First, for fluxons, because of Eq.~\eqref{string_orthogonal} it is obvious that only if $k=k^{\prime}\in H$, $M_{(k,x),(k^{\prime},x^{\prime})}(\pmb{\alpha})$ can be nonzero, so $M(\pmb{\alpha})$ has the block diagonal forms shown in Eqs.~\eqref{full_deconfinement} and \eqref{partial_deconfinement}. For chargons, in general $M(\pmb{\alpha})$ is not block diagonal. But we can block diagonalize it by applying a unitary transformation $U$ to transform the irrep of $G$ to the special form shown in Eq.~\eqref{subduced_rep}, it is equivalent to transform $M(\pmb{\alpha})$ to $\hat{M}(\pmb{\alpha})=(U\otimes U)M(\pmb{\alpha})(U^{\dagger}\otimes U^{\dagger})$. The reason is that after applying $U$, $E^{\pmb{\alpha}}_{pq}$ with different $p$ transform under the different irreps of $H$:
\begin{equation}
  R_h E^{\pmb{\alpha}}_{pq} R^\dagger_h=I^G_{\pmb{\alpha}}(h^{-1})_{pp}E^{\pmb{\alpha}}_{pq}=I^H_{f(p)}(h^{-1})E^{\pmb{\alpha}}_{pq},
\end{equation}
where $E^{\pmb{\alpha}}_{pq}$ is defined in Eq.~\eqref{end_tensor_QD} and $f$ is defined below Eq.~\eqref{subduced_rep}.  Therefore $P^{\pmb{x}}E^{\pmb{\alpha}}_{pq}=\delta_{x,f(p)}E^{\pmb{\alpha}}_{pq}$, where $P^{\pmb{x}}$ is the idempotent for the irrep $x$ of $H$. Following the steps shown from Eq.~\eqref{pull_through_env} to Eq.~\eqref{Derive_block_structure_3}, it can be proven that
$\hat{M}_{pq,p^\prime q^\prime}(\pmb{\alpha})\propto\delta_{f(p),f(p^\prime)}$ and $\hat{M}(\pmb{\alpha})$ is block diagonal for chargons $\pmb{\alpha}$. Combining the analyses for chargons and fluxons, it can be found that $M(\pmb{\alpha})$ for dyons can also be block diagonalized after the phase transition.

In the last, let us consider the structure of $M(\pmb{\alpha})$ after a phase transition, assuming that the symmetry of the fixed points $\rho$ enhances to $G\boxtimes Q$, where $Q$ is an Abelian normal subgroup of $G$. For chargons, because $\rho$ still has the diagonal symmetry $G$, the steps shown from Eq.~\eqref{pull_through_env} to Eq.~\eqref{Derive_block_structure_3} are still valid, so  $M(\pmb{\alpha})=\mathbbm{1}_{\text{str}}\otimes B_{\text{int}}(\pmb{\alpha})$. But because of the off-diagonal symmetry $Q$, the diagonal blocks can be zero if $E^{\pmb{\alpha}}_{st}$ transforms under a nontrivial irrep $y$ of $Q$. The reason is that the fixed point will be annihilated by the corresponding nontrivial idempotent of $Q$:
\begin{equation}
 \frac{1}{|Q|}\sum_{q\in Q}I^{Q}_{y}(q)R^{\otimes N}_q \rho=\frac{\rho}{|Q|}\sum_{q\in Q}I^{Q}_{y}(q)=\delta_{y,1}.
\end{equation}
So $B_{\text{int}}(\pmb{\alpha})$ for the chargons $\pmb{\alpha}$ is zero, and the chargon $\pmb{\alpha}$ are confined.
For a fluxon, we assume that its conjugacy class $K \subset hQ$ where $h\in H\simeq G/Q$ (this is always true if $H$ is an Abelian subgroup). It means that all strings $k\in K$ are identified and their actions on the fixed points $\rho$ are identical to $h$. In this case, we can defined a new set of MPO strings $O_{k}=\sum_{k\in K}F_{kk^{\prime}}R^{\otimes N}_{k^{\prime}}$ attached to the fluxon $\pmb{\alpha}$ using the discrete Fourier transformation $F$, such that Eq.~\eqref{string_orthogonal} is satisfied for these new strings:
\begin{eqnarray}
 &&\lim_{N\rightarrow +\infty}\frac{\text{tr}[(\sum_{l,l^{\prime}\in K}F_{kl}\mathbb{T}_l^{l^\prime}F^{*}_{l^{\prime}k^{\prime}})^N]}{\text{tr}[(\mathbb{T})^N]}\notag\\
 &=&\lim_{N\rightarrow +\infty}\frac{\text{tr}(O_{k}\rho O_{k^\prime}^{\dagger}\rho^\dagger)}{\text{tr}(\rho \rho^{\dagger})}=\delta_{k,k^{\prime}}.
\end{eqnarray}
 It is equivalent to apply the Fourier transformation on $\hat{M}(\pmb{\alpha})$: $\hat{M}(\pmb{\alpha})=(F\otimes\mathbbm{1})M(\pmb{\alpha})(F^{\dagger}\otimes\mathbbm{1})$, and $\hat{M}_{kx,k^{\prime}y}(\pmb{\alpha})\propto\delta_{kk^\prime}$, which means $\hat{M}(\pmb{\alpha})$ is block diagonal. Combining the analyses for chargons and fluxons, it can be found that $M(\pmb{\alpha})$ for dyons can also be block diagonalized after the phase transition.

There is also a remark for the string-net models. Although we haven't proven the structure of $M(\pmb{\alpha})$ for the string-net models, we numerically test that for the anyons $\pmb{B}$, $\pmb{F}$ and $\pmb{G}$ in the Rep$(S_3)$ string-net model as well as the anyon $\pmb{\sigma}\bar{\pmb{\sigma}}$ in the DIsing string-net model, their matrices $M(\pmb{\alpha})$ still have the form shown in Eq.~\eqref{form_of_M_matrix} before phase transitions. And we believe that all conclusions derived from the quantum double models are still valid for the string-net models.
\section{Nonsplitting anyon after the phase transition}\label{what_if}
In this Appendix, using the quantum double models, we show that if the anyon $\pmb{\alpha}$ does not split after a phase transition, $M(\pmb{\alpha})$ still has the tensor product structure. The main reason is that the idempotents and nilpotents always exist.

First we assume that the chargon $\pmb{\alpha}=(K_1=\{e\},\alpha)$ does not split after a phase transition to the $H\boxtimes Z_1$ phase, which means that the restricted representation $x$ of $H$ from the irrep $\alpha$ of $G$:
\begin{equation}
I^{H}_{x}(h)=I^{G}_{\alpha}(h),\quad h\in H
\end{equation}
is an irrep of $H$. We apply the simple idempotent or nilpotent $P^{\pmb{x}}_{tu}$ of $\pmb{x}=(K_1,x)$ on the end tensor $E^{\pmb{\alpha}}_{sx}$ carrying the chargon $\pmb{\alpha}$:
\begin{eqnarray}
  &&\frac{d_{x}}{|H|}\sum_{h\in H}I^H_{x}(h)_{tu}R_h\sum_{g\in G}I^G_{\alpha}(g^{-1})_{sx}|g)( g|R_h^{\dagger}\notag\\
  &=&\frac{d_{x}}{|H|}\sum_{h\in H}I^H_{x}(h)_{tu}\sum_{g\in G}I^G_{\alpha}(g^{-1})_{sx}|gh^{-1})( gh^{-1}|\notag\\
  &=&\frac{d_{x}}{|H|}\sum_{h\in H}I^H_{x}(h)_{tu}\sum_{g\in G}\sum_{m}I^G_{\alpha}(h^{-1})_{sm}I^G_{\alpha}(g^{-1})_{mx}|g)( g|\notag\\
   &=&\delta_{us}\sum_{g\in G}\sum_{m}\delta_{tm}I^G_{\alpha}(g^{-1})_{mx}|g)( g|\notag\\
    &=&\delta_{us}\sum_{g\in G}I^G_{\alpha}(g^{-1})_{tx}|g)( g|.
\end{eqnarray}
So we find that
\begin{equation}\label{general_relation_between_P_and_E}
  P^{\pmb{x}}_{tu}E^{\pmb{\alpha}}_{sx}=\delta_{us}E^{\pmb{\alpha}}_{tx},
\end{equation}
which generalizes Eq.~\eqref{relation_between_idempotents_and_exitations}.

Then we consider the case that the fluxons $\pmb{\alpha}$ does not split after a phase transition to the $H\boxtimes Z_1$ phase, which means that the conjugacy class $K$ of $G$ is still a conjugacy class of $H$. We denote $\pmb{\alpha}=(K,\pmb{1})$ in the $G\boxtimes Z_1$ phase and $\pmb{x}=(K,\pmb{1})$ in the $H\boxtimes Z_1$ phase. The end tensor carrying a fluxon $\pmb{\alpha}$ with an internal state $nln^{-1}$ and a string $l$ is the rank three tensor:
\begin{equation}
  E^{\pmb{\alpha}}_{l,nln^{-1}}=|l)\otimes\sum_{g\in C^G_l}|ng)(ng|.
\end{equation}
 The simple idempotent or nilpotent of $\pmb{x}$ is
\begin{equation}
 P_{mk m^{-1},k}^{\pmb{x}}=\frac{1}{|C_k^H|}\sum_{h\in mC_k^H }|mkm^{-1})(k|\otimes R^{\otimes N}_{h}.
\end{equation}
We apply the simple idempotent or nilpotent $P_{mkm^{-1},k}^{\pmb{x}}$ ($N=2$) on $E^{\pmb{\alpha}}_{l,nln^{-1}}$, which gives rise to
\begin{eqnarray}\label{D_6}
   && \delta_{kl}\frac{1}{|C_k^H|}\sum_{h\in mC_k^H}\sum_{g\in C_l^G}|mlm^{-1})\otimes |ngh^{-1})( ngh^{-1}| \notag\\
   &=& \delta_{kl}\frac{1}{|C_k^H|}\sum_{h\in C_k^H}\sum_{g\in C_l^G}|mlm^{-1})\otimes |ngh^{-1}m^{-1})( ngh^{-1}m^{-1}|\notag \\
   &=&\delta_{kl}|mlm^{-1})\otimes \sum_{g\in C_l^G} |ngm^{-1})( ngm^{-1}|\notag \\
   &=& \delta_{kl}|mlm^{-1})\otimes \sum_{g\in C_{mlm^{-1}}^G}|nm^{-1}g)( nm^{-1}g|\notag\\
   &=&\delta_{kl} E^{\pmb{\alpha}}_{mlm^{-1},nln^{-1}}.
\end{eqnarray}
So Eq.~\eqref{general_relation_between_P_and_E} is still satisfied for fluxons.

Combining the procedures for chargons and fluxons, the same result can be derived for dyons, and Eq.~\eqref{general_relation_between_P_and_E} is valid for all anyons in the quantum double models.
So we find that through a phase transition to the $H\boxtimes Z_1$ phase, if a non-Abelian anyon $\pmb{\alpha}$ does not split, we have a generalized relation between the simple idempotents/nilpotents and the end tensors, shown in Eq.~\eqref{general_relation_between_P_and_E}, which allow us to prove $M=\mathbbm{1}_{\text{str}}\otimes B_{\text{int}}$ by simply replacing $H$ with $G$ in Appendix~\ref{prove_block_structure}.

Let us also talk about what if anyon does not split in the case of partial deconfinement. Since chargons cannot be confined when the transfer operator fixed points have the symmetry $H\boxtimes Z_1$, we only consider fluxons. We assume that $K$ is not a conjugacy class of $H$ but $K^{\prime}=K\cap H$ is a conjugacy class of $H$. Since $C^H_k\subset C^G_k$, the derivation in Eq.~\eqref{D_6} is still valid, and we still have the relation
\begin{equation}
  P_{mkm^{-1},k}^{\pmb{x}}E^{\pmb{\alpha}}_{l,nln^{-1}}=\delta_{kl}E^{\pmb{\alpha}}_{l,nln^{-1}},
\end{equation}
where $k,l\in K^\prime, m\in H, n\in G$. Hence, in this case $M=P_{\text{str}}\otimes B_{\text{int}}$, where $(P_{\text{str}})_{k,k^\prime}=1$ if $k=k^\prime\in K^{\prime}$, and $(P_{\text{str}})_{k,k^\prime}=0$ otherwise. Moreover, if the symmetry of the transfer operator fixed points is off-diagonal, $P_{\text{str}}$ is not diagonal but it is still a projector.

\section{Map to the classical models}
In this Appendix we map the deformed PEPS to the known statistical mechanics models. We map the deformed $D(S_3)$ PEPS along the $k_1$ or $k_2$ axis to the 3-state Potts model, and along the $k_3$ axis it can be mapped to the Ising model. We also map the Rep$(S_3)$ PEPS to the $D(S_3)$ PEPS.  Moreover, we also relate the deformed DIsing string-net model to the Ashkin-Teller model.

\subsection{Deformed $D(S_3)$ PEPS in the $k_1$-$k_3$ plane}\label{map_D_S3_to_stat_model_1}

At first, let us consider the case that $k_2=0$ so that $Q=Q(k_1,0,k_3)$ is a diagonal matrix. By contracting the physical DOFs of the tensor \eqref{tensor_of_S3} together with $Q^2$, we obtain the double tensor,  which can be further written as
\begin{equation}
  \includegraphics[width=1.8cm,valign=c,page=1]{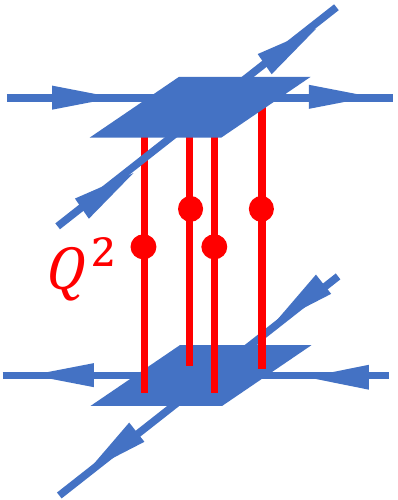}=\sum_{h\in S_3} \includegraphics[width=1.5cm,valign=c,page=2]{Bra_ket_relation.pdf}.
\end{equation}
 The green ring is a diagonal matrix whose diagonal entries are
\begin{equation}\label{green_ring}
  \includegraphics[width=1.4cm,valign=c]{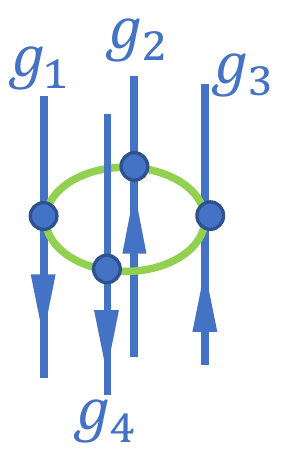}=W_{g_1g_2^{-1}}W_{g_2g_3^{-1}}W_{g_3g_4^{-1}}W_{g_4g_1^{-1}},
\end{equation}
where $W_g=Q^2_{g,g}$. The green ring commutes with $R_h^{\otimes 4}$.
Therefore, the norm of the PEPS can be written as $\sum_{h\in S_3}\mathcal{Z}_h$,
where $\mathcal{Z}_h$ is a tensor network generated by
\begin{equation}
 T_h=\includegraphics[width=1.5cm,valign=c,page=2]{Bra_ket_relation.pdf}.
\end{equation}
In the tensor network $\mathcal{Z}_h$, the DOFs in the bra and ket layers are locked with by $R_h$. Notice that the tensor networks $\mathcal{Z}_h, \forall h$ are equal, because $T_h$ and $T_{h^\prime}$ can be transformed to each other by a gauge transformation $R_h\otimes \mathbbm{1}$. So the norm of the deformed PEPS is $6\times \mathcal{Z}_e$.

In the tensor network $\mathcal{Z}_e$, the tensor $T_e$ is just the green ring \eqref{green_ring}, and it can be found that $W$
is the Boltzmann weight matrix
\begin{equation}
W=W_{\text{Ising}}(k_3)\otimes  W_{\text{Potts}}(k_1),
\end{equation}
where
\begin{eqnarray}\label{Boltzmann_weight_matrix}
  W_{\text{Ising}}(k)&=&\left(
                 \begin{array}{cc}
                   1 & e^{-2k} \\
                   e^{-2k} & 1 \\
                 \end{array}
               \right),\nonumber\\
               W_{\text{Potts}}(k)&=&\left(
   \begin{array}{ccc}
     1 & e^{-2k} & e^{-2k} \\
     e^{-2k} & 1 & e^{-2k} \\
     e^{-2k} & e^{-2k} & 1 \\
   \end{array}
 \right)
\end{eqnarray}
are the Boltzmann weight matrices of the Ising model and the 3-state Potts model. Hence the tensor $T_e$ can be written as $T_e=T_{\text{Ising}}\otimes T_{\text{Potts}}$, where $T_{\text{Ising}}$($T_{\text{Potts}}$) is obtained by replacing $W$ with $W_{\text{Ising}}$($W_{\text{potts}}$) in Eq. (\ref{green_ring}). So the tensor network $\mathcal{Z}_e$ can be decomposed into two parts:
\begin{equation}
  \mathcal{Z}_e
  =\text{tTr}
  \left[{\bigotimes}T_{\text{Ising}}(2 k_3)\right]\text{tTr}\left[{\bigotimes}T_{\text{Potts}}(2 k_1)\right].\\
\end{equation}
The tensor network generated by $T_{\text{Ising}}(k)$ $(T_{\text{Potts}}(k))$ is the partition function $\mathcal{Z}_{\text{Ising}}(k)$($\mathcal{Z}_{\text{Potts}}(k)$) of the Ising (3-state Potts) model. Since $\mathcal{Z}_{\text{Ising}}(0)=2^{\#\text{sites}}$, $\mathcal{Z}_{\text{Potts}}(0)=3^{\#\text{sites}}$, the norm of the deformed PEPS along the $k_1$($k_3$) axis is equivalent to the partition function of the 3-state Potts (Ising) model. For the Ising (3-state Potts) model, the critical point is $k_c=\log(\sqrt{1+\sqrt{2}})$$(k_c=\log(\sqrt{1+\sqrt{3}}))$. Because the PEPS norm splits into a product of two partition functions, the phase transition lines in the $k_1-k_3$ plane are straight.

\subsection{Deformed $D(S_3)$ model in the $k_2-k_3$ plane}

Next let us consider that the case $k_1=0$ so that $Q=Q(0,k_2,k_3)$ is a block diagonal matrix. Since $S_3$ can be written as a semidirect product of $Z_2=\{e,s\}$ and $Z_3=\{e,r,\bar{r}\}$, we have $g_i=m_ih_i\in S_3$, where $m_i\in Z_2$ and $h_i\in Z_3$. So a $\delta$ tensor in Eq.~\eqref{tensor_of_S3} can be rewritten as
\begin{equation}
  \delta_{g_1,g_2g_3^{-1}}= \delta_{m_1,m_2e}\delta_{h_1,h_2h_3^{-1}}+\delta_{m_1,m_2s} \delta_{h_1,h_2^{-1}h_3}.
\end{equation}
The tensors $\delta_{h_1,h_2h_3^{-1}}$ and $\delta_{h_1,h_2^{-1}h_3}$ can transform with each other:
\begin{equation}
  \delta_{h_1,h_2h_3^{-1}}=\sum_{h_1^{\prime}}X_{h_1,h_1^{\prime}}\delta_{h_1^{\prime},h_2^{-1}h_3}
\end{equation}
where $X_{h_1,h_1^{\prime}}=\delta_{h_1^{-1},h_1^{\prime}}$.
Then we contract the physical DOFs on a lattice edge together with $Q^2$:
\begin{widetext}
\begin{eqnarray}\label{double_tensor_2}
&&\sum_{g_1,g_1^\prime}\includegraphics[width=2cm,valign=c]{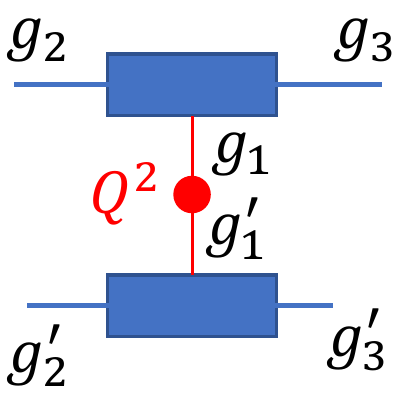}\propto\sum_{g_1,g_1^{\prime}}\delta_{g_1,g_2g_3^{-1}}Q_{g_1,g_1^{\prime}}(0,2k_2,2k_3) \delta_{g_1^{\prime},g^{\prime}_2g^{\prime-1}_3} = \exp(-2k_3\delta_{m_2m_3,e})\nonumber\\
  &\times&\left\{\delta_{m_2,m_2^\prime}\delta_{ m_3,m_3^\prime}\sum_{h_1,h_1^\prime}\delta_{h_1,h_2h_3^{-1}}[q(k_2)^2]_{h_1,h_1^{\prime}} \delta_{h_1^{\prime},h^{\prime}_2h^{\prime-1}_3}
  +\delta_{sm_2,m_2^\prime}\delta_{sm_3,m_3^\prime}\sum_{h_1,h_1^\prime}\delta_{h_1,h_2h_3^{-1}}[Xq(k_2)^2]_{h_1,h_1^{\prime}} \delta_{h_1^{\prime},h^{\prime}_2h^{\prime-1}_3}\right\},\nonumber
\end{eqnarray}
\end{widetext}
where
\begin{equation}
  q(k)=\frac{1}{3}\left(
\begin{array}{ccc}
 1+2e^{-k} & 1-e^{-k} & 1-e^{-k} \\
 1-e^{-k} & 1+2e^{-k} & 1-e^{-k} \\
 1-e^{-k} & 1-e^{-k} & 1+2e^{-k} \\
\end{array}\right).
\end{equation}
It can be checked that
\begin{eqnarray}\label{Boltzmann_weight_potts}
&&\sum_{h_1,h_1^\prime}\delta_{h_1,h_2h_3^{-1}} (q^2)_{h_1,h_1^{\prime}} \delta_{h_1^{\prime},h_2^{\prime}h_3^{\prime-1}}= (q^2)_{11}e^{-k_2^*\delta_{h_2^{\prime}h_2^{-1},h_3^{\prime}h_3^{-1}}},\nonumber\\
&&\sum_{h_1,h_1^\prime}\delta_{h_1,h_2h_3^{-1}} (Xq^2)_{h_1,h_1^{\prime}} \delta_{h_1^{\prime},h_2^{\prime}h_3^{\prime-1}}
    = (q^2)_{11}e^{-k_2^*\delta_{h_2^{\prime}h_2,h_3^{\prime}h_3}},\nonumber\\
\end{eqnarray}
where $k_2^\star=\log[1+3/(e^{2k_2}-1)]$. Defining $n_i=h_i^{\prime}h_i^{-1}$ or $n_i=h_i^\prime h_i$, the right-hand side of the above equations becomes a Boltzmann weight matrix of the 3-state Potts model shown in Eq. (\ref{Boltzmann_weight_matrix}). Therefore
\begin{equation}
 \sum_{g-1,g_1^\prime}\includegraphics[width=1.5cm,valign=c]{edge_contraction.pdf}\propto [W_{\text{Ising}}(2k_3)]_{m_2,m_3}[W_{\text{Potts}}(k_2^\star)]_{n_2,n_3},
\end{equation}
and the PEPS norm on the $k_2-k_3$ plane will split into a product of two parts and be proportional to $\mathcal{Z}_{\text{Potts}}(k^\star_2)\times \mathcal{Z}_{\text{Ising}}(2k_3)$. This is the reason why the phase transition lines in the $k_2-k_3$ plane are straight.

\subsection{The $\text{Rep}\left(S_3\right.$) model\\}\label{Map_Rep(S3)_to stat_model}
\begin{figure}
  \centering
  \includegraphics[width=8cm]{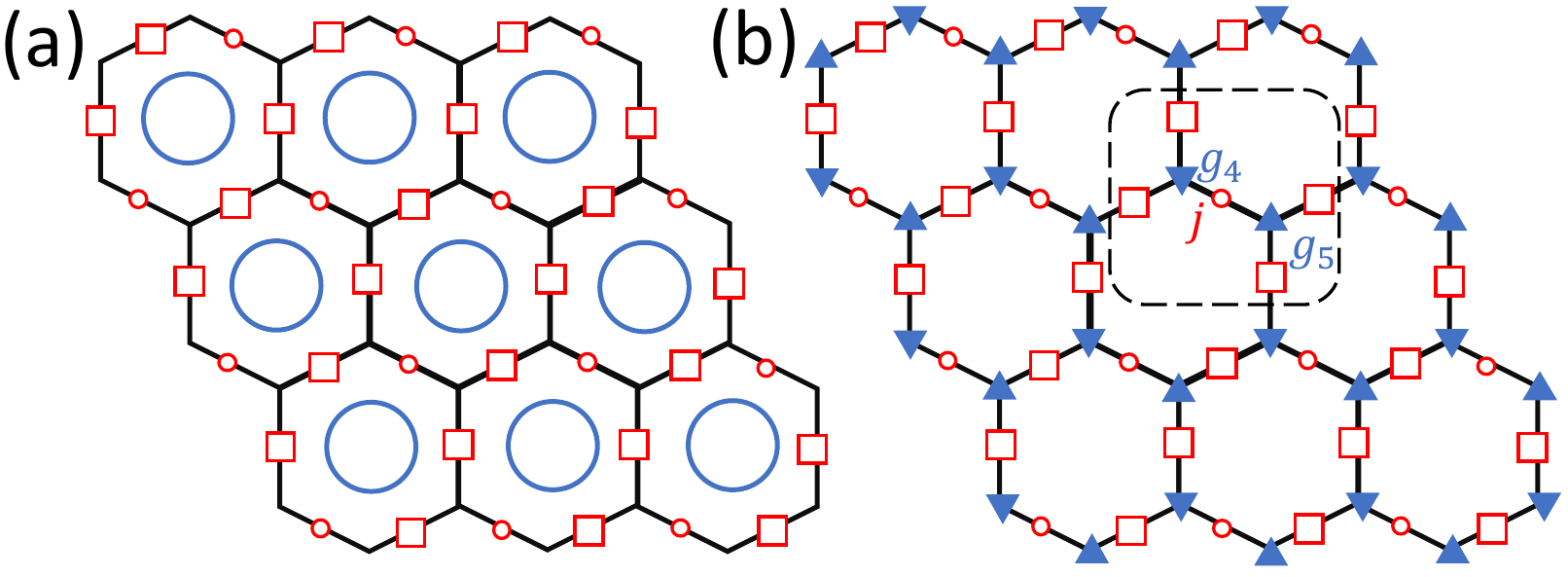}
  \caption{ (a) A string-net model and (b) a quantum double model on a honeycomb lattice. Red squares represent the physical DOFs on which the deformation operators act, and red circles are the physical DOFs on which deformation operators do not act. Blue circles are the virtual loops of the string-net PEPS. Blue triangles are the virtual DOFs of the quantum double PEPS.}\label{QD_honey}
\end{figure}

In this subsection, we map the deformed Rep$(S_3)$ PEPS shown in Eq.~\eqref{deformed_PEPS_Rep_S3} to the 3-state Potts model.
First, we only apply the deformation operators on a part of physical DOFs on the string-net PEPS, as shown in Fig.~\ref{QD_honey} (a). Then, there is a Fourier transformation which maps between the Rep$(G)$ model and the $D(G)$ model\cite{Map_QD_to_string_net}:
\begin{equation}
  |\alpha_{pq}\rangle=\sqrt{\frac{\text{dim}{\alpha}}{|G|}}\sum_{g}I^{G}_{\alpha}(g)_{pq}|g\rangle.
\end{equation}
Specifically, when $G=S_3$, the transformation is
\begin{equation}
 V=\frac{1}{\sqrt{6}}\left(
  \begin{array}{cccccc}
    1 & 1 & 1 & 1 & 1 & 1 \\
    1 & 1 & 1 & -1 & -1 & -1 \\
    \sqrt{2} & \sqrt{2}\bar{\omega} & \sqrt{2}\omega & 0 & 0 &0  \\
    0 & 0 & 0 & \sqrt{2} & \sqrt{2}\omega & \sqrt{2}\bar{\omega}  \\
    0 & 0 & 0 & \sqrt{2} & \sqrt{2}\bar{\omega} & \sqrt{2}\omega \\
    \sqrt{2} & \sqrt{2}\omega & \sqrt{2}\bar{\omega} & 0 & 0 &0  \\
  \end{array}
\right),
\end{equation}
where the bases of columns are the irreps: $|1\rangle, |\epsilon\rangle, |\pi_{11}\rangle,|\pi_{12}\rangle,|\pi_{21}\rangle,|\pi_{22}\rangle$, the bases of rows are group elements: $|e\rangle, |r\rangle, |\bar{r}\rangle,|s\rangle,|sr\rangle,|s\bar{r}\rangle$ and $\omega=\exp(2 \pi i/3)$. For the Rep$(S_3)$ model, the string tension operator $Q_\text{rep}(k)=\text{diag}(1,1,e^{-k})$ in the basis $\{|1\rangle, |\epsilon\rangle, |\pi\rangle\}$ becomes $\tilde{Q}(k)=\text{diag}(1,1,e^{-k},e^{-k},e^{-k},e^{-k})$ in the basis $\{|1\rangle, |\epsilon\rangle, |\pi_{11}\rangle,|\pi_{12}\rangle,|\pi_{21}\rangle,|\pi_{22}\rangle\}$. Under the Fourier transformation $V$, $\tilde{Q}$ becomes
\begin{equation}
  Q(0,k_2,0)\propto V^{\dagger}\tilde{Q}V=\mathbbm{1}_2\otimes q(k).
\end{equation}
Therefore the Rep$(S_3)$ PEPS deformed by $Q_\text{rep}(k)$ in Eq.~\eqref{deformed_PEPS_Rep_S3} is equivalent to $Q(0,k_2,0)^{\otimes N}|\Psi_0\rangle$, where $|\Psi_0\rangle$ is a fixed point $D(S_3)$ PEPS on a honeycomb lattice, and $Q(0,k_2,0)$ acts on a subset of the honeycomb lattice edges shown in Fig.~\ref{QD_honey} (b).

Since there are edges in which $Q(0,k_2,0)$ does not act, we need to consider the contraction of the physical DOFs on these edges
\begin{eqnarray}
 && \sum_{j\in S_3}\includegraphics[width=1.5cm,valign=c]{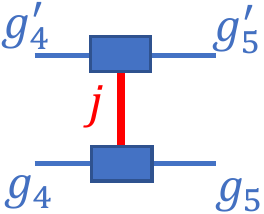}\notag\\
  &=&
  2\delta_{m_4,m_4^\prime}\delta_{h_4h_5^{-1},h_4^{\prime}h_5^{\prime-1}}+2\delta_{m_4,m_4^\prime s}\delta_{h_4h_5^{-1},h_4^{\prime-1}h_5^{\prime}}\notag\\
 & =&2\delta_{m_4,m_4^\prime}\delta_{h_4^{\prime}h_4^{-1},h_5^{\prime}h_5^{-1}}+2\delta_{m_4,m_4^\prime s}\delta_{h_4^{\prime}h_4,h_5^{\prime}h_5},
\end{eqnarray}
 which implies that $n_4$ and $n_5$ ($n_i=h_i^{\prime}h_i^{-1}$ or $n_i=h_i^\prime h_i$) are equal and they can be viewed as the same DOF. So, after contracting these physical DOFs, the remaining virtual DOFs locate on a square lattice, and the norm of the deformed PEPS \eqref{deformed_PEPS_Rep_S3} is equivalent to the norm of the deformed $D(S_3)$ PEPS along the $k_2$ axis. Therefore, the deformed Rep$(S_3)$ PEPS can be mapped to the 3-state Potts model.

 \subsection{The DIsing model}\label{Map_DIsing_to stat_model}
\begin{figure}[hbp]
  \centering
  \includegraphics[width=8.5cm]{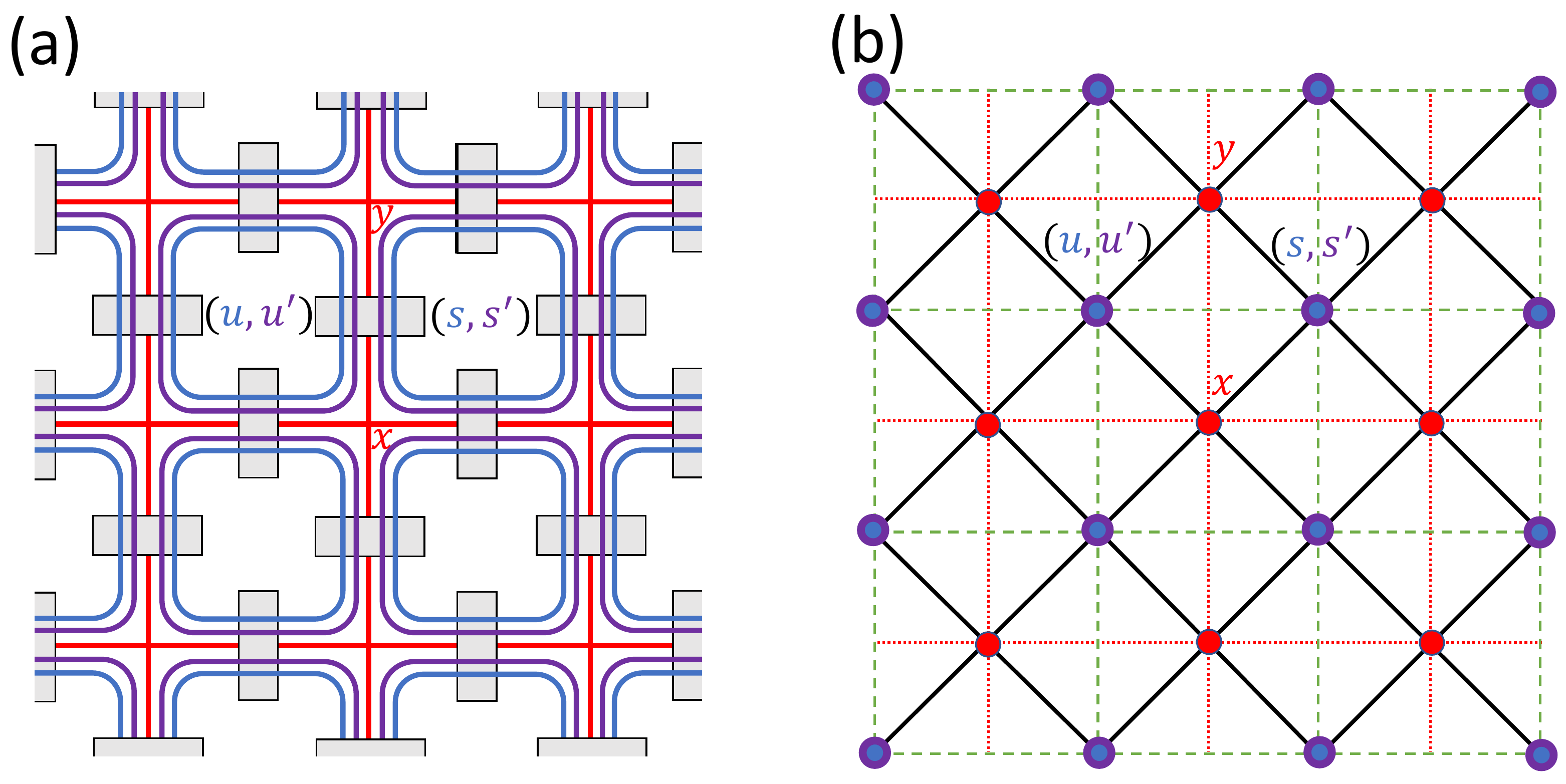}
  \caption{(a) A tensor network generated by the tensor in Eq.~\eqref{contract_edge_tensors}. (b) The DOFs of the tensor network are represented by dots. The red, green, and black lines form the primal, dual, and medial lattice, respectively.}\label{RSOS}
\end{figure}

Using the method in our previous work\cite{Xu_Schuch_2021}, we related the deformed DIsing PEPS to the Ashkin-Teller (AT) model. At first we consider a local double tensor for the norm of the DIsing PEPS without the deformation, which is obtained by contracting the physical DOFs of the tensor in Eq.~\eqref{double_tensor_string_net}:
 \begin{equation}\label{double_tensor}
\sum_{k}d_k\includegraphics[width=2.5cm,valign=c]{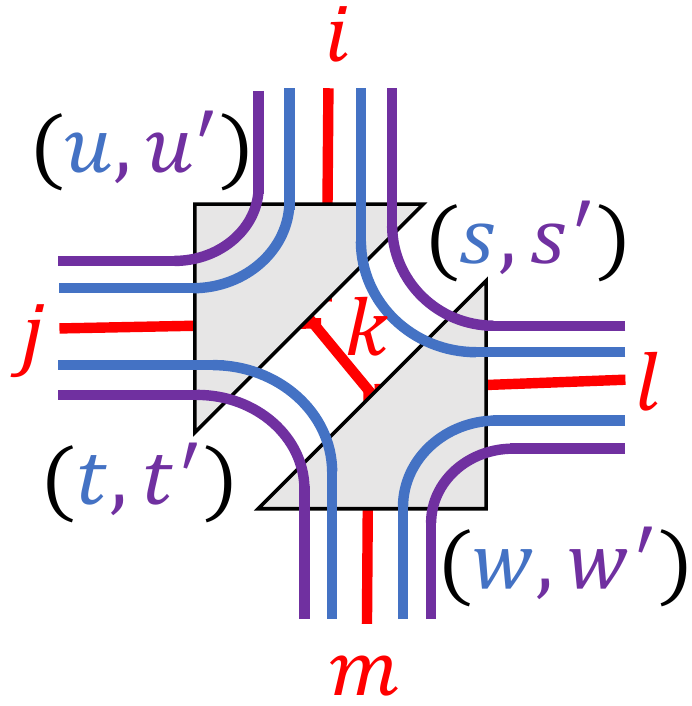}
=\sum_{k}d_kG^{ijk}_{tsu}G^{klm}_{wts}G^{ijk}_{t^\prime s^\prime u^\prime}G^{klm}_{w^\prime t^\prime s^\prime}.
\end{equation}
This double tensor at a vertex of the lattice can be decomposed into four $G$ tensors living separately on the four edges\cite{Xu_Schuch_2021}:
 \begin{equation}\label{decomposition}
\sum_{k}d_k\includegraphics[width=2.5cm,valign=c]{double_tensor_1.pdf}
=\sum_{x}d_{x}\includegraphics[width=2.5cm,valign=c]{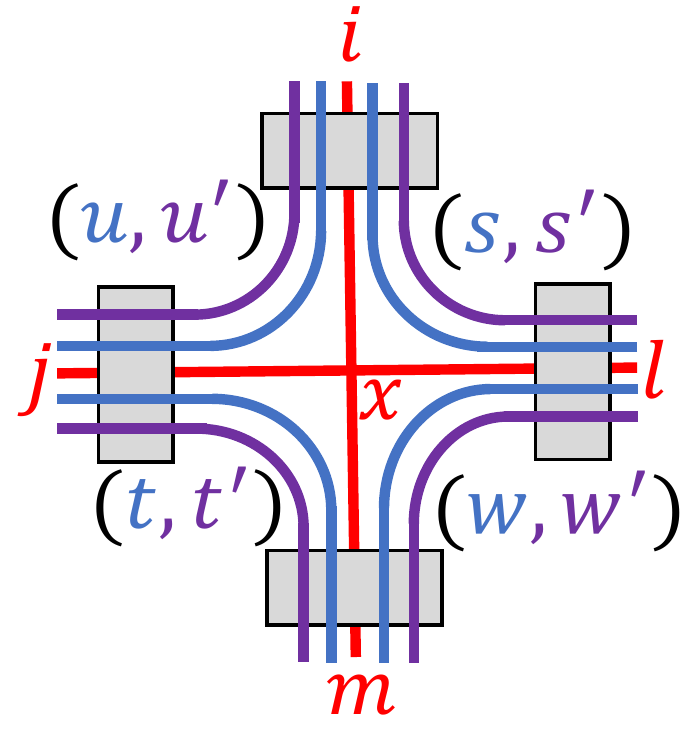}.
\end{equation}
Namely
 \begin{eqnarray}
&&\sum_{k}d_kG^{ijk}_{tsu}G^{klm}_{wts}G^{ijk}_{t^\prime s^\prime u^\prime}G^{klm}_{w^\prime t^\prime s^\prime}\notag\\
&=&\sum_{x}d_x G^{utj}_{t^\prime u^\prime x}G^{twm}_{w^\prime t^\prime x}G^{wsl}_{s^\prime w^\prime x}G^{sui}_{u^\prime s^\prime x}.
\end{eqnarray}
Since there are two $G$ tensors on each edge after decomposition \eqref{decomposition}, we contract the two $G$ tensors on the same edge together with the deformation matrix $Q=\text{diag}(1, e^{-k_{\sigma}},e^{-k_{\psi}})$:
 \begin{equation}\label{contract_edge_tensors}
\sum_{i}d_iQ_i^2\includegraphics[width=5cm,valign=c]{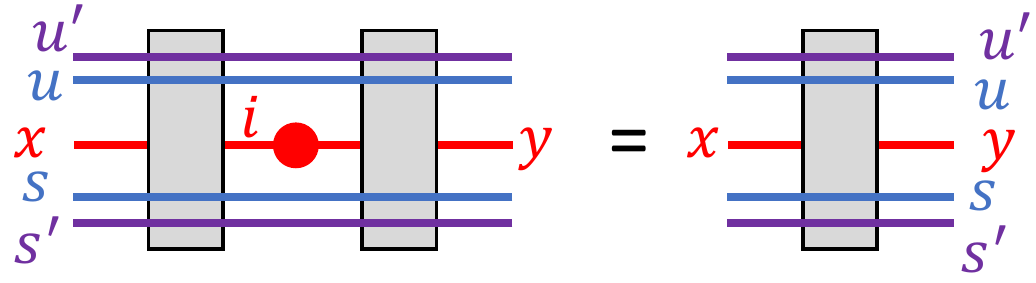}.
\end{equation}
It generates a tensor network shown in Fig. \ref{RSOS} (a). As displayed in Fig. \ref{RSOS} (b), the red lines in Fig. \ref{RSOS} (a) are the DOFs living on the sites of the primal square lattice and they take three values $1$, $\sigma$ and $\psi$. The blue and purple lines in Fig. \ref{RSOS} (a) are the DOFs $(\alpha,\beta), \alpha,\beta\in\{1,\sigma,\psi\}$ living on the sites of the dual square lattice. All of these DOFs  live on the site of the medial lattice.

Since $G^{ijk}_{tsu}\propto N^{i}_{jk}N^{i}_{su}N^{j}_{tu}N^{k}_{st}$, the nearest neighboring DOFs are constrained by the $N$ tensors, i.e., $x,y,(u,u^\prime),(s,s^\prime)$ in Fig.~\ref{RSOS} (b) are constrained by $N^{x}_{ss^\prime}$ and $N^{y}_{uu^\prime}$. The constraint can be represented by the following $\hat{D}_4$ and $\hat{D}_6$ Dynkin diagrams\cite{Gils_2009_Ashkin_Teller_DIsing}:
 \begin{equation}\label{Dynkin_diagram}
\includegraphics[width=7cm,valign=c]{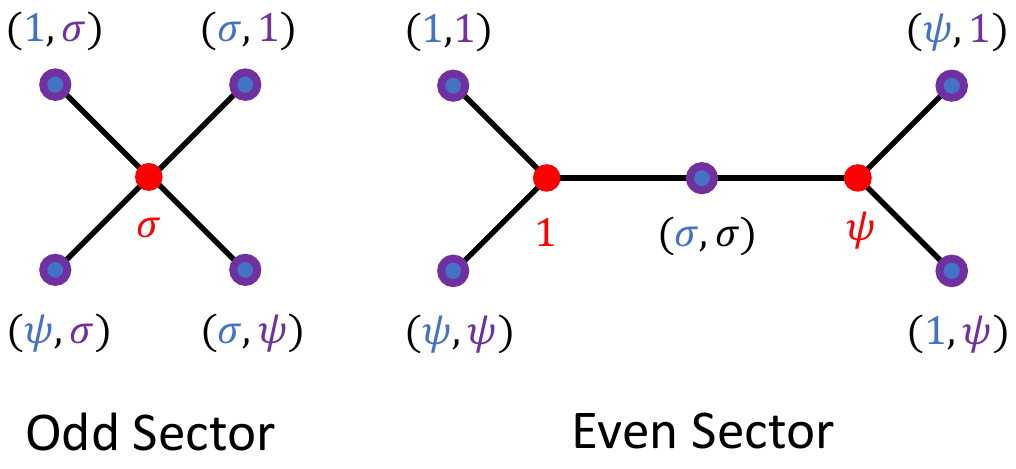}.
\end{equation}
Two DOFs can be nearest neighboring if and only if they are adjacent in the Dynkin diagrams, and the tensor network of the PEPS norm is decomposed into two sectors, the even (odd) sector contains DOFs in the $\hat{D}_6$ ($\hat{D}_4$) Dynkin diagram.

Next we map the odd sector to the AT model. Notice that $x$ and $y$ are fixed to $\sigma$ in the odd sector. First, the DOFs $(s,s^\prime)$ can be mapped to a pair of $Z_2$ classical spin variables $(z_s,z_s^\prime)$:
\begin{eqnarray}\label{mapping}
  (1,\sigma)&\rightarrow& (1,1), \quad\quad(\sigma,1)\rightarrow (1,-1), \nonumber\\ (\sigma,\psi)&\rightarrow& (-1,1), \quad (\psi,\sigma)\rightarrow (-1,-1).
\end{eqnarray}
Then Eq.~\eqref{contract_edge_tensors} can be expressed in terms of $Z_2$ spins:
\begin{equation}\label{edge_tensor}
\frac{\exp[k_2 (z_uz_s+z^\prime_uz^\prime_s)+k_4z_uz_sz^\prime_uz^\prime_s]}{\sqrt{2}\exp(2 k_2+k_4)},
\end{equation}
where $k_2=k_\psi/2$ and $k_4=k_\sigma-k_\psi/2$.
Eq.~\eqref{edge_tensor} is nothing but a local Boltzmann weight of the AT model. The contraction of the tensor network in the odd sector gives rise to the partition function of the AT model.
Under the parameter transformation  from $(k_\sigma,k_\psi)$ to $(k_2,k_4)$, the DIsing phase is mapped to the paramagnetic phase of the AT model, the toric code phase is mapped to the partially ordered phase of the AT model,
and the trivial phase is mapped to the ferromagnetic phase of the AT model.
By the way, when $(k_\sigma=k_\psi)(k_2=k_4)$, the AT model is equivalent to the 4-state Potts model.

In the following let us consider the even sector. Interestingly, the even sector is equivalent to the norm of a symmetry enriched toric code state $|\Psi_\text{SET}\rangle$, which is obtained from the DIsing string-net model by promoting virtual loops to the physical level\cite{williamson_2017_SET}:
 \begin{equation}\label{SET_tensor}
\includegraphics[width=3cm,valign=c]{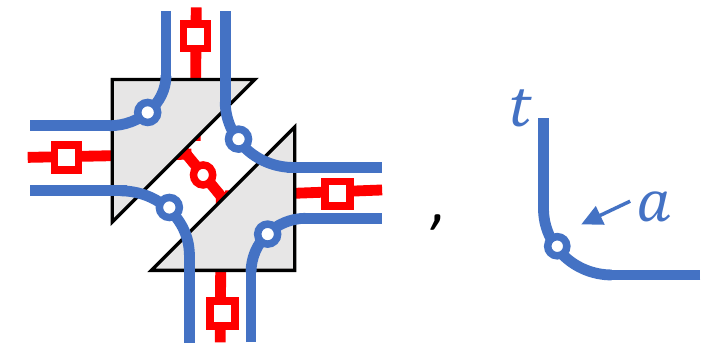}=T_{at}.
\end{equation}
where the new physical DOFs denoted by blue circles are $Z_2$ spins $\uparrow$ and $\downarrow$, and virtual loops $1$ and $\psi$ are mapped to $\uparrow$ while virtual loop $\sigma$ is mapped to $\downarrow$, so $T_{\uparrow1}=T_{\uparrow\psi}=T_{\downarrow\sigma}=1$, otherwise $T_{at}=0$. The symmetry enriched toric code has the $Z_2$ symmetry $\bigotimes_i X_i$ acting on the new physical DOFs represented by the blue dots, where $X_i$ is the Pauli $X$ operator. With the symmetry operator, the even sector can be viewed as $\langle\Psi_\text{SET}|\bigotimes_i X_i|\Psi_\text{SET}\rangle$, and the odd sector is $\langle\Psi_\text{SET}|\Psi_\text{SET}\rangle$. Since $\bigotimes_i X_i|\Psi_\text{SET}\rangle=|\Psi_\text{SET}\rangle$, the even and odd sector are equivalent.

\end{document}